\documentclass[twocolumn,showpacs,preprintnumbers,amsmath,amssymb]{revtex4}
\usepackage{epsfig}
\usepackage{dcolumn}
\usepackage{bm}

\newcommand{\remove}[1]{}
\newcommand{\dd}{\mathrm{d}}

\def\be{\begin{equation}}
\def\ee{\end{equation}}

\newcommand{\beq}{\begin{equation}}
\newcommand{\eeq}{\end{equation}}
\newcommand{\beqa}{\begin{eqnarray}}
\newcommand{\eeqa}{\end{eqnarray}}

\renewcommand{\pl}{\partial}
\newcommand{\lag}{\langle}
\newcommand{\rag}{\rangle}

\newcommand{\ii}{{\rm i}}

\newcommand{\vn}{{\bf n}}

\newcommand{\vv}{{\bf v}}

\newcommand{\vx}{{\bf x}}
\newcommand{\vk}{{\bf k}}
\newcommand{\vq}{{\bf q}}
\renewcommand{\vr}{{\bf r}}

\newcommand{\vF}{{\bf F}}

\newcommand{\tdelta}{{\tilde{\delta}}}

\newcommand{\tg}{{\tilde{g}}}

\newcommand{\tpsi}{{\tilde{\psi}}}

\newcommand{\tu}{{\tilde{u}}}
\newcommand{\tv}{{\tilde{v}}}
\newcommand{\tW}{{\tilde{W}}}

\newcommand{\trho}{{\tilde{\rho}}}
\newcommand{\tR}{{\tilde{R}}}
\newcommand{\tS}{{\tilde{S}}}
\newcommand{\talpha}{{\tilde{\alpha}}}

\newcommand{\cF}{{\cal F}}
\newcommand{\cG}{{\cal G}}
\newcommand{\cH}{{\cal H}}

\newcommand{\cO}{{\cal O}}
\newcommand{\cP}{{\cal P}}
\newcommand{\cS}{{\cal S}}
\newcommand{\cV}{{\cal V}}

\newcommand{\rhob}{\overline{\rho}}
\newcommand{\Om}{\Omega_{\rm m}}
\newcommand{\Ode}{\Omega_{\rm de}}
\newcommand{\wde}{w}

\newcommand{\bea}{\begin{array}}
\newcommand{\ea}{\end{array}}

\newcommand{\MPl}{M_{\rm Pl}}
\newcommand{\phib}{\overline{\varphi}}
\newcommand{\Rb}{\overline{R}}

\begin{document}

\title{Impact on the power spectrum of Screening in Modified Gravity Scenarios}

\author{Philippe Brax}
\affiliation{Institut de Physique Th\'eorique,\\
CEA, IPhT, F-91191 Gif-sur-Yvette, C\'edex, France\\
CNRS, URA 2306, F-91191 Gif-sur-Yvette, C\'edex, France}
\author{Patrick Valageas}
\affiliation{Institut de Physique Th\'eorique,\\
CEA, IPhT, F-91191 Gif-sur-Yvette, C\'edex, France\\
CNRS, URA 2306, F-91191 Gif-sur-Yvette, C\'edex, France}
\vspace{.2 cm}

\date{\today}
\vspace{.2 cm}

\begin{abstract}
We study the effects of screened modified gravity of the $f(R)$, dilaton and symmetron types on structure formation, from the quasilinear to the nonlinear regime,
using semi-analytical methods. For such models, where the range of the new scalar field is typically within the Mpc range and below in the cosmological context, nonlinear techniques are required to understand the deviations of the power spectrum of the matter density contrast compared to the $\Lambda$-CDM template. This is nowadays commonly tackled using extensive N-body simulations. Here we present new results combining  exact perturbation theory at the one-loop level (and a partial resummation of the perturbative series) with a halo model. The former allows one to extend the linear perturbative analysis up to $k\lesssim  0.15 h{\rm Mpc}^{-1}$ at the perturbative level while the latter leads to a reasonable, up to a few percent, agreement with numerical simulations for $k\lesssim 3 h{\rm Mpc}^{-1}$ for large-curvature $f(R)$ models, and $k\lesssim 1 h{\rm Mpc}^{-1}$ for dilatons and symmetrons, at $z=0$. We also discuss
how the behaviors of the perturbative expansions and of the spherical collapse differ
for $f(R)$, dilaton, and symmetron models.

\end{abstract}

 \vskip2pc

\maketitle

\section{Introduction}
\label{Introduction}

The acceleration of the expansion of the Universe\cite{Perlmutter:1998np,Riess:1998cb} can be accommodated with General Relativity (GR) by introducing a finely tuned cosmological constant. It could also be
that the laws of gravity on sufficiently large scales are not well understood and need to be reexamined (see for instance \cite{Khoury:2010xi}). Scalar field models with a coupling to  matter density lead to the existence of a fifth force depending on the gradient of the scalar field profile in the vicinity of dense bodies. Of course, the existence of long-range scalar forces in the Solar System is tightly constrained by the Cassini probe (fifth-force test)\cite{Bertotti:2003rm} and the Lunar Ranging experiment (test of the strong equivalence principle)\cite{Williams:2004qba}. All in all, such a scalar force in the Solar System must be highly suppressed, implying that screening mechanisms must be introduced to guarantee the compatibility of the Solar System tests with long-range scalar forces on cosmological scales.
Three types of screening have been unraveled: the chameleon mechanism\cite{Khoury:2003aq,Khoury:2003rn,Brax:2004qh} where the scalar field mass grows with the matter density and Yukawa suppresses the fifth force in dense environments, the Damour-Polyakov property\cite{Damour:1994zq} where the coupling to matter itself is attracted towards zero in dense matter, and finally the Vainshtein mechanism\cite{Vainshtein:1972sx} whereby the normalized scalar fluctuations have a reduced coupling in dense environments. The first two are describable as scalar-tensor theories with a nonlinear potential $V(\varphi)$ and a coupling function $A(\varphi)$. In the presence of matter, the effective potential has a minimum where the mass and/or the coupling of the scalar field are large and/or small enough. The Vainshtein mechanism is characteristic of models with nonlinear kinetic terms. In the rest of this paper, we will solely focus on scalar-tensor models with the chameleon or the Damour-Polyakov property, the latter for the dilatons\cite{Brax:2010gi} and symmetrons\cite{Pietroni:2005pv,Olive:2007aj,Hinterbichler:2010es}. Interestingly, a seemingly unrelated model, the $f(R)$ theories\cite{Carroll:2003wy}, are viable when possessing the chameleon property\cite{Hu:2007nk}. We will describe the behavior of the large-curvature $f(R)$ models too.

Models with chameleon or  Damour-Polyakov signatures can all be described by a tomographic method\cite{Brax:2011aw} whereby the potential and the coupling function of $\varphi$ can be reconstructed from
the time behavior of the mass function $m(a)$ and the coupling $\beta (a)$ at the minimum of the effective potential for $\varphi$, for the matter density $\rhob(a)$, as a function of the cosmological scale factor $a$. This is a very physical way of defining models as it determines properties affecting the growth of structures. Indeed, the scalar force in the cosmological background modifies the geodesics of the Cold Dark Matter (CDM) particles implying a change of the growth rate of linear perturbations\cite{Brax:2004px}. This modification depends only on the range of the fifth force cosmologically, $\lambda (a)= m^{-1} (a)$, and on the coupling of matter particles to the scalar field, $\beta (a)$. The tomographic mapping allows one to relate these features of linear perturbation theory to the full nonlinear Lagrangian description of the model, which is determined by the shape of $V(\varphi)$ and $\beta(\varphi)$. Explicit examples are known for the dilatons, symmetrons and large-curvature $f(R)$ models\cite{Brax:2012gr}. Here we shall consider  $f(R)$ models in their Jordan frame, although an Einstein-frame treatment would be just as adequate.
For all these models, we perform a cosmological perturbative analysis, using the $m(a)-\beta (a)$ parametrization, which can be pursued to all orders, but we stop at the third order to take into account effects up to  one loop in the power spectrum. We also tackle nonperturbative properties of the screened models with the same parametrization in a spherical collapse approach. This $m(a)-\beta (a)$ way of defining screened models goes beyond the effective descriptions of linear perturbations which has been recently developed by several groups\cite{Baker:2011jy,Gubitosi:2012hu,Battye:2013er,Gleyzes:2013ooa} in as much as higher orders in perturbation theory and nonperturbative features are readily available and calculable as exemplified in this paper.

The analysis of these models in the nonlinear regime of structure formation has been recently tackled using the tomographic mapping and large N-body simulations with the ECOSMOG code for dilatons and symmetrons\cite{Brax:2012nk}. Earlier numerical simulations of $f(R)$ models are also available\cite{Oyaizu2008}. It turns out that the typical scale for the scalar range cosmologically must be lower than 1 Mpc\cite{Brax:2011aw,Wang:2012kj}, which is already in the mildly nonlinear regime, implying that such simulations are indeed necessary. They all reveal features on scales up to a few Mpc's which can be qualitatively understood as follows. On large and linear scales, hardly any modification of gravity occurs while on very small scales, where modified gravity is screened, the usual GR behavior is recovered. In between these two limits, the power spectrum of the matter density contrast is largely model dependent. For large-curvature $f(R)$ models, the deviation from the $\Lambda$-CDM result  has a ``bump'' at a scale corresponding to the range of the scalar force. For dilatons and symmetrons, there is a flattening of the discrepancy over the same range of nonlinear scales. In general, the screening of $f(R)$ models is less strong than that of the dilatons and symmetrons.

In this paper, we combine perturbation theory and a partial resummation of the perturbative series (more precisely, a regularization of the perturbative expansion) with a halo model in order to probe the same scales as the N-body simulations.
At the perturbative level, we expand the scalar contribution to the effective Newtonian potential up to third order in the nonlinear density contrast. Whereas the first order
enhances the gravitational clustering (for the models we consider here) and is
similar to a scale-dependent Newton constant, the second and third orders
are the first signs of the chameleon or screening mechanisms that make these theories
consistent with Solar-System constraints.
This allows us to calculate exactly the one-loop power spectrum in the presence of a scalar modification of gravity.
This provides a reasonable agreement with the simulations up to quasilinear scales around $0.15 h{\rm Mpc}^{-1}$ at $z=0$.
We discuss how the behavior of this perturbative expansion differs for the $f(R)$,
dilaton, and symmetron models.
We also give the recipe to extend this treatment up to an arbitrary number of loops, but this approach becomes computationally increasingly complex.

Next, we go beyond the one-loop approximation by using an improved halo model\cite{Valageas2013}.
This includes both a Lagrangian-space regularization of the one-loop expansion,
which automatically generates a partial account of higher orders (so that the probability distribution of relative particle displacements
is well behaved, i.e., positive and well normalized, which is not
the case in the sharp truncation associated with the standard one-loop result),
and an account of nonperturbative terms, associated with pancake formation
(large-scale structures on mildly nonlinear scales) and halo formation (which
governs the high-$k$ behavior).

This necessitates describing the spherical collapse in $f(R)$ models and scalar-tensor modifications of gravity\cite{Hu:2007nk,Brax:2010tj,Li:2011qda}. This is done by assuming a profile ansatz (identical to the typical profile
in the linear regime for Gaussian fields)
for the calculation of the scalar force on a spherical shell and using the $m(a)-\beta (a)$ parametrization of dilatons and symmetrons.
We find that the $f(R)$ models differ from the dilatons and symmetrons in as much as the linear density threshold, which becomes mass dependent here due to the scale dependence of the scalar force, converges to the linear weak-field limit for small masses, whereas it converges to the $\Lambda$-CDM value
for dilatons and symmetrons.

Using this combined approach for the matter density power spectrum,
we find a good agreement with numerical simulations up to
$3h {\rm Mpc}^{-1}$ for $f(R)$ models, and $1h{\rm Mpc}^{-1}$ for dilatons and symmetrons. For some cases, the difference with N-body simulations is up to a factor of 2: this occurs for certain symmetron models where the perturbative expansion does not converge rapidly enough (more precisely, the screening mechanism has not converged
on weakly nonlinear scales at third order in the scalar and density fields)
or the scalar field quickly relaxes to a singular value in spherically symmetric
large overdensities.
In the other cases, where the third-order terms in the perturbative expansion are smaller than the second order ones and the scalar field remains in a regular (analytic) domain,
 we find a good agreement.

In summary, our semi-analytical treatment of the power spectrum of screened models provides a much faster description of the nonlinear power spectrum than N-body simulations.
As such it can be used as a sieve to distinguish interesting models where deviations from $\Lambda$-CDM could be large enough to be within reach of future large galaxy surveys from
less constrained ones which are out of reach. Indeed, we have found that the semi-analytical results always reproduce the correct order of magnitude given by N-body simulations and even correspond to the simulated results up to a few percent
on large scales, when the perturbative series of the screened models converges fast enough. Increasing the accuracy of our method would certainly require us to better understand the shape of the halos in modified gravity and the halo mass function. This is left for future work.

In section~\ref{Equations of motion}, we recall the main features of modified gravity models that are relevant for large-scale structures and briefly discuss the
quasistatic approximation.
In section III, we analyse the perturbative series and the power spectrum in the single-stream approximation at the one-loop order. In section IV, we consider the spherical collapse of $f(R)$ and scalar-tensor models. In section V, we define the halo model and give our results on the power spectrum of $f(R)$, dilaton and symmetron models. Finally we conclude before an appendix where the construction of the halo model and its combination
with perturbation theory is recalled.

\section{Modified Gravity Models}
\label{Equations of motion}

We describe here the two classes of models that we consider in this article:
i) ``$f(R)$'' theories, where the Einstein-Hilbert action is complemented by a term which
depends on the Ricci curvature, and
ii) scalar-tensor theories, where an additional scalar field $\varphi$ is conformally
coupled to the ordinary matter.

\subsection{$f(R)$ models}
\label{fR-models}

The first class of models that we consider in this paper corresponds to ``$f(R)$ theories'',
where the Einstein-Hilbert action is supplemented by a term that depends on the Ricci scalar,
which we choose to have the form \cite{Carroll:2003wy,Sotiriou:2008rp,Hu:2007nk}
\beq
f(R) = - 2 \Lambda - \frac{f_{R_0} c^2}{n} \frac{R_0^{n+1}}{R^n} .
\label{fR-def}
\eeq
This involves two parameters: the normalization $f_{R_0}$ and the exponent $n>0$.
This is also the large-curvature regime of the model proposed in \cite{Hu:2007nk},
which is consistent with Solar-System and Milky-Way constraints thanks to the
chameleon mechanism, for $|f_{R_0}| \lessapprox 10^{-5}$.
This modification of the Einstein-Hilbert action leads to a modified Poisson equation,
which reads as \cite{Oyaizu2008}
\beq
\nabla^2 \Psi = \frac{16 \pi {\cal G}}{3} a^2 \, \delta \rho - \frac{a^2}{6} \delta R ,
\label{Poisson-fR}
\eeq
where $\Psi$ is the modified Newtonian potential, whose gradient governs the motion of
particles, and $\delta \rho$ is the matter density fluctuation.
Here and in the following, we use comoving coordinates, $\vx=\vr/a$ (and
$\nabla=\nabla_{\vx}$), where $a$ is the cosmological scale factor, while we use the
physical matter density $\rho$ (its mean $\bar\rho$ decreases with time as $a^{-3}$).

The fluctuation of the Ricci scalar, $\delta R=R - \Rb$, is determined by the constraint
\cite{Oyaizu2008}
\beq
\nabla^2 \delta f_R = \frac{a^2}{3} \; \left[ \delta R - 8 \pi {\cal G} \, \delta \rho \right]
\label{fR-delta} ,
\eeq
where we have introduced:
\beq
\delta f_R = f_R(R) - f_R(\Rb) ,
\label{dfR}
\eeq
and
\beq
 f_R(R) = \frac{\dd f}{\dd R} = f_{R_0} c^2 \frac{R_0^{n+1}}{R^{n+1}} .
 \label{fR}
\eeq
Here we have used the quasistatic approximation and have discarded a negligible $f_R R$ term.
Then, Eqs.(\ref{Poisson-fR}) and (\ref{fR-delta}) govern the dynamics of the system.

Even though Eq.(\ref{fR-delta}) is nonlinear in $\delta R$, we can check that
it is self-averaging, in the sense that on large scales we recover $\delta R \rightarrow 0$.
This is due to the fact that nonlinearities enter through the Laplacian $\nabla^2$
on the left-hand side of Eq.(\ref{fR-delta}). Integrating over a large volume $\cV$ of
boundary $\cS$ and using Ostrogradsky's theorem gives
\beq
\int_{\cV} \frac{\dd V}{\cV} \, \delta R = \frac{3}{a^2} \int_{\cS} \frac{\dd S}{\cV}
(\vn \cdot \nabla \delta f_R) + 8\pi\cG \frac{\delta M}{\cV} ,
\eeq
where $\vn$ is the  normal unit vector to $\cS$.
Thanks to the conservation of matter and the finite amplitude of particle motions with respect
to the Hubble flow, the last term $\delta M/\cV$ goes to zero for $\cV\rightarrow\infty$,
while the surface term also goes to zero (typically faster than the inverse of the radius
of the volume $\cV$).
Therefore, the average over a large volume of $\delta R$ goes to zero, which means
that there is no cumulative contribution to the potential (\ref{Poisson-fR}) on large scales
due to small-scale nonlinearities, and we recover the background Hubble flow on large
scales\footnote{In contrast, if nonlinearities such as $\delta R^2$, without a Laplacian
prefactor, entered Eq.(\ref{fR-delta}), this would yield contributions such as
$\lag (\delta\rho)^2\rag$ for large volume averages which do not vanish, whence
a component $\propto |\vx|^2$ to the potential (\ref{Poisson-fR}) and a modification
of the Hubble expansion rate itself, that is, a backreaction on large scales from the
cumulative effect of small-scale nonlinearities.}.

In a perturbative approach to the formation of large-scale structures,
we expand in the fluctuations $\delta R$ and $\delta\rho$ with respect to the
background, which gives rise to successive derivatives of $f_R(R)$. Thus, we define the
quantities
\beq
n \geq 1 : \;\; \kappa_n(a) = H^{2n-2} \; \frac{\dd^n f_R}{\dd R^n}(\Rb) ,
\label{kappa-fR}
\eeq
(so that all coefficients $\kappa_n$ have the dimension of a length squared),
while the background Ricci scalar is given by
\beq
\Rb(a) =  3 H_0^2 \; \left[ \Omega_{\rm m 0} \, a^{-3} + 4 \, \Omega_{\Lambda 0} \right].
\eeq
To be consistent with previous works which focused on linear theory, we also define
\beq
m(a) = \frac{1}{\sqrt{3\kappa_1}} .
\label{m-fR}
\eeq

In this paper, we perform numerical computations for the three $f(R)$ theories of the
power-law form (\ref{fR-def}) with $n=1$ and $f_{R_0}=-10^{-4}$, $-10^{-5}$, and
$-10^{-6}$, to compare our results with numerical simulations from
\cite{Oyaizu2008}.

\subsection{Scalar field models}
\label{scalar field}

\subsubsection{Klein-Gordon and modified Poisson equations}
\label{K-G-P}

We now turn to scalar-tensor theories, where
the action  defining the system in the Einstein frame has the general form \cite{Will:2001mx}
\beqa
S & = & \int \dd^4 x \; \sqrt{-g} \left[ \frac{\MPl^2}{2} R - \frac{1}{2}
(\nabla \varphi)^2 - V(\varphi) \right] \nonumber \\
&& + \int \dd^4 x \; \sqrt{-\tg} \, {\cal L}_{\rm m}(\psi^{(i)}_{\rm m},\tg_{\mu\nu}) ,
\label{S-def}
\eeqa
where $g$ is the determinant of the metric tensor $g_{\mu\nu}$ and
$\psi^{(i)}_{\rm m}$ are various matter fields. The additional scalar field $\varphi$ is explicitly
coupled to matter through the Jordan-frame metric $\tg_{\mu\nu}$, which is given by the
conformal rescaling
\beq
\tg_{\mu\nu} = A^2(\varphi) \, g_{\mu\nu} ,
\eeq
and $\tg$ is its determinant.

This coupling implies that matter particles of mass $m$ are sensitive to a ``fifth force'',
$\vF=  - m c^2 \nabla \ln A$ \cite{Khoury:2003aq}. This can be written as an additional contribution $\Psi_{\rm A}$
to the Newtonian term $\Psi_{\rm N}$ in the total gravitational potential,
\beq
\Psi = \Psi_{\rm N} + \Psi_{\rm A} ,
\label{Psi-N+A}
\eeq
with
\beq
\frac{1}{a^2} \nabla^2 \Psi_{\rm N} = 4 \pi {\cal G} \, \delta \rho , \hspace{1cm}
\Psi_{\rm A} = c^2 (A - \bar{A}) ,
\label{Psi-N-A}
\eeq
were we assumed $A(\varphi) \simeq 1$, as required by experimental constraints on the
variation of fermion masses.

On the other hand, the coupling also means that the equation of motion of the scalar field
explicitly depends on the matter environment (which also enables screening mechanisms
to appear), and in the quasistatic limit the Klein-Gordon equation reads as
\beq
\frac{c^2}{a^2} \nabla^2 \varphi = \frac{\dd V}{\dd\varphi} + \rho \;
\frac{\dd A}{\dd\varphi} .
\label{Klein-Gordon}
\eeq

Thus, Eqs.(\ref{Psi-N+A}) and (\ref{Klein-Gordon}) play the same role as
Eqs.(\ref{Poisson-fR}) and (\ref{fR-delta}) encountered in $f(R)$ theories and fully
determine the dynamics of the system and the formation of large-scale structures.
From the point of view of the matter dynamics (e.g., after integration over the field $\varphi$),
this is indeed a ``modified gravity'' theory because the contribution $\Psi_{\rm A}$ appears as
a modification to the Poisson equation.
Finally, each scalar field model will be specified by the choice of the scalar field potential
$V(\varphi)$ and of the coupling function $A(\varphi)$.

Again, we can check that small-scale nonlinearities are self-averaging. Indeed, assuming
for instance periodic boundary conditions for the matter density $\rho$, we can look for
periodic solutions to the nonlinear Klein-Gordon equation (\ref{Klein-Gordon}).
Then, the potential $\Psi_{\rm A}$ is periodic and does not show a cumulative growth
with $|\vx|$, so that we recover the background Hubble flow on large scales.

\subsubsection{Einstein frame}
\label{Einstein-frame}

We will exclusively work in the Einstein frame where matter particles feel both the effect of the metric $g_{\mu\nu}$ appearing in the Einstein-Hilbert  action
and the scalar field. In the Jordan frame, where matter particles couple directly to the metric $\tilde g_{\mu\nu}$, the matter density  $\rho_J$ is related to the conserved matter density $\rho$ by
\beq
\rho_J=A^{-3}(\varphi) \rho ,
\eeq
implying that at linear order the perturbed energy density contrasts are related by
\beq
\delta_J= \delta -3 \beta (\phib) \frac{\delta\varphi}{M_{\rm Pl}}
\eeq
where $\beta(\phib)= M_{\rm Pl} \dd A/\dd \phib \simeq M_{\rm Pl} \frac{d\ln A}{d\phib}$ is the coupling to matter (we always have $A \simeq 1$).
To leading order the perturbed Klein-Gordon equation gives
\beq
\frac{\delta\varphi}{M_{\rm Pl}}= - \frac{3\beta \Om a^2 H^2}{c^2(a^2 m^2+k^2)} \; \delta
\eeq
where $m^2(\phib)$ is the mass of the scalar field in the background $\phib$. This
implies that
\beq
\delta_J=  \left[1+ \frac{9 \beta^2 \Om a^2 H^2}{c^2(a^2 m^2+k^2)} \right] \; \delta .
\eeq
For couplings of order one, the energy density perturbation in the Jordan frame differs from $\delta$ by a term of order $H^2/(c^2m^2)$ at most. The background
 $\phib$ is stable provided that $m^2(\phib)\gg H^2/c^2$ \cite{Brax:2012gr} and Solar System tests of gravity imply that $m(\phib) \ge 10^3 H/c$ \cite{Brax:2011aw}. This leads to a very small correction term of order less than $10^{-6}$,
which is negligible when considering the effects of modified gravity at the percent level and we shall neglect it in the following. Hence we will always calculate perturbations in the
Einstein frame and we do not need to distinguish between Jordan-frame and Einstein-frame densities.

\subsubsection{Derived functions and tomography}
\label{Tomography}

In a perturbative approach, we expand in powers of the fluctuations $\delta\varphi$
and $\delta\rho$ of the scalar field and of the matter density  field, with respect to the
uniform background $(\phib,\rhob)$.
More generally,
we can expand the potential $V(\varphi)$ and the coupling function $A(\varphi)$,
and we are led to define the successive derivatives
\beq
n \geq 1 : \;\;\; \beta_n(\phib) = \MPl^n \, \frac{\dd^n A}{\dd \varphi^n}(\phib) ,
\label{beta-n-def-scalar}
\eeq
\beqa
n \geq 2 : \;\;\; \kappa_n(\phib,\rhob) & = & \frac{\MPl^{n-2}}{c^2}
\frac{\pl^n V_{\rm eff}}{\pl \varphi^n} \nonumber \\
&& \hspace{-1.5cm} = \frac{\MPl^{n-2}}{c^2} \left[  \frac{\dd^n V}{\dd\varphi^n}(\phib)
+ \rhob \; \frac{\dd^n A}{\dd\varphi^n}(\phib) \right] ,
\label{kappa-n-def-scalar}
\eeqa
where $V_{\rm eff}=V+\rhob (A-1)$ is the effective potential which enters the Klein-Gordon
equation (\ref{Klein-Gordon}).
Thus, the coefficients $\beta_n$ are dimensionless while the coefficients $\kappa_n$ have the
dimension of a wave number squared.
To use consistent notations with previous works which focused on linear theory,
we also define
\beq
\beta = \beta_1 \;\;\; \mbox{and} \;\;\; m^2 = \kappa_2 .
\label{beta1-kappa2}
\eeq

Following \cite{Brax2012a,Brax2012b},
it is convenient to write these functions in terms of the scale factor $a(t)$,
by defining $\beta_n(a) \equiv \beta_n[\phib(a)]$ and
$\kappa_n(a) \equiv \kappa_n[\phib(a),\rhob(a)]$
(we use the same notations, to avoid introducing too many functions).
Through the two functions $\beta(a)$ and $m(a)$ it is possible
(at least in some regular domain) to reconstruct the two functions $V(\varphi)$ and
$A(\varphi)$, so that one can also define each scalar field model through the former
functions $\beta(a)$ and $m(a)$. This allows one to build for instance models which satisfy
a specific pattern for the growth of large-scale structures at linear order.
In any case, as in \cite{Brax:2011aw}, we note that Eq.(\ref{Klein-Gordon}) reads at zeroth order (i.e., for the
uniform background) as
\beq
\frac{\dd V}{\dd\phib} + \rhob \; \frac{\dd A}{\dd\phib} = 0 ,
\label{KG-0}
\eeq
and taking the derivative with respect to the scale factor $a$ yields
\beq
\left( \frac{\dd^2 V}{\dd\phib^2} + \rhob \; \frac{\dd^2 A}{\dd\phib^2} \right) \;
\frac{\dd\phib}{\dd a} + \frac{\dd\rhob}{\dd a} \; \frac{\dd A}{\dd\phib} = 0 .
\eeq
Using $\rhob \propto a^{-3}$ and Eqs.(\ref{beta-n-def-scalar})-(\ref{beta1-kappa2}) we
obtain
\beq
\frac{\dd\phib}{\dd a} = \frac{3\beta\rhob}{c^2\MPl m^2 a}  .
\label{phi-a}
\eeq
Then, we easily obtain high-order derivatives $\beta_n$ and $\kappa_n$ by recursion,
as
\beq
\beta_{n+1}(a) = \MPl \, \frac{\dd\beta_n}{\dd\phib} =
\frac{c^2 \MPl^2 m^2 a}{3\beta\rhob} \, \frac{\dd\beta_n}{\dd a} ,
\label{beta-n+1}
\eeq
and
\beq
\kappa_{n+1}(a) = \frac{c^2 \MPl^2 m^2 a}{3\beta\rhob} \, \frac{\dd\kappa_n}{\dd a}
+ \frac{m^2\beta_n}{\beta} .
\label{kappa-n+1}
\eeq

\paragraph{Generalized dilaton models}
\label{dilaton}

The original dilaton model corresponds to the coupling function \cite{Damour:1994zq}
\beq
A(\varphi) = 1 + \frac{1}{2} \; \frac{A_2}{\MPl^2} \; (\varphi-\varphi_*)^2
\label{A-dilaton}
\eeq
and the potential $V(\varphi) = V_0 \; \exp(-\varphi/\MPl)$. This can be generalized \cite{Brax:2012nk} by keeping
the coupling function as in (\ref{A-dilaton}) but specifying the mass $m(a)$ instead of the
potential $V(\varphi)$. Thus, the model is determined by the parameters
$\{ m_0, r, A_2, \beta_0 \}$, with
\beq
m(a) = m_0 \; a^{-r} ,
\eeq
and $\beta_0$ is the value of $\beta(a)$ today ($a_0=1$). Then, using Eqs.(\ref{KG-0}) and
(\ref{phi-a}) we obtain
\beq
\beta(a) = \beta_0 \; e^{-s (a^{2 r-3}-1)/(3-2 r)} , \;\; \mbox{with} \;\;
s= \frac{9 A_2 \Omega_{\rm m 0} H_0^2}{c^2 m_0^2} .
\eeq
Using Eqs.(\ref{beta-n+1})-(\ref{kappa-n+1}), the derivatives needed for second-order
computations read as
\beq
\beta_2 = A_2, \;\;\;
\kappa_3 = \frac{m^2 A_2}{\beta} \left( 1 - \frac{2 r}{s} \, a^{3-2 r} \right) ,
\eeq
and at third order,
\beq
\beta_3 = 0 , \;\;\;
\kappa_4 = - \frac{m^2 A_2^2}{\beta^2} \left( 1+ \frac{2r (3-4r)}{s^2} a^{6-4r} \right) .
\eeq
To compare our results with the numerical simulations from \cite{Brax2012}, we consider
the same set of parameters, which we recall in Table~\ref{table-dilaton}.

\begin{table}[h!]
\caption{List of dilaton models considered in this paper. They are the same as
in \cite{Brax2012}.}
\begin{center}
\begin{tabular}{|c||c|c|c|c|}
\hline
model name & \; $m_0 [h/\mbox{Mpc}]$ \; & \;\;\; $r$ \;\;\; & \;\; $\beta_0$ \;\; & \;\;\; $s$ \;\;\; \\
\hline\hline
A1 & 0.334 & 1 & 0.5 & 0.6 \\
\hline
A2 & 0.334 & 1 & 0.5 & 0.24 \\
\hline
A3 & 0.334 & 1 & 0.5 & 0.12 \\
\hline\hline
B1 & 0.334 & 1 & 0.25 & 0.24 \\
\hline
B3 & 0.334 & 1 & 0.75 & 0.24 \\
\hline
B4 & 0.334 & 1 & 1 & 0.24 \\
\hline\hline
C1 & 0.334 & 1.33 & 0.5 & 0.24 \\
\hline
C3 & 0.334 & 0.67 & 0.5 & 0.24 \\
\hline
C4 & 0.334 & 0.4 & 0.5 & 0.24 \\
\hline\hline
D1 & 0.667 & 1 & 0.5 & 0.06 \\
\hline
D3 & 0.167 & 1 & 0.5 & 0.96 \\
\hline
D4 & 0.111 & 1 & 0.5 & 2.16 \\
\hline
\end{tabular}
\end{center}
\label{table-dilaton}
\end{table}

\paragraph{Generalized symmetron models}
\label{symmetron}

The symmetron model \cite{Pietroni:2005pv,Olive:2007aj,Hinterbichler:2010es} corresponds to a phase transition from a single-well to a double-well
effective potential, so that the modifications to gravity only appear after a finite time
(with respect to the background), at a scale factor $a_s$ where the field $\phib$ moves
from the initial single minimum $\phib=0$, in high-density environments with
$\rho>\rhob(a_s)$,  to one of the two new minima $\pm \varphi_c(\rhob)$ which appear
in low-density environments with $\rho<\rhob(a_s)$.
Following \cite{Brax2012}, we consider a generalization \cite{Brax:2012nk} defined by the functions
\beqa
\beta(a) & = & \beta_0 \left[ 1 - \left(\frac{a_s}{a}\right)^3 \right]^{\hat{n}} ,
\label{beta-symmetron} \\
m(a) & = & m_0 \left[ 1 - \left(\frac{a_s}{a}\right)^3 \right]^{\hat{m}} ,
\label{m-symmetron}
\eeqa
for $a>a_s$, and $\beta=0$ and $m=0$ for $a \leq a_s$.
Thus, the model is now defined by the parameters $\{ \beta_0,\hat{n},m_0,\hat{m} \}$,
with $\hat{n}>0$, $\hat{m}>0$, and $\hat{n}-2\hat{m}+1>0$ (which arises from the
requirement that $\phib(a_s)$ be finite).
Using again Eqs.(\ref{beta-n+1})-(\ref{kappa-n+1}), we obtain
\beq
\beta_2 = \frac{\hat{n} c^2 a_s^3 m^2}{3\Omega_{\rm m 0}H_0^2}
\left[ 1 - \left(\frac{a_s}{a}\right)^3 \right]^{-1} ,
\eeq
\beq
\kappa_3 =  \frac{\hat{n}+2\hat{m}}{\hat{n}} \, \frac{m^2 \beta_2}{\beta} ,
\eeq
and
\beq
\beta_3 = \frac{2\hat{m}-1}{\hat{n}} \, \frac{\beta_2^2}{\beta} ,
\eeq
\beq
\kappa_4 =  \frac{8\hat{m}^2-\hat{n} (2+\hat{n})+\hat{m}(4\hat{n}-2)}{\hat{n}^2} \,
\frac{m^2 \beta_2^2}{\beta^2} .
\eeq
To compare our results with the numerical simulations from \cite{Brax2012}, we consider
the same set of parameters, which we recall in Table~\ref{table-symmetron}.

\begin{table}[h!]
\caption{List of symmetron models considered in this paper. They are the same as
in \cite{Brax2012}.}
\begin{center}
\begin{tabular}{|c||c|c|c|c|c|}
\hline
model name & \;\; $a_s$ \;\; & \, $m_0 [h/\mbox{Mpc}]$ \, & \;\;\; $\hat{m}$ \;\;\; & \; $\beta_0$ \; & \;\;\; $\hat{n}$ \;\;\; \\
\hline\hline
A1 & 0.5 & 0.033 & 0.5 & 1 & 0.5 \\
\hline
A2 & 0.5 & 0.033 & 0.5 & 1 & 0.25 \\
\hline
A3 & 0.5 & 0.017 & 0.5 & 1 & 0.25 \\
\hline
A4 & 0.5 & 0.017 & 1 & 1 & 1.5 \\
\hline
\hline
B1 & 0.33 & 0.033 & 0.5 & 1 & 0.5 \\
\hline
B2 & 0.33 & 0.033 & 0.5 & 1 & 0.25 \\
\hline
B3 & 0.33 & 0.017 & 0.5 & 1 & 0.5 \\
\hline
B4 & 0.33 & 0.017 & 1 & 1 & 1.5 \\
\hline
\end{tabular}
\end{center}
\label{table-symmetron}
\end{table}

\subsubsection{Quasistatic approximation}
\label{Quasi-static}

As in most published works, throughout this article we use the quasistatic approximations
(\ref{fR-delta}) and (\ref{Klein-Gordon}). However, for the symmetron models described
above, the singularity of the functions $\beta(a)$ and $m(a)$ of
Eqs.(\ref{beta-symmetron})-(\ref{m-symmetron}) at $a_s$ could be expected to give rise to
significant transients. We investigate here the magnitude of this effect, at the linear level over
the fluctuation $\delta\varphi$ \cite{Brax:2011pk}, due to the singularity of $\beta(a)$.
Without the quasistatic approximation, the Klein-Gordon equation (\ref{Klein-Gordon})
becomes
\beq
\ddot{\varphi} + 3 H \dot{\varphi} - \frac{c^2}{a^2} \nabla^2 \varphi =
- \frac{\dd V}{\dd\varphi} - \rho \frac{\dd A}{\dd\varphi} .
\label{KG-t1}
\eeq
Then, the equation of motion for the field fluctuation, $\delta\varphi=\varphi-\phib$, reads
at linear order in $\delta\varphi$ and $\delta\rho$ as
\beq
\delta\ddot{\varphi} + 3 H \delta\dot{\varphi} - \frac{c^2}{a^2} \nabla^2 \delta\varphi =
- c^2 m^2 \delta\varphi - \frac{\beta}{\MPl} \delta\rho ,
\label{KG-t2}
\eeq
where $\delta\rho=\rho-\rhob$.
Here we have absorbed possible transients of the background $\phib$, with respect to its
quasistatic approximation, into a redefinition of the derivatives $m^2$ and $\beta$.
Introducing the rescaled field $v=a \, \delta\varphi$ and the conformal time
$\tau=\int \dd t/a$, the Klein-Gordon equation (\ref{KG-t2}) becomes
\beq
v'' - \frac{a''}{a} v - c^2 \nabla^2 v  = - c^2 m^2 a^2 v - \frac{\beta a^3 \delta\rho}{\MPl} ,
\label{KG-v1}
\eeq
where primes denote derivatives with respect to $\tau$.
This reads in Fourier space as
\beq
\tv'' + \omega^2(\tau) \,  \tv = \tS(\tau) ,
\label{KG-v2}
\eeq
with
\beq
\omega^2 = k^2 c^2 + c^2 a^2 m^2 - \frac{a''}{a} , \;\;\;
\tS = - \frac{\beta a^3}{\MPl} \delta\trho .
\eeq
The quasistatic approximation is recovered by neglecting the time derivatives, which yields
$\tv=\tS/\omega^2$ (at this linear order in $\delta\varphi$).
Here we only investigate the impact of sudden changes or singularities of the coupling
function $\beta(a)$, whence of the source $\tS$. For this purpose, we can go beyond
the quasistatic approximation by keeping the term $\tv''$ in the linearized
Klein-Gordon equation (\ref{KG-v2}) but neglecting the time dependence of $\omega^2$
(this applies to cases where $\beta(\tau)$ and $\tv(\tau)$ vary on a shorter time scale
than the scale factor $a(\tau)$ and the mass $m^2(a)$). This yields the approximation
\beq
\tv(\tau) \simeq \int_0^{\tau} \dd\tau' \, \tS(\tau') \,
\frac{\sin[\omega (\tau-\tau')]}{\omega} ,
\eeq
and an integration by parts gives
\beq
\tv(\tau) \simeq \frac{\tS}{\omega^2} - \int_0^{\tau} \dd\tau' \, \tS'(\tau') \,
\frac{\cos[\omega(\tau-\tau')]}{\omega^2} ,
\label{v-S-1}
\eeq
where we assumed that the source decays for $\tau\rightarrow 0$.
The first term in Eq.(\ref{v-S-1}) is the quasistatic approximation, and further
integrations by parts yield terms of increasing order in $1/\omega$.
However, for singular coupling functions $\beta(\tau)$ this stops at the order where
the integral over the $n$-derivative $\tS^{(n)}$ becomes divergent.
In particular, for singular coupling functions of the form (\ref{beta-symmetron})
we must stop at Eq.(\ref{v-S-1}) if $\hat{n}<1$. Let us consider the case
\beq
\tau > \tau_s : \;\; \tS(\tau) = S_s \, (\tau-\tau_s)^{\hat{n}} , \;\;\; \mbox{with} \;\;\; \hat{n} < 1 ,
\eeq
and $\tS=0$ for $\tau<\tau_s$. Then, Eq.(\ref{v-S-1}) gives at late times
\beq
\tv(\tau) \simeq \frac{\tS(\tau)}{\omega^2} - \frac{S_s \Gamma(\hat{n}+1)}{\omega^{2+\hat{n}}} \,
\cos[\hat{n}\frac{\pi}{2} + \omega (\tau_s-\tau)] ,
\eeq
and the quasistatic approximation is valid if $(\omega \tau)^{\hat{n}} \gg 1$.
The modified gravity effects that we consider in this paper appear on scales where
$k \gtrapprox a m$, whence $\omega \sim k c$, and a ten percent accuracy on
$\delta\varphi$ requires
\beq
k > \frac{10^{1/\hat{n}}}{c\tau} \sim 3 \times 10^{-4+1/\hat{n}} \, h \mbox{Mpc}^{-1} .
\eeq
Thus, we obtain $k > 0.03 h$Mpc$^{-1}$ for $\hat{n}=0.5$ and
$k > 3 h$Mpc$^{-1}$ for $\hat{n}=0.25$.
Therefore, on the scales of interest for modified gravity probes,
$k \gtrapprox 0.1 h$Mpc$^{-1}$, the quasistatic approximation is only valid up to a slightly
lower accuracy than ten percent if $\hat{n}=0.25$, and to better accuracy for higher values
of $\hat{n}$. In particular, for $\hat{n} > 1$ or for regular coupling functions as in dilaton
models, or the generic case which includes the $f(R)$ theories, the correction to the
quasistatic approximation is suppressed by a factor $1/(\omega\tau)$ and a ten percent
accuracy (at least) is reached as soon as $k > 3 \times 10^{-3} h$Mpc$^{-1}$ (and better
at higher $k$). For small $k$ or the homogeneous background, the accuracy of the
quasistatic approximation is set by $1/(\omega\tau) \simeq 1/(c a m \tau) \ll 1$.
Thus, the quasistatic approximation is sufficient for our purposes throughout this paper.

A different issue is the fact that the symmetron models arise from a double-well potential
and that different domains may fall within different minima
\cite{Llinares2013}. As seen above, soon after
$a_s$ the quasistatic approximation should become valid within each domain. However,
at the boundaries between different regions, new phenomena associated with these
domain walls take place and are not described in this paper. They would require specific
methods suited to such topological defects.

\section{Perturbative approach}
\label{Perturbative}

The equations of motion are nonlinear and there are no explicit solutions in the general case. As in the usual $\Lambda$-CDM cosmology we can look for perturbative solutions,
where we expand in fluctuations with respect to the uniform expanding background.
We describe in this section this perturbative approach to the equations of motion,
up to any order in all field fluctuations. We give explicit expressions up to third
order.
For scalar-tensor models, this is carried out in full generality using the $m(a)-\beta (a)$ parametrization.

\subsection{Expansion of the modified ``gravitational potential''}
\label{Expansion-Psi}

To compute the dynamics of the matter particles we need the modified gravitational potential
$\Psi$ given by either Eq.(\ref{Poisson-fR}) or Eq.(\ref{Psi-N+A}).
In the quasistatic approximation, $\Psi$ is a mere functional of the matter density fluctuation
$\delta\rho$ (i.e., it does not depend on the past evolution) and it is convenient to first
solve for $\Psi[\delta\rho]$. Next, this expression can be used in the equation of motion of the
matter particles (the Euler equation in the single-stream approximation), which can be
solved as in the standard $\Lambda$-CDM case by a perturbative expansion of the density and
velocity fields in powers of the linear growing mode.
A similar approach was already used in \cite{Koyama2009} for DGP \cite{Dvali2000}
and $f(R)$ models, up to one-loop order, and in \cite{BraxPV2012}
(where only the linear order was kept in the modified gravitational potential).
Here we describe how this perturbative approach, which relies on two
successive expansions, applies in a similar fashion to $f(R)$ theories and
scalar-tensor models, with an explicit coupling to the matter density in the
Klein-Gordon equations that governs this additional scalar field.
We also show how the tomographic approach determines the higher-order
terms from derivatives of the two coupling functions that appear at linear order.

\subsubsection{$f(R)$ models}
\label{fR-PT}

In the $f(R)$ theories, the modified potential $\Psi$ is given by equation (\ref{Poisson-fR})
(in the quasistatic approximation), which involves the fluctuation $\delta R$
of the Ricci scalar.
Therefore, we first need to solve the constraint equation
(\ref{fR-delta}) to obtain the functional $\delta R[\delta\rho]$.
Expanding the function $f_R$, with the help of the derivatives $\kappa_n$ introduced in
(\ref{kappa-fR}), and moving linear terms in $\delta R$ to the left-hand side,
Eq.(\ref{fR-delta}) becomes
\beq
\left( 1 - \frac{\nabla^2}{a^2m^2} \right) \cdot \delta R = \frac{\delta\rho}{\MPl^2}
+ \sum_{n=2}^{\infty} \frac{3H^{2-2n}\kappa_n}{a^2 \, n!} \, \nabla^2 \, (\delta R)^n .
\label{cL-R-fR}
\eeq
Then, we can solve Eq.(\ref{cL-R-fR}) for $\delta R$ by looking for a perturbative
expansion in powers of the nonlinear density fluctuation $\delta\rho$.
Going to Fourier space, with the normalization
$\delta R(\vx) = \int\dd\vk \; e^{\ii\vk\cdot\vx} \; \delta\tR(\vk)$, we write
this expansion as
\beqa
\delta\tR(\vk) & = & \sum_{n=1}^{\infty} \int \dd\vk_1 .. \dd\vk_n \;
\delta_D(\vk_1+..+\vk_n-\vk) \nonumber \\
&& \times \, h_n(\vk_1,..,\vk_n) \, \delta\trho(\vk_1) \dots \delta\trho(\vk_n) .
\label{hn-R}
\eeqa
As the linear operator in the left-hand side in Eq.(\ref{cL-R-fR}) is diagonal in Fourier space,
with the inverse $a^2m^2/(a^2m^2+k^2)$, we easily obtain the kernels $h_n$ by recursion,
after substituting the expansion (\ref{hn-R}) into Eq.(\ref{cL-R-fR}). This yields for instance
for the first two kernels
\beq
h_1(\vk) = \frac{a^2m^2}{\MPl^2(a^2m^2+k^2)} ,
\eeq
\beq
h_2(\vk_1,\vk_2) = \frac{-3a^4m^6\kappa_2 k^2}{2H^2\MPl^4 (a^2m^2 \!+\! k_1^2)
(a^2m^2 \!+\! k_2^2) (a^2m^2 \!+\! k^2)} .
\eeq

The expansion (\ref{hn-R}) in powers of the nonlinear density fluctuation $\delta\rho$
should be distinguished from the expansion in powers of the linear density fluctuation
$\delta\rho_L$ (or $\psi_L$) that we introduce below to solve the Euler equation.
In particular, the order and the range of validity of these two expansions are not
necessarily identical.
For instance, if the high-order derivatives $\kappa_n$ are very small (or if $f(R)$ is
a polynomial) the expansion (\ref{hn-R}) may be truncated at a low order, even on
scales where the density field is highly nonlinear.

Next, substituting into Eq.(\ref{Poisson-fR}) we obtain the expansion in the
nonlinear density fluctuation $\delta\rho$ of the modified gravitational potential,
\beqa
\tilde{\Psi}(\vk) & = & \sum_{n=1}^{\infty} \int \dd\vk_1 .. \dd\vk_n \;
\delta_D(\vk_1+..+\vk_n-\vk) \nonumber \\
&& \times \, H_n(\vk_1,..,\vk_n) \, \delta\trho(\vk_1) \dots \delta\trho(\vk_n) .
\label{Psi-n}
\eeqa
This gives
\beq
H_1(\vk) = - \frac{a^2 (3a^2m^2+4k^2)}{6\MPl^2k^2 (a^2m^2+k^2)} ,
\label{H1-fR}
\eeq
and
\beq
n \geq 2: \;\; H_n = \frac{a^2}{6k^2} \, h_n .
\eeq

\subsubsection{Scalar field models}
\label{scalar field-PT}

In scalar-tensor theories, the modified potential $\Psi$ depends on the scalar field
$\varphi$, hence we first need to solve the Klein-Gordon equation (\ref{Klein-Gordon}),
to obtain the functional $\delta\varphi[\delta\rho]$.
Subtracting from Eq.(\ref{Klein-Gordon}) the uniform background (\ref{KG-0}) and
expanding in $\delta\varphi$, using the derivatives
(\ref{beta-n-def-scalar})-(\ref{kappa-n-def-scalar}), we obtain
\beqa
\left( \frac{\nabla^2}{a^2} - m^2 \right) \cdot \delta\varphi & = &
\frac{\beta\,\delta\rho}{c^2\MPl} + \frac{\beta_2\,\delta\rho}{c^2\MPl^2} \delta\varphi
\nonumber \\
&& \hspace{-1.5cm}+ \sum_{n=2}^{\infty}
\left( \frac{\kappa_{n+1}}{\MPl^{n-1}}+\frac{\beta_{n+1}\,\delta\rho}
{c^2\MPl^{n+1}} \right) \frac{(\delta\varphi)^n}{n!} .
\label{KG-n}
\eeqa
Again, we solve this Klein-Gordon equation as a perturbative expansion in the
nonlinear matter density fluctuation $\delta\rho$,
\beqa
\delta\tilde{\varphi}(\vk) & = & \sum_{n=1}^{\infty} \int \dd\vk_1 .. \dd\vk_n \;
\delta_D(\vk_1+..+\vk_n-\vk) \nonumber \\
&& \times \, h_n(\vk_1,..,\vk_n) \, \delta\trho(\vk_1) \dots \delta\trho(\vk_n) ,
\label{hn-phi}
\eeqa
using the fact that the linear operator on the left-hand side in Eq.(\ref{KG-n}) is diagonal
in Fourier space and easily inverted as $-a^2/(a^2m^2+k^2)$.
This yields for instance for the first two kernels
\beq
h_1(\vk) = \frac{-a^2\beta}{c^2\MPl(a^2m^2+k^2)} ,
\eeq
\beq
h_2(\vk_1,\vk_2) = \frac{a^4\beta( -a^2\beta\kappa_3+2\beta_2 (a^2m^2+k_1^2))}
{2c^4\MPl^3 (a^2m^2 \!+\! k_1^2)
(a^2m^2 \!+\! k_2^2) (a^2m^2 \!+\! k^2)} .
\eeq

Next, substituting into Eq.(\ref{Psi-N+A}) we obtain the expansion over
$\delta\rho$ of the modified gravitational potential, as in Eq.(\ref{Psi-n}), writing the
fifth-force contribution as
\beq
\Psi_{\rm A} = \sum_{n=1}^{\infty} \frac{c^2 \beta_n}{\MPl^n \, n!} \,  ( \delta\varphi )^n .
\eeq
This yields, for instance,
\beq
H_1(\vk) = - \frac{a^2 (a^2m^2+k^2 (1+2\beta^2))}{2\MPl^2 k^2 (a^2m^2+k^2)} .
\eeq

\subsection{Single-stream approximation for the matter fluid}
\label{single-stream}

\subsubsection{Hydrodynamical equations of motion}

We have described in the previous sections how to compute the modified gravitational
potential up to any order in $\delta\rho$ for $f(R)$ theories and scalar-tensor
models.
From expansions like Eq.(\ref{Psi-n}) (which may also be associated with other models),
we now derive the dynamics of large-scale structures in the perturbative regime.
In the single-stream approximation which is valid on large scales, the dynamics of the
matter fluid is given by the continuity and Euler equations,
\beqa
\frac{\pl \delta}{\pl\tau} + \nabla \cdot [ (1+\delta) \vv ] & = & 0 ,
\label{continuity-1} \\
\frac{\pl \vv}{\pl\tau} + {\cal H} \vv + (\vv\cdot\nabla) \vv & = & - \nabla \cdot \Psi ,
\label{Euler-1}
\eeqa
where $\tau = \int \dd t/a$ is the conformal time, $\cH=aH=\dot{a}$ the conformal expansion
rate, $\delta=\delta\rho/\rhob$ the matter density contrast, and $\vv$ the peculiar velocity.
Introducing the time variable $\eta=\ln a$ and the two-component vector $\psi$,
\beq
\psi \equiv \left(\bea{c} \psi_1 \\ \psi_2 \ea \right) \equiv
\left( \bea{c} \delta \\ -(\nabla\cdot\vv)/\dot{a} \ea \right) ,
\label{psi-def}
\eeq
Eqs.(\ref{continuity-1})-(\ref{Euler-1}) read in Fourier space as
\beqa
\frac{\pl\tpsi_1}{\pl\eta} - \tpsi_2 & = & \int \dd\vk_1\dd\vk_2 \;
\delta_D(\vk_1\!+\!\vk_2\!-\!\vk) \hat{\alpha}(\vk_1,\vk_2) \nonumber \\
&& \times \; \tpsi_2(\vk_1) \tpsi_1(\vk_2) ,
\label{continuity-2}
\eeqa
\beqa
\hspace{-1cm} \frac{\pl\tpsi_2}{\pl\eta} + \frac{k^2}{a^2H^2} \, \tilde{\Psi}
+ \frac{1-3\wde \Ode}{2} \, \tpsi_2 & =  & \nonumber \\
&& \hspace{-5cm} \int\!\! \dd\vk_1\dd\vk_2 \; \delta_D(\vk_1\!+\!\vk_2\!-\!\vk)
\hat{\beta}(\vk_1,\vk_2) \tpsi_2(\vk_1) \tpsi_2(\vk_2) ,
\label{Euler-2}
\eeqa
with
\beq
\hat{\alpha}(\vk_1,\vk_2)= \frac{(\vk_1\!+\!\vk_2)\cdot\vk_1}{k_1^2} ,
\hat{\beta}(\vk_1,\vk_2)= \frac{|\vk_1\!+\!\vk_2|^2(\vk_1\!\cdot\!\vk_2)}{2k_1^2k_2^2} .
\label{alpha-beta-def}
\eeq

In the standard $\Lambda$-CDM cosmology, where the Newtonian gravitational potential
is linear in the density field, the continuity and Euler equations
(\ref{continuity-2})-(\ref{Euler-2}) are quadratic. In modified gravity models, such as those
studied in this paper, the potential $\Psi$ is nonlinear and contains terms of all orders
in $\delta\rho$. Therefore, we must introduce vertices of all orders and we write
Eqs.(\ref{continuity-2})-(\ref{Euler-2}) under the more concise form
\beq
\cO(x,x') \cdot \tpsi(x') = \sum_{n=2}^{\infty} K_n^s(x;x_1,..,x_n) \cdot
\tpsi(x_1) \dots \tpsi(x_n) ,
\label{O-Ks-def}
\eeq
where we have introduced the coordinates $x=(\vk,\eta,i)$, $i=1,2$ is the discrete index
of the two-component vector $\tpsi$, and repeated coordinates are integrated over.
The matrix $\cO$ reads as
\beqa
\hspace{-0.3cm} \cO(x,x') & \! = \! & \delta_D(\eta' \!-\! \eta) \, \delta_D(\vk' \!-\! \vk) \nonumber \\
&& \hspace{-1.5cm} \times \left( \bea{cc} \frac{\pl}{\pl\eta} & -1 \\  & \\
- \frac{3}{2} \Om(\eta) (1\!+\!\epsilon(k,\eta)) &  \frac{\pl}{\pl\eta} \!+\!
\frac{1\!-\! 3 \wde \Ode(\eta)}{2} \ea \right) ,
\label{O-mod}
\eeqa
where $\epsilon(k,\eta)$, which measures the deviation from the Newtonian gravitational
potential at linear order, is given by
\beq
1+\epsilon(k,\eta) = - 2 \MPl^2 a^{-2} k^2 \, H_1(k,\eta) .
\eeq
The vertices $K_n^s$ are equal-time vertices of the form
\beqa
K_n^s(x;x_1,..,x_n) & = & \delta_D(\eta_1 \!-\! \eta) .. \delta_D(\eta_n \!-\! \eta)
\nonumber \\
&& \hspace{-2.5cm} \times \; \delta_D(\vk_1 \!+..+\! \vk_n \!-\! \vk) \;
\gamma_{i;i_1,..,i_n}^s(\vk_1,..,\vk_n;\eta) .
\label{Ks-def}
\eeqa
The nonzero vertices are the usual $\Lambda$-CDM ones,
\beq
\gamma_{1;1,2}^s(\vk_1,\vk_2) = \frac{\hat{\alpha}(\vk_2,\vk_1)}{2} , \;\;
\gamma_{1;2,1}^s(\vk_1,\vk_2) = \frac{\hat{\alpha}(\vk_1,\vk_2)}{2} , \nonumber
\eeq
\beq
\gamma_{2;2,2}^s(\vk_1,\vk_2) = \hat{\beta}(\vk_1,\vk_2) ,
\eeq
which are of order $n=2$ and do not depend on time, and the new vertices associated with
the modified gravitational potential (\ref{Psi-n}),
\beqa
n \geq 2: \;\;\; \gamma_{2;1,..,1}^s(\vk_1,..,\vk_n;\eta) & = & - \frac{k^2}{a^2H^2} \,
(3\Om H^2 \MPl^2)^n \nonumber \\
&& \hspace{-1.5cm} \times \; \frac{1}{n!} \sum_{\rm perm.} H_n(\vk_1,..,\vk_n;\eta) ,
\eeqa
where we sum over all permutations of $\{\vk_1,..,\vk_n\}$ to obtain symmetrized
kernels $\gamma^s$.

From the analysis in the previous sections, and in particular from Eqs.(\ref{cL-R-fR})
and (\ref{KG-n}), we can check that at all orders the vertices decay as $k^2$ at low $k$,
\beq
n \geq 2 , \;\;\; k \rightarrow 0 : \;\;\;  \gamma^s_{2;1,..,1}(\vk_1,..,\vk_n) \sim k^2 ,
\eeq
where the limit is taken by letting the sum $\vk=\vk_1+..+\vk_n$ go to zero while the
individual wave numbers $\{\vk_1,..,\vk_n\}$ remain finite.
This is related to the lack of backreaction on large scales from small-scale nonlinearities,
as noted in Sec.~\ref{Equations of motion}.
As for the usual Newtonian gravity, this means that if the initial conditions had very little
power on large scales (i.e., the linear power spectrum $P_L(k)$ would decay faster than
$k^4$ at low $k$), nonlinearities would only generate a $k^4$ tail at low $k$.
For CDM initial conditions, where $P_L(k) \sim k^{0.96}$ at low $k$, this ensures that
we recover the linear theory on large scales.

\paragraph{$f(R)$ theories:}
From Sec.~\ref{fR-PT} we obtain for the first three kernels the expressions
\beq
\epsilon(k,\eta) = \frac{k^2}{3(a^2m^2+k^2)} ,
\label{eps-fR}
\eeq
\beq
\gamma^s_{2;1,1}(\vk_1,\vk_2) = \frac{9 a^4 \Om^2 m^6 \kappa_2 k^2}
{4(a^2m^2 \!+\! k^2)(a^2m^2 \!+\! k_1^2)(a^2m^2 \!+\! k_2^2)} ,
\eeq
and
\beqa
\hspace{-1cm} \gamma_{2;1,1,1}(\vk_1,\vk_2,\vk_3) & = & \frac{9 a^6 m^8 \Om^3 k^2}
{4(a^2m^2+k_1^2)(a^2m^2+k_2^2)} \nonumber \\
&& \hspace{-2.6cm} \times \frac{a^2m^2\kappa_3+(\kappa_3-9m^2\kappa_2^2)
|\vk_2 \!+\! \vk_3|^2}{(a^2m^2+k_3^2)(a^2m^2+|\vk_2\!+\!\vk_3|^2)(a^2m^2+k^2)} ,
\eeqa
where we give an expression for the nonsymmetrized kernel $\gamma_{2;1,1,1}$ as
it is more compact.

\paragraph{Scalar-tensor models:}
From Sec.~\ref{scalar field-PT} we obtain
\beq
\epsilon(k,\eta) = \frac{2 \beta^2 k^2}{a^2m^2+k^2} ,
\label{eps-scalar}
\eeq
\beqa
\hspace{-1cm} \gamma^s_{2;1,1}(\vk_1,\vk_2) & = & \frac{9 a^2 H^2 \Om^2 \beta^2 k^2}
{2 c^2 (a^2m^2 \!+\! k^2)} \nonumber \\
&& \hspace{-0.7cm} \times \frac{a^2 \beta \kappa_3 - \beta_2 (k^2+k_1^2+k_2^2+3 a^2m^2)}
{(a^2m^2 \!+\! k_1^2)(a^2m^2 \!+\! k_2^2)} ,
\eeqa
and
\beqa
\lefteqn{ \gamma_{2;1,1,1}(\vk_1,\vk_2,\vk_3) = 9 a^4 H^4 \Om^3 \beta^2 k^2 } \nonumber \\
&& \times \biggl \lbrace  6 \beta_2^2 (a^2m^2+k_2^2) (2 a^2m^2+k_1^2+k^2) \nonumber \\
&& - 3a^2\beta\beta_2\kappa_3(4 a^2m^2+k_1^2+2k_2^2+k^2) \nonumber \\
&& + \beta \; [ 3a^4\beta\kappa_3^2-a^2\beta\kappa_4(a^2m^2+|\vk_2+\vk_3|^2) \nonumber \\
&& +\beta_3(a^2m^2+|\vk_2+\vk_3|^2) (4a^2m^2+3k_1^2+k^2) ] \biggl \rbrace
\nonumber \\
&& \times \biggl \lbrace 2c^4(a^2m^2+k_1^2)(a^2m^2+k_2^2)(a^2m^2+k_3^2) \nonumber \\
&& \times (a^2m^2+|\vk_2+\vk_3|^2) (a^2m^2+k^2) \biggl \rbrace^{-1} .
\eeqa

\subsubsection{One-loop matter power spectrum}
\label{Pk-one-loop}

From the equation of motion (\ref{O-Ks-def}) we can now compute the matter density
power spectrum up to the required order in perturbation theory.
In this paper, we only go up to third order in the fields, which corresponds to one-loop
diagrams.

Thus, as in the standard perturbation theory, we look for a solution of the nonlinear
equation of motion (\ref{O-Ks-def}) as a perturbative expansion in powers of the
linear growing mode $\psi_L$,
\beq
\psi(x) = \sum_{n=1}^{\infty} \psi^{(n)}(x) , \;\;\; \mbox{with} \;\;\;
\psi^{(n)} \propto \psi_L^n .
\label{psi-n-def}
\eeq

At linear order, the equation of  motion (\ref{O-Ks-def}) becomes $\cO\cdot\tpsi_L=0$ and
we obtain two linear growing and decaying modes, $D_{\pm}(k,\eta)$, which are solutions
of
\beq
\frac{\pl^2 D}{\pl\eta^2} + \frac{1- 3 \wde \Ode}{2} \,
\frac{\pl D}{\pl\eta} - \frac{3}{2} \Om (1+\epsilon) D =  0 .
\label{D-pm}
\eeq
Because at early times we recover the Einstein-de Sitter universe (the dark energy
component also becomes negligible) we have the usual behaviors:
\beq
t\rightarrow 0: \;\; D_+ \rightarrow a = e^{\eta} , \;\;
D_- \propto a^{-3/2} = e^{-3\eta/2} .
\label{D+-asymp}
\eeq
However, at finite redshift, because of the $k$-dependent factor $\epsilon(k,\eta)$, the
linear modes $D_{\pm}(k,\eta)$ now depend on the wave numbers $k$.
In any case, assuming as usual that the decaying mode has had time to become negligible
we can write the first-order solution as
\beq
\tpsi^{(1)} = \tpsi_L = \tdelta_{L0}(\vk) \;
\left( \bea{c} D_+(k,\eta) \\ \frac{\pl D_+}{\pl\eta}(k,\eta) \ea \right)   .
\label{psi-1}
\eeq
Hence the initial conditions are fully defined by the linear density field $\tdelta_{L0}(\vk)$.
We refer the reader to \cite{BraxPV2012} for a detailed analysis of the linear
growing and decaying modes.

\begin{figure}
\begin{center}
\epsfxsize=7.5 cm \epsfysize=3.6 cm {\epsfbox{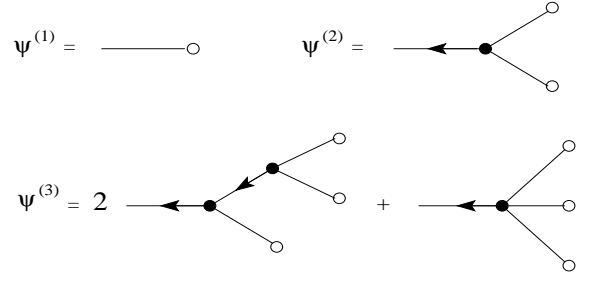}}
\end{center}
\caption{Diagrammatic expansion of the field $\tpsi$ up to third order, as in Eqs.(\ref{psi-1}),
(\ref{psi-2}), and (\ref{psi-3}). The white circles are the linear solution $\tpsi_L$, the black
dots are the vertices $K_n^s$, and the lines with an arrow are the retarded propagator $R_L$,
with time increasing along the direction of the arrow.}
\label{fig-psi-n}
\end{figure}

Next, to compute the higher orders $\psi^{(n)}$ by recursion from Eq.(\ref{O-Ks-def}),
we introduce the retarded Green function $R_L$ of the linear operator $\cO$,
also called the linear propagator or response function, which obeys:
\beq
\cO(x,x') \cdot R_L(x',x'') = \delta_D(x-x'') ,
\label{RL-def}
\eeq
\beq
\eta_1 < \eta_2 : \;\;\; R_L(x_1,x_2)=0 ,
\eeq
and reads as
\beqa
\hspace{-0.5cm} R_L(x_1,x_2) & = & \frac{\Theta(\eta_1-\eta_2) \, \delta_D(\vk_1-\vk_2)}
{D_{+2}'D_{-2}-D_{+2}D_{-2}'}  \nonumber \\
&& \hspace{-1.8cm} \times \left( \! \bea{lr} D_{+2}'D_{-1}\!-\!D_{-2}'D_{+1} \;\;
& \;\;  D_{-2}D_{+1}\!-\!D_{+2}D_{-1} \\ & \\ D_{+2}'D_{-1}'\!-\!D_{-2}'D_{+1}'
&  D_{-2}D_{+1}'\!-\!D_{+2}D_{-1}' \ea \! \right)
\label{RL-1}
\eeqa
It involves both the linear growing and decaying modes, $D_+$ and $D_-$, and
$\Theta(\eta_1-\eta_2)$ is the Heaviside function, which ensures causality.
Then, from Eq.(\ref{O-Ks-def}) we obtain at second and third order
\beq
\tpsi^{(2)} = R_L \cdot K_2^s \cdot \tpsi^{(1)} \tpsi^{(1)} ,
\label{psi-2}
\eeq
\beq
\tpsi^{(3)} = 2 R_L \cdot K_2^s \cdot \tpsi^{(2)} \tpsi^{(1)} + R_L \cdot K_3^s \cdot
\tpsi^{(1)} \tpsi^{(1)} \tpsi^{(1)} .
\label{psi-3}
\eeq
We show the diagrams associated with Eqs.(\ref{psi-1}), (\ref{psi-2}), and (\ref{psi-3})
in Fig.~\ref{fig-psi-n}.
The last diagram, associated with the last term in Eq.(\ref{psi-3}), does not appear in the
standard $\Lambda$-CDM case. It is due to the vertex $\gamma^s_{2;1,1,1}$ associated
with the term of order $(\delta\rho)^3$ of the nonlinear modified gravitational potential
$\Psi$.

\begin{figure}
\begin{center}
\epsfxsize=7.8 cm \epsfysize=2.5 cm {\epsfbox{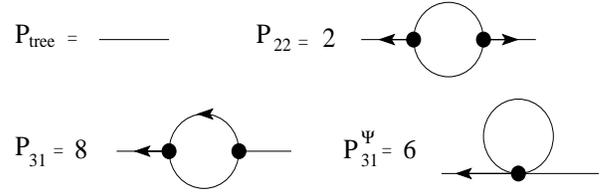}}
\end{center}
\caption{Diagrammatic expansion of the matter density power spectrum $P(k)$ up to order
$P_L^2$.
The black dots are the vertices $K_n^s$, the lines with an arrow are the retarded propagator
$R_L$, with time increasing along the direction of the arrow, and the lines without an arrow are the
linear correlation $C_L$.}
\label{fig-P1loop}
\end{figure}

Then, the two-point correlation $C_2$ of the field $\psi$ reads up to order $\psi_L^4$ as
\beqa
C_2(x_1,x_2) & \equiv & \lag \tpsi(x_1) \tpsi(x_2) \rag \nonumber \\
& = & \lag \tpsi^{(1)} \tpsi^{(1)} \rag + \lag \tpsi^{(2)}\tpsi^{(2)}\rag
+ \lag\tpsi^{(3)}\tpsi^{(1)}\rag \nonumber \\
&& + \lag\tpsi^{(1)}\tpsi^{(3)}\rag + ..
\label{C2-def}
\eeqa
Defining the equal-time matter density power spectrum as
\beq
\lag \tdelta(\vk_1,\eta) \tdelta(\vk_2,\eta) \rag = \delta_D(\vk_1+\vk_2) \; P(k_1,\eta) ,
\eeq
substituting the expressions (\ref{psi-1}), (\ref{psi-2}), and (\ref{psi-3}) into Eq.(\ref{C2-def})
and using Wick's theorem, we obtain up to order $P_L^2$,
\beq
P(k) = P_{\rm tree}(k) + P_{\rm 1loop}(k) ,
\label{Ptree+1loop}
\eeq
which corresponds to the ``tree'' and ``one-loop'' diagrams shown in Fig.~\ref{fig-P1loop}.
The tree contribution is simply the linear power spectrum,
\beq
P_{\rm tree} = P_L(k) ,
\label{P-tree}
\eeq
while the one-loop contribution corresponds to three diagrams,
\beq
P_{\rm 1loop} = P_{22} + P_{31} + P_{31}^{\Psi} .
\label{P-1loop}
\eeq
The diagrams $P_{22}$ (which arises from the average $\lag \tpsi^{(2)}\tpsi^{(2)}\rag$
in Eq.(\ref{C2-def}), by gluing together two diagrams $\psi^{(2)}$ of Fig.~\ref{fig-psi-n})
and $P_{31}$ (which arises from the average $\lag \tpsi^{(3)}\tpsi^{(1)}\rag$
in Eq.(\ref{C2-def}), by gluing together the first diagram $\psi^{(3)}$ of Fig.~\ref{fig-psi-n}
with the diagram $\psi^{(1)}$) already appear in the $\Lambda$-CDM cosmology
(but with different linear propagators and vertices).
The diagram $P_{31}^{\Psi}$ is a new term which arises from the second diagram $\psi^{(3)}$
in Fig.~\ref{fig-psi-n}. It is again due to the new vertex $\gamma^s_{2;1,1,1}$ associated
with the term of order $(\delta\rho)^3$ of the nonlinear modified gravitational potential
$\Psi$.
To be more explicit, the contribution $P_{31}$ becomes
\beqa
P_{31}(k,\eta) & = & 8 \int \! \dd\vk_1 \dd\vk_2 \; \delta_D(\vk_1 \!+\! \vk_2 \!-\! \vk)
\int_{-\infty}^{\eta} \! \dd\eta_1 \int_{-\infty}^{\eta_1} \! \dd\eta_2 \nonumber \\
&& \hspace{-1.3cm} \times \sum_{i_1,j_1,m_1} \sum_{i_2,j_2,m_2}
R_{L,1 i_1}(k;\eta,\eta_1) \, R_{L,j_1 i_2}(k_1;\eta_1,\eta_2) \nonumber \\
&& \hspace{-1.3cm} \times C_{L,m_1 m_2}(k_2,\eta_1,\eta_2) \, C_{L,1 j_2}(k;\eta,\eta_2)
\nonumber \\
&& \hspace{-1.3cm} \times \gamma^s_{i_1;j_1 m_1}(\vk_1,\vk_2;\eta_1) \,
\gamma^s_{i_2;j_2 m_2}(\vk,-\vk_2;\eta_2) ,
\label{P31-1}
\eeqa
while the new contribution $P_{31}^{\Psi}$ reads as
\beqa
P_{31}^{\Psi}(k,\eta) & \! = \! & 6 \! \int \! \dd\vk_1\! \int_{-\infty}^{\eta} \! \dd\eta_1 \,
R_{L,12}(k;\eta,\eta_1) \, C_{L,11}(k;\eta,\eta_1) \nonumber \\
&& \hspace{-0.3cm} \times C_{L,11}(k_1;\eta_1,\eta_1) \,
\gamma^s_{2;1,1,1}(\vk_1,-\vk_1,\vk;\eta_1) ,
\label{P31-Psi-1}
\eeqa
where we focus on the equal-time power spectrum, so that both ends of the diagrams
in Fig.~\ref{fig-P1loop} are taken at the same time $\eta$.
Here $R_{L,i_1 i_2}(k;\eta_1,\eta_2)$ is the linear propagator, given by
Eq.(\ref{RL-1}), while $C_{L,i_1 i_2}(k;\eta_1,\eta_2)$ is the linear correlation, given by
\beqa
\hspace{-0.5cm} C_L(x_1,x_2) & = & \lag \tpsi_L(x_1) \tpsi_L(x_2)\rag \nonumber \\
&& \hspace{-1.5cm} = \delta_D(\vk_1\!+\!\vk_2) P_{L0}(k_1)
\left( \! \bea{cc} D_{+1} D_{+2} & D_{+1} D_{+2}' \\ & \\ D_{+1}' D_{+2} &
D_{+1}' D_{+2}' \ea \! \right) .
\label{CL-1}
\eeqa

Thus, Eq.(\ref{P-1loop}) provides the expression of the matter density power spectrum
up to one-loop order (i.e., up to $P_L^2$), using Fig.~\ref{fig-P1loop}.
Apart from the new diagram $P_{31}^{\Psi}$, the difference from the $\Lambda$-CDM
cosmology is that the linear propagator $R_L$ also depends on wave number while the
vertices also depend on time, through functions which depend on the details of the modified
gravity theory as described in the previous sections.
This means that it is not possible to compute analytically the integrals over time
and the summations over indices which appear in the diagrams. In particular, there is
no factorization of the form $P_{\rm 1loop}(k,\eta) = D(\eta)^4 P_{\rm 1loop ; 0}(k)$.

Our perturbative approach, illustrated up to one-loop order in
Fig.~\ref{fig-P1loop}, differs from the usual computation of the standard perturbative
expansion (see \cite{Bernardeau2002} for the $\Lambda$-CDM cosmology,
\cite{Koyama2009} for the DGP model and \cite{Bernardeau:2011sf,BraxPV2012} for
modified gravity in the weak-field limit).
In the usual presentation,
the expansion (\ref{psi-n-def}) is written as $\tpsi = \sum_n F_n^s \cdot \tpsi_L .. \tpsi_L$,
in a fashion similar to Eq.(\ref{Psi-n}), and the kernels $F_n^s$ are explicitly
computed by substituting this expansion into the equation of motion
(\ref{O-Ks-def}). Then, the power spectrum is obtained as in
Eq.(\ref{C2-def}).
In our framework we do not explicitly compute these kernels $F_n^s$
[but $F_2^s$ and $F_3^s$ are implicitly determined by Eqs.(\ref{psi-2}) and
(\ref{psi-3})] and we go directly from the diagrams of Fig.~\ref{fig-psi-n} for $\psi$
to the diagrams of Fig.~\ref{fig-P1loop} for $P(k)$ (the standard approach yields
a different type of diagrams and there is not always a one-to-one correspondence
between these two diagrammatic expansions, in particular for higher-order correlations).
The practical advantage of our formulation is that the diagrams of Fig.~\ref{fig-P1loop}
only involve two-point functions, $C_L$ and $R_L$, and the vertices $\gamma^s$.
This avoids the need to compute kernels $F_n^s(\vk_1,..,\vk_n)$ with an increasing
number of dependent wave numbers $\vk_i$ as we go to higher orders.

A difference with the ``closure method'' used in \cite{Koyama2009} for $f(R)$
models is that we obtain explicit expressions for the power spectrum,
as in Eqs.(\ref{P31-1}) and (\ref{P31-Psi-1}), instead of differential equations
over time. This is because the integration over time of the equation of motion
(\ref{O-Ks-def}) has already been performed at the level of the expansion
(\ref{psi-n-def}), through the Green functions $R_L$ in Eqs.(\ref{psi-2}) and
(\ref{psi-3}).
Another difference with the closely related ``steepest-descent'' expansion
described in \cite{BraxPV2012} is that we do not compute ``self-energy''
diagrams in intermediate steps.
(A description of different perturbative expansions and diagrams may be
found in \cite{Valageas2008} for the $\Lambda$-CDM cosmology,
our current approach being equivalent to the one described in Sec.4 of that paper
but with a different derivation.)

All these perturbative approaches coincide when they are eventually expanded
up to a given order over $P_L$. Our approach, associated with
Figs.~\ref{fig-psi-n} and \ref{fig-P1loop}, is simple (no more complex than the
standard perturbative expansion) and convenient for numerical computations
(it only involves the linear two-point functions $R_L$ and $C_L$ and the
bare vertices $\gamma^s$).
In particular, it can be readily applied to any equation of motion of the form
(\ref{O-Ks-def}), whatever the dependence on the index, scale, or time, and the
order of the new vertices $\gamma^s$.
This would hold for any modified gravity model with the quasistatic approximation,
where the new degree of freedom can be written in terms of the density and
velocity fields.

If the quasistatic approximation is not valid, one can still use the perturbative
approach described in this section. However, instead of first looking for
an expansion in $\delta\rho$ for the modified gravitational potential $\Psi$,
we extend the doublet (\ref{psi-def}) to a triplet
$(\delta,-(\nabla\cdot\vv)/\dot{a},\delta\varphi/M_{\rm Pl})$ and we treat on the same
footing the density and velocity fields and the new scalar field
$\delta\varphi$ (the modified gravitational potential $\Psi$ being written in terms
of both $\delta\rho$ and $\delta\varphi$).

\subsection{Numerical results}
\label{numerical-1loop}

\subsubsection{$f(R)$ theories}
\label{1loop-fR}

\begin{figure}
\begin{center}
\epsfxsize=8.5 cm \epsfysize=6.7 cm {\epsfbox{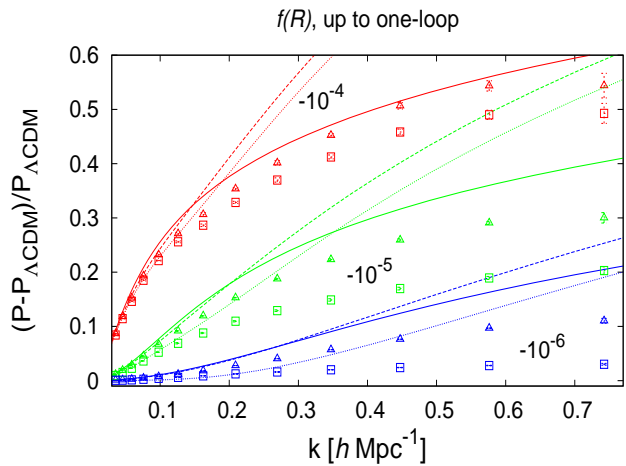}}
\end{center}
\caption{Relative deviation from $\Lambda$-CDM of the power spectrum in $f(R)$
theories, at redshift $z=0$, for $n=1$ and $f_{R_0}=-10^{-4}, -10^{-5}$, and $-10^{-6}$.
In each case, the triangles and the squares are the results of the ``no-chameleon'' and
``with-chameleon'' simulations from \cite{Oyaizu2008}, respectively.
We plot the relative deviation of the linear power (solid line),
of the one-loop power without ``chameleon'' effect
($\gamma^s_{2;1,1}=\gamma^s_{2;1,1,1}=0$) (dashed line), and with lowest-order
``chameleon'' effect ($\gamma^s_{2;1,1}\neq 0$, $\gamma^s_{2;1,1,1}=0$) (dotted line).}
\label{fig-dPk_fR_Hu_PT_z0}
\end{figure}

We show in Fig.~\ref{fig-dPk_fR_Hu_PT_z0} the relative deviation from $\Lambda$-CDM
of the matter power spectrum obtained in $f(R)$ theories at $z=0$, using perturbation
theory.
The triangles correspond to the ``no-chameleon'' simulations of \cite{Oyaizu2008},
where the constraint equation (\ref{fR-delta}) is linearized in $\delta R$.
This corresponds to truncating the expansions (\ref{hn-R}) and (\ref{Psi-n})
at the first order, $n=1$, and to discard the new vertices $\gamma^s_{2;1,..,1}$ in
Eq.(\ref{O-Ks-def}), so that the only modification from $\Lambda$-CDM enters through
the factor $\epsilon$ in the matrix $\cO$ of Eq.(\ref{O-mod}). The squares are the
fully nonlinear simulations of \cite{Oyaizu2008}, where the constraint equation
(\ref{fR-delta}) is exactly solved.
Because of the ``chameleon'' effect, which only appears through the nonlinear terms
of Eq.(\ref{fR-delta}), the squares lie somewhat below the triangles in
Fig.~\ref{fig-dPk_fR_Hu_PT_z0}, for the same value of $f_{R_0}$.
As is well known, this effect is somewhat larger for lower values of $|f_{R_0}|$.

The solid lines are the relative difference of the linear power spectra,
$(P_L-P_{L,\Lambda \rm CDM})/P_{L,\Lambda \rm CDM}$.
By definition, this can only include the effect of the factor $\epsilon$ in the matrix
$\cO$ of Eq.(\ref{O-mod}).
We can check that this recovers the deviation of the full nonlinear power spectrum
measured in the simulations on large scales, $k \leq 0.1 h\mbox{Mpc}^{-1}$.

The dashed lines are the relative difference of the one-loop power spectra,
$(P_{\rm tree+1loop}-P_{\rm tree+1loop,\Lambda CDM})
/P_{\rm tree+1loop,\Lambda CDM}$,
when we only take into account the factor $\epsilon$ for the modification of gravity,
as in the ``no-chameleon'' simulations.
In terms of the relative deviation of the matter power spectrum, this does not significantly
improve the range of validity of the predictions as compared to linear theory
(and fares worse at $k > 0.2 h$Mpc$^{-1}$).

The dotted lines are the relative difference of the one-loop power spectra when we
also take into account the first nonlinear vertex $\gamma^s_{2;1,1}$ associated with
modified gravity. This corresponds to truncating the modified gravitational potential
(\ref{Psi-n}) at second order $(\delta\rho)^2$ and neglecting the new contribution
$P^{\Psi}_{31}$ in Fig.~\ref{fig-P1loop}.
As expected, we can see the first clue of the chameleon effect and the one-loop power
spectrum becomes closer to its $\Lambda$-CDM counterpart, in agreement with the
trend shown by the simulations. This extends somewhat the range of validity of
the predictions (in terms of the relative deviation for $P(k)$), up to
$k \sim 0.2 h\mbox{Mpc}^{-1}$.
These results agree with \cite{Koyama2009}.

We have  also computed the results obtained at one-loop order when we go up to order
$(\delta\rho)^3$ for the modified gravitational potential (\ref{Psi-n}), that is, when we
take into account  the new diagram $P^{\Psi}_{31}$ in Fig.~\ref{fig-P1loop}.
It happens that for these models the curves would not be distinguishable from
the dotted lines in Fig.~\ref{fig-dPk_fR_Hu_PT_z0} (hence they are not plotted in the
figure). Thus, for $f(R)$ theories the new contribution $P^{\Psi}_{31}$ is negligible.

\subsubsection{Scalar-tensor models}
\label{1loop-scalar}

\begin{figure}
\begin{center}
\epsfxsize=8.5 cm \epsfysize=6.7 cm {\epsfbox{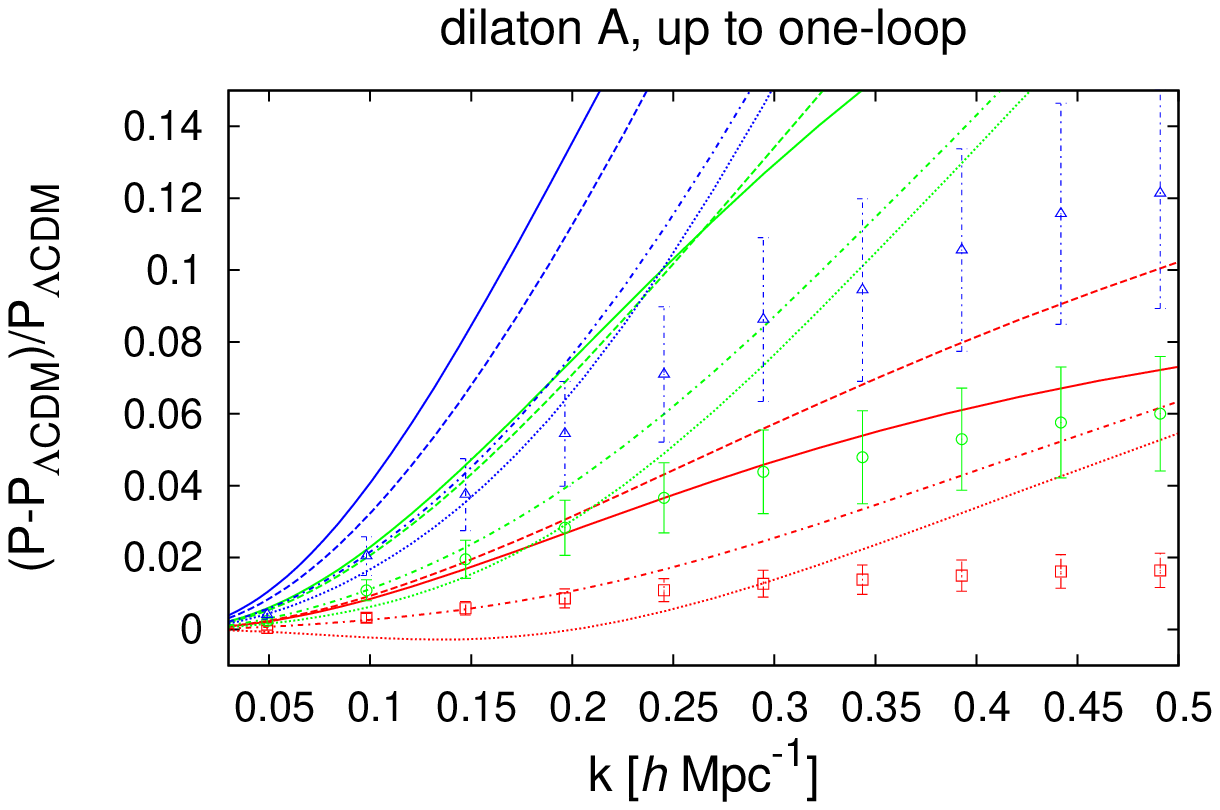}}
\epsfxsize=8.5 cm \epsfysize=6.7 cm {\epsfbox{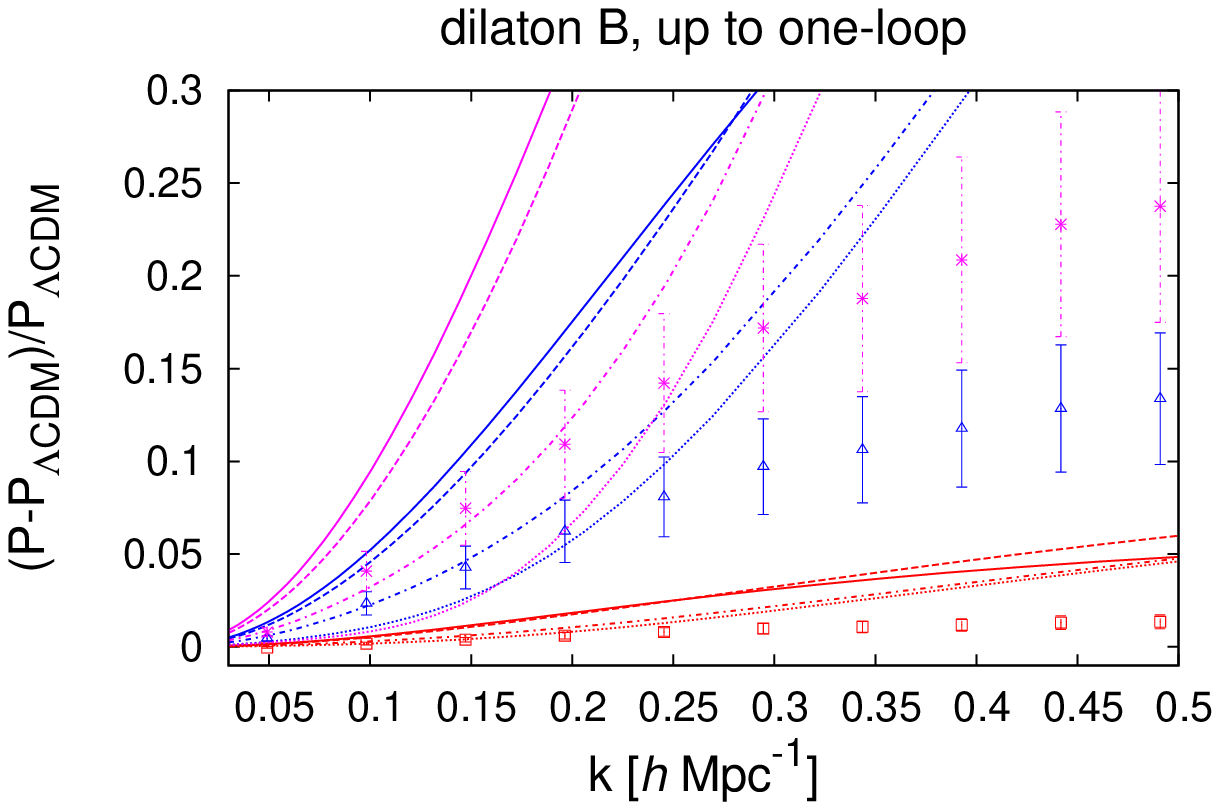}}
\end{center}
\caption{Relative deviation from $\Lambda$-CDM of the power spectrum in dilaton models,
at redshift $z=0$, for models of type A (upper panel, A1, A2, A3 from bottom to top) and
type B (lower panel, B1, B3, B4, from bottom to top).
The symbols are the results for the fully nonlinear power spectrum from the simulations
in \cite{Brax2012}.
We plot the relative deviation of the linear power (solid line),
of the one-loop power without ``screening'' effect
($\gamma^s_{2;1,1}=\gamma^s_{2;1,1,1}=0$) (dashed line), with lowest-order
``screening'' effect ($\gamma^s_{2;1,1}\neq 0$, $\gamma^s_{2;1,1,1}=0$) (dotted line),
and with third-order in $\delta\rho$ ``screening'' effect ($\gamma^s_{2;1,1}\neq 0$,
$\gamma^s_{2;1,1,1} \neq 0$) (dot-dashed line).}
\label{fig-dPk_dilaton_PT_z0}
\end{figure}

We show our results for dilaton models in Fig.~\ref{fig-dPk_dilaton_PT_z0}, for the
power spectrum up to one-loop order.
Here the simulations include the ``screening'' effect to all orders, as they exactly
solve the Klein-Gordon equation (\ref{Klein-Gordon})
(in contrast with the $f(R)$ theories, here we do not have simulation results which follow the
nonlinear evolution of the density field while keeping the Klein-Gordon equation at
the linear order in $\delta\varphi$).

Let us describe our results for the case ``A1'': four lower lines and symbols in the upper
panel.
The solid line is again the relative deviation from $\Lambda$-CDM for the linear
power spectra and it only matches the simulations on very large linear scales,
$k<0.1 h\mbox{Mpc}^{-1}$. The other three lines are the relative deviations of the
one-loop power spectrum when we take into account the effect of modified gravity
on the gravitational potential $\Psi$ up to first, second, and third order over
$\delta\rho$.

The dashed line corresponds to truncation at first order (i.e., only the factor
$\epsilon$ is taken into account in the equation of motion (\ref{O-Ks-def})), which implies
that no screening occurs.
The location with respect to the linear curve depends on the model, because the two
curves correspond to different quantities (linear or one-loop power spectra).

The dotted line takes into account the term of order $(\delta\rho)^2$ in the modified
potential, that is, the new vertex $\gamma^s_{2;11}$.
This nonlinearity corresponds to the lowest order of the screening mechanism and as
we can see in the figure it yields a power spectrum which becomes closer to the
$\Lambda$-CDM one, as compared to the previous dashed line.
In the case of model A1 this even leads to a power spectrum which is smaller than the
$\Lambda$-CDM one for $k \simeq 0.15 h\mbox{Mpc}^{-1}$.
In fact, a numerical computation of the spherical collapse using the equation of motion
truncated at this order for the modified potential $\Psi$ shows that the collapse is
slowed down and even stops before reaching very high densities.
Indeed, whereas the term of order $\delta\rho$ associated to the modification of the
potential $\Psi$ speeds up the collapse (like a scale-dependent amplification of the
Newton constant), the term of order $(\delta\rho)^2$ shows the first sign of the convergence
back to General Relativity in dense environments and corresponds to a slowing down.
If we truncate at this order, as densities become large this quadratic term may become
dominant and halt the collapse. Of course, in the truly nonlinear dynamics higher orders
come into play at this stage and ensure that we actually recover the Newtonian force.

Next, the dot-dashed line includes in addition the term of order $(\delta\rho)^3$ of the
modified potential $\Psi$, that is, the new vertex $\gamma^s_{2;111}$ and the new
diagram $P_{31}^{\Psi}$ in Fig.~\ref{fig-P1loop}.
As could be expected from the discussion above, this higher-order term partly corrects
the ``over-screening'' associated with the previous term of order $(\delta\rho)^2$
and we obtain a result which is slightly above the previous one and in better agreement
with the simulations. This shows the gradual convergence of the results as higher orders
of the screening mechanism are included, over large perturbative scales
(but as for the $\Lambda$-CDM case the standard perturbation theory in powers of
$\tpsi_L$ is not expected to converge very well).
Thus, in contrast with the $f(R)$ theories shown in Fig.~\ref{fig-dPk_fR_Hu_PT_z0},
it appears that the new diagram $P_{31}^{\Psi}$ cannot be neglected and
significantly improves the results.
This again extends the validity of the predictions up to $k \sim 0.2 h\mbox{Mpc}^{-1}$
at $z=0$.
We do not plot the models C and D here because their deviation from $\Lambda$-CDM
is very small on these scales and they show the same behaviors.

\begin{figure}
\begin{center}
\epsfxsize=8.5 cm \epsfysize=6.7 cm {\epsfbox{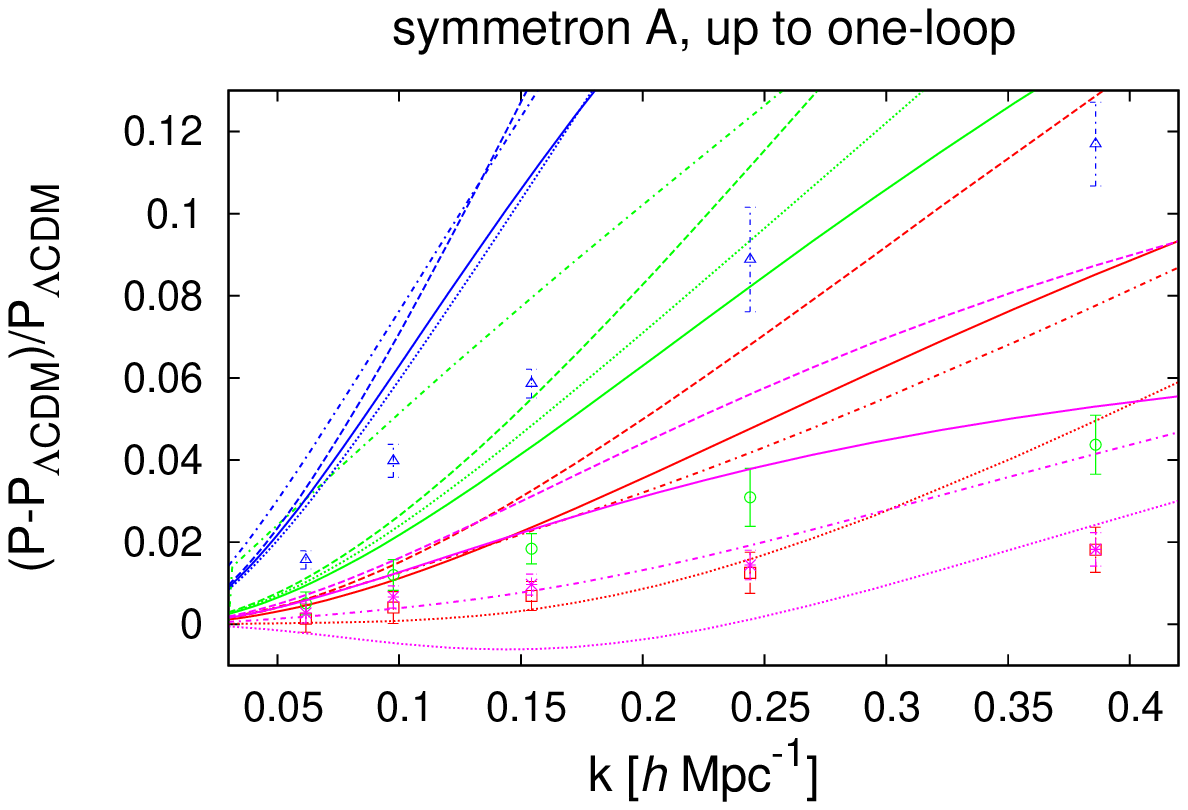}}
\epsfxsize=8.5 cm \epsfysize=6.7 cm {\epsfbox{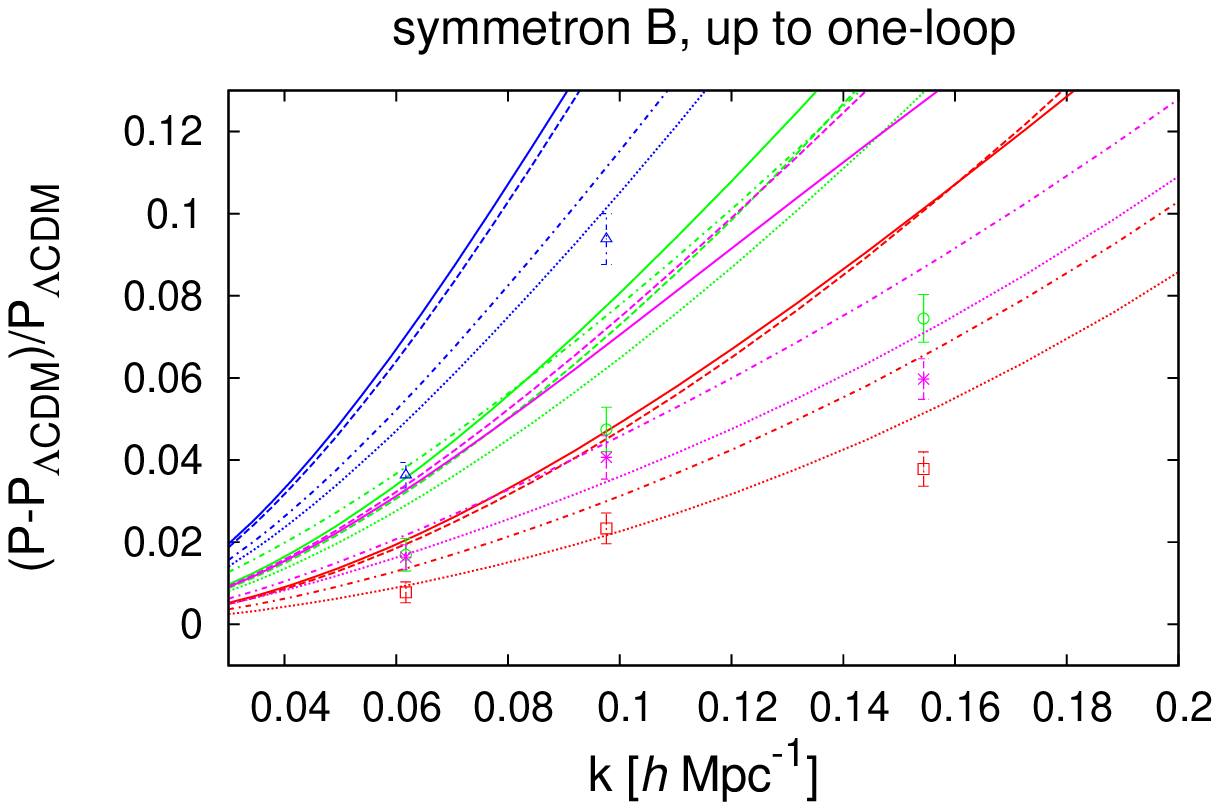}}
\end{center}
\caption{Relative deviation from $\Lambda$-CDM of the power spectrum in symmetron
models, at redshift $z=0$, for models of type A (upper panel, A1: squares,
A2: circles, A3: triangles, A4: stars, which almost coincide with the squares)
and type B (lower panel, B1: squares, B2: circles, B3: triangles, B4: stars).
The symbols are the results for the fully nonlinear power spectrum from the simulations
in \cite{Brax2012}.
We plot the relative deviation of the linear power (solid line),
of the one-loop power without ``screening'' effect
($\gamma^s_{2;1,1}=\gamma^s_{2;1,1,1}=0$) (dashed line), with lowest-order
``screening'' effect ($\gamma^s_{2;1,1}\neq 0$, $\gamma^s_{2;1,1,1}=0$) (dotted line),
and with third-order in $\delta\rho$ ``screening'' effect ($\gamma^s_{2;1,1}\neq 0$,
$\gamma^s_{2;1,1,1} \neq 0$) (dot-dashed line).}
\label{fig-dPk_symmetron_PT_z0}
\end{figure}

We show our perturbative results for the symmetron models in
Fig.~\ref{fig-dPk_symmetron_PT_z0}, using the same line styles as in
Fig.~\ref{fig-dPk_dilaton_PT_z0}.
Again, at one-loop order, including the quadratic term in $(\delta\rho)$ gives a
first screening correction, with a decrease of the one-loop power spectrum with respect
to the one obtained when we only take into account the linear factor $\epsilon$,
whereas the next cubic term in $(\delta\rho)$ partly corrects this screening.
This works best for the cases A1, A4, B1, B3, and B4, where these successive orders
seem to converge, in the sense that the results obtained with the three new factors
$\epsilon$, $\gamma^s_{2;1,1}$, and $\gamma^s_{2;1,1,1}$, lie in-between the
curves obtained with only $\epsilon$ (no screening) and only $\epsilon$ and
$\gamma^s_{2;1,1}$ (over-screening).
There, although we tend to overestimate the deviation from $\Lambda$-CDM, there is
a reasonable agreement with simulations on very large scales (but not as good as for
the dilaton models).
For the models A2, A3, and B2, we find on the contrary that the results obtained with the
three new factors $\epsilon$, $\gamma^s_{2;1,1}$, and $\gamma^s_{2;1,1,1}$, lie
above the curves obtained with only $\epsilon$ (no screening). This means that, at this
order, the expansion in $\delta\rho$ of the screening mechanism has not yet started to
converge.
As could be expected, these cases are those with the lowest values for
the exponents $\{\hat{n},\hat{m}\}$, see Table~\ref{table-symmetron},
whence the most singular functions $\beta(a)$ and
$m(a)$ from Eqs.(\ref{beta-symmetron})-(\ref{m-symmetron}).
Then, their higher-order derivatives $\beta_n$ and $\kappa_n$ diverge faster for
$a\rightarrow a_s$ and the perturbative expansion (\ref{KG-n}) of the Klein-Gordon
equation shows a smaller range of validity.
More generally, the symmetron scenario is associated with a phase transition, from
a single-well potential $V_{\rm eff}=V+\rho (A-1)$ for $a<a_s$ (i.e., for high densities),
to a double-well potential for $a>a_s$ (i.e., for low densities).
Then, it is clear that perturbative approaches, which are best suited for cases where the
background is at the unique minimum of a deep and  isolated potential well, cannot handle
very well epochs close to the transition time.

Therefore, the validity of the perturbative approach depends on the modified gravity
scenarios. Among the three models studied in this paper, the most favorable case is
the $f(R)$ theory, where (at one-loop order) the screening mechanism converges
very fast and the modified potential $\Psi$ can be truncated at quadratic order
$(\delta\rho)^2$.

The dilaton model remains within reach of this perturbative approach, as the expansion
in $\delta\rho$ of the screening converges (at this order) and we find a gradual
improvement as we go from first to third order in $\delta\rho$ for $\Psi$, with
a good match to simulations at this order over the scales described by one-loop
standard perturbation theory.

The symmetron model is the most difficult case, because of the singularity of the
potentials and the coupling functions near the transition $a_s$, which limits the validity
of a perturbative approach. Then, depending on the value of the parameters of the model,
the expansion may have started to converge or not at order $(\delta\rho)^3$.

In any case, these results show that it is important to take into account nonlinear
effects of the modified gravity model. This allows one to extend somewhat the linear
regime, which is limited to very large scales where the deviations from $\Lambda$-CDM
are very small, by going to one-loop (or higher) order and including the first effects of
screening mechanisms. By comparing the results obtained at different orders, one
may also estimate the range of validity of the perturbative expansion, although a
more direct and reliable approach is to compare the perturbative and nonperturbative
contributions within a halo-model framework as described in Sec.~\ref{2-halo-1-halo}
below.
One drawback is that in a fully general parametrization of modified gravity, where
for instance one considers all possible operators or degrees of freedom at a given
order \cite{Baker:2011jy,Gubitosi:2012hu,Battye:2013er}, the number of combinations
increases at higher order and most works have focused on the linear regime.
Therefore, it remains useful to consider specific but still rather broad classes of models,
such as the $f(R)$ and scalar-tensor models studied in this paper. Indeed, using for
instance the tomographic approach described in Sec.~\ref{Equations of motion},
which also applies to the fully nonlinear spherical collapse described in the next
section, the model is fully defined at the nonlinear level. This allows us to compute
the power spectrum on a broad range of scales, as shown in Sec.~\ref{power-spectrum}
below, and to go beyond linear theory, which has a rather limited application.

\section{Spherical collapse}
\label{spherical-collapse}

To go beyond the large scales described by one-loop standard perturbation
theory, we wish to combine the perturbative expansion described in the previous
section with a halo model. This requires a description of the halo mass function
and density profiles. Unfortunately, even for the $\Lambda$-CDM cosmology,
there is no well-controlled modelization of the low-mass tail of the halo mass function
and of the halo density profiles. Therefore, as in \cite{BraxPV2012} we only include
the effects of modified gravity on the large-mass tail of the halo mass function,
which must fall as $e^{-\delta_L^2/(2\sigma_M^2)}$, where $\sigma_M^2$ is the
linear density variance at mass $M$ and $\delta_L$ is the linear density threshold
required to reach a given nonlinear density contrast, which we take as $200$ to
define virialized halos.
This property derives from the Gaussian initial conditions and this rare-event
tail is governed by spherical density fluctuations (because we define halos by
a spherical overdensity criterion).
Therefore, to compute the linear threshold $\delta_L(M)$ we first study the
spherical dynamics in this section.

\subsection{Spherical dynamics}

If the initial conditions are spherically symmetric, the equation of motion of the physical
radius $r(t)$ of a given particle reads as usual as
\beq
\ddot{r} = - \frac{\pl \Psi}{\pl r} = - \frac{\pl\Psi_{\rm N}}{\pl r} - \frac{\pl\Psi_{\rm A}}{\pl r} ,
\eeq
where as in Eq.(\ref{Psi-N+A}) we split the modified gravitational potential as
$\Psi=\Psi_{\rm N}+\Psi_{\rm A}$ and $\Psi_{\rm N}$ is the Newtonian potential given by
the first equation in (\ref{Psi-N-A}) (for the $f(R)$ theories we also define
$\Psi_{\rm A}\equiv \Psi-\Psi_{\rm N}$).
Introducing the comoving Lagrangian coordinate $q$ of each shell, which would enclose
the same initial mass $M$ in a uniform universe, and its normalized radius $y(t)$,
\beq
y(t) = \frac{r(t)}{a(t) q}  \;\;\; \mbox{with} \;\;\;
q= \left( \frac{3M}{4\pi \rhob_0} \right)^{1/3} , \;\;\; y(t \!=\! 0) = 1 ,
\label{y-def}
\eeq
we obtain the equation of motion
\beq
\frac{\pl^2 y}{\pl\eta^2} + \frac{1 \!-\! 3 w \Ode}{2} \frac{\pl y}{\pl\eta}
+ \frac{\Om}{2} (y^{-3} \!-\! 1) y = \frac{-3\Om y}{8\pi\cG\rhob r} \, \frac{\pl\Psi_{\rm A}}{\pl r} .
\label{yt}
\eeq
This equation gives the evolution with time of the field $y(q,\eta)$, and because
the fifth force $-\pl\Psi_{\rm A}/\pl r$ usually depends on the shape of the density profile
we must simultaneously follow the dynamics of all shells, $0<q<\infty$.

In Eq.(\ref{yt}), to write the contribution associated with the Newtonian potential
as $\Om (y^{-3}-1) y/2$, we have assumed that the mass within the shell $q$ is constant.
In principle, it would be possible to write the spherical dynamics without using this
assumption, by following the crossings of different shells. However, it would be
very time-consuming to follow the fast oscillations of the inner shells and not sufficient to
reach a high accuracy because in these collapsed regions a strong radial orbit instability
develops and leads to virialization (the dynamics are singular and infinitesimal deviations
from spherical symmetry are amplified up to the magnitude of the radial motions
\cite{Valageas2002b}).

To bypass this problem and the need to compute the motion of all shells, we
follow \cite{BraxPV2012} and we simplify the equation of motion (\ref{yt}) by focusing
on the shell associated with the mass $M$ of interest and using an ansatz for the
shape of the density profile. In other words, for a given mass $M$, we follow the
dynamics of the radius $r_M(t)$ which contains this mass $M$, using Eq.(\ref{yt}) as the
equation of motion for $y_M(t)$. However, in contrast to the $\Lambda$-CDM case, the
fifth force cannot be written in terms of $y_M$ only, because it depends on the shape
of the profile, and to compute the right-hand side in Eq.(\ref{yt}) we use the density
profile ansatz
\beqa
\hspace{-0.5cm} \delta(x) & = & \frac{\delta_M}{\sigma^2_{x_M}}
\int_{V_M} \frac{\dd\vx'}{V_M} \, \xi_L(\vx,\vx') \\
& = & \frac{\delta_M}{\sigma^2_{x_M}} \int_0^{\infty} \! \dd k \, 4\pi k^2 P_L(k) \tW(k x_M)
\frac{\sin(kx)}{kx} .
\label{profile-ansatz}
\eeqa
Here $\xi_L$ is the linear density correlation function, $\sigma^2_{x_M}$ the
variance of the linear density contrast at the comoving radius $x_M$, which defines the
sphere of volume $V_M$,
$\delta_M=y_M^{-3}-1$ the nonlinear density contrast at radius $x_M$,
and $\tW(kx_M)=3 [\sin(k x_M)-k x_M\cos(k x_M)]/(k x_M)^3$ the Fourier transform
of the top hat of radius $x_M$.
By definition, this profile is normalized so that the density contrast within radius $x_M$
is equal to $\delta_M$. It is also the typical profile of rare events in the linear regime
\cite{Bernardeau1994a,Valageas2002b}, and governs the large-mass tail of the
halo mass function \cite{Valageas2009} (when we neglect the nonlinear distortion of the
profile).
As recalled above, this procedure only applies until the nonlinear density contrast
reaches about $200$, because at higher densities shell crossings modify the
Newtonian force itself.

This approximation transforms Eq.(\ref{yt}) into an ordinary differential equation for
$y_M(t)$, and this defines a function $\delta_M=\cF_M[\delta_{L}]$ which maps the
linear density contrast $\delta_L$ (which defines the initial amplitude of the density
fluctuation) to the nonlinear density contrast $\delta_M$. In contrast with the
$\Lambda$-CDM cosmology, this function $\cF_M$ now depends on the mass $M$
because of the scale dependence of the fifth force.
Next, we can invert this function to obtain the linear density contrast,
$\delta_L=\cF_M^{-1}(\delta)$, associated with a given nonlinear threshold $\delta$.
In particular, defining as in \cite{Valageas2009,BraxPV2012} virialized halos by
a nonlinear density threshold of $200$, we obtain the associated linear threshold
$\delta_L(M)=\cF^{-1}_M(200)$.
This function describes how the formation of massive halos is made easier by the
fifth force, as a smaller linear threshold $\delta_L$ is required as compared to the
$\Lambda$-CDM case.
We describe below our results for this characteristic function for $f(R)$ theories and
scalar-field models.

\subsection{$f(R)$ theories}

In the case of $f(R)$ theories, the fifth force potential reads from Eq.(\ref{Poisson-fR})
as
\beq
\nabla_x^2 \Psi_{\rm A} = \frac{4\pi\cG}{3} \rhob a^2 \, \delta - \frac{a^2}{6} \, \delta R .
\eeq
Introducing the normalized fluctuation $\alpha(x)$ of the Ricci scalar,
\beq
\delta R = 8\pi\cG\rhob \, \alpha(x) ,
\label{alpha-def-fR}
\eeq
we obtain in spherical symmetry
\beq
\frac{\pl\Psi_A}{\pl x} = \frac{4\pi\cG\rhob a^2}{3 x^2} \int_0^x \dd x' \, x'^2 \, (\delta-\alpha) ,
\label{PsiA-fR}
\eeq
where as usual $x=r/a$ is the comoving coordinate.
Then, Eq.(\ref{yt}) writes as
\beqa
\hspace{-0.5cm} \frac{\dd^2 y_M}{\dd\eta^2} + \frac{1 \!-\! 3 w \Ode}{2} \frac{\dd y_M}{\dd\eta}
+ \frac{\Om}{2} (y_M^{-3} \!-\! 1) y_M & = & \nonumber \\
&& \hspace{-4.5cm} \frac{-\Om y_M}{2} \, \int_0^{x_M} \frac{\dd x \; x^2}{x_M^3} \, (\delta-\alpha) ,
\label{yt-fR}
\eeqa
where we focus on the shell associated with a given mass $M$.
On the other hand, the field $R(x)$ (whence $\alpha(x)$) is given by the constraint
equation (\ref{fR-delta}). In spherical symmetry, for the power-law models (\ref{fR-def}),
this yields,
\beqa
\hspace{-0.5cm} \frac{\dd^2\alpha}{\dd x^2} + \frac{2}{x} \frac{\dd \alpha}{\dd x}
- \frac{(n+2) \Omega_{\rm m0}}{\Omega_{\rm m0} (1+\alpha) + 4 \Omega_{\Lambda 0} a^{-3}}
\left( \frac{\dd\alpha}{\dd x} \right)^2 & = & \nonumber \\
&& \hspace{-7cm}  a^2 m_0^2 \left( \frac{\Omega_{\rm m0} a^{-3} (1+\alpha) + 4 \Omega_{\Lambda 0}}{\Omega_{\rm m0} + 4 \Omega_{\Lambda 0}} \right)^{n+2}  (\alpha-\delta ) ,
\label{alpha-fR}
\eeqa
with
\beq
m_0 = \frac{H_0}{c} \sqrt{ \frac{\Omega_{\rm m0}+4\Omega_{\Lambda 0}}
{(n+1) |f_{R_0}|} } .
\eeq
Thus, to compute the spherical dynamics we numerically solve Eqs.(\ref{yt-fR}) and
(\ref{alpha-fR}). At each time step we solve the constraint equation
(\ref{alpha-fR}), using a multigrid relaxation algorithm and the density profile
(\ref{profile-ansatz}), normalized by $\delta_M$ at radius $x_M$, and we advance
over time with Eq.(\ref{yt-fR}).

It is interesting to consider the ``weak-field'' regime, which has been studied in many
previous works \cite{Oyaizu2008,BraxPV2012},
where the constraint equations (\ref{fR-delta}) or (\ref{alpha-fR})
are linearized in $\delta R$ or $\alpha$.
This gives in Fourier space the weak-field expressions
\beq
\talpha_{\rm w.f.} = \frac{a^2 m^2}{a^2m^2+k^2} \; \tdelta , \;\;\;\;
\tilde{\Psi}_{\rm A, w.f.} = \epsilon(k) \tilde{\Psi}_{\rm N} ,
\label{fR-w.f.}
\eeq
where $\epsilon(k)$ was given in Eq.(\ref{eps-fR}).

Equation (\ref{alpha-fR}) is nonlinear and clearly shows the ``chameleon'' mechanism
which ensures convergence to General Relativity in dense environments.
Indeed, the term $(\alpha-\delta)$ tends to make $\alpha$ converge to
$\delta$, so that the fifth force vanishes as seen in Eqs.(\ref{PsiA-fR}) and (\ref{yt-fR})
This happens on large scales, where the spatial derivatives in Eq.(\ref{alpha-fR}) can
be neglected, which corresponds to $k\rightarrow 0$ in the weak-field expression
(\ref{fR-w.f.}), and in very dense regions, where both $\alpha$ and $\delta$ are large.
This latter chameleon mechanism cannot be seen in the linearized solution
(\ref{fR-w.f.}) and is due to the nonlinear character of Eq.(\ref{alpha-fR}).
For large $\alpha$ and $\delta$, the left-hand side scales linearly with $\alpha$ whereas
the right-hand side scales as $\alpha^{n+3}$, so that in sufficiently dense environments
we recover $\alpha \simeq \delta$, up to corrections of order $\delta^{-n-1}$.

\begin{figure}
\begin{center}
\epsfxsize=8.5 cm \epsfysize=6. cm {\epsfbox{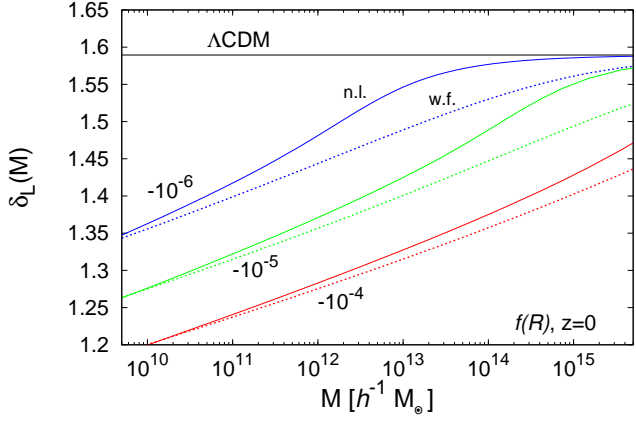}}
\end{center}
\caption{Linear density threshold $\delta_L(M)$, associated with a nonlinear density
contrast $\delta=200$, for $f(R)$ theories at $z=0$. The dotted lines (w.f.) correspond to the
weak-field limit (\ref{fR-w.f.}) and the solid lines (n.l.) to the fully nonlinear constraint
(\ref{alpha-fR}).}
\label{fig-deltaLM_fR_Hu_z0}
\end{figure}

\begin{figure}
\begin{center}
\epsfxsize=8.5 cm \epsfysize=6. cm {\epsfbox{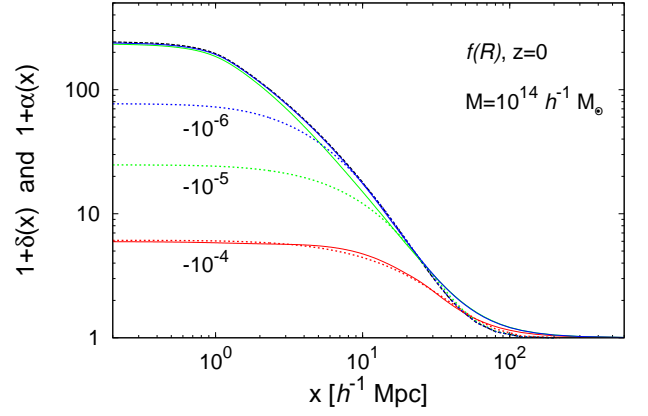}}
\epsfxsize=8.5 cm \epsfysize=6. cm {\epsfbox{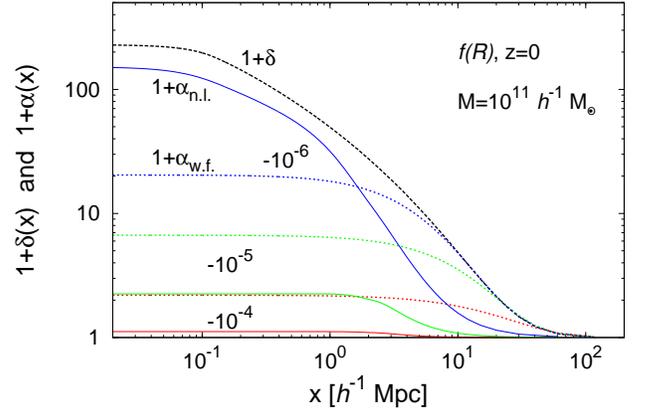}}
\end{center}
\caption{Radial profile of the nonlinear density contrast $\delta(x)$ (upper black dashed line),
and of the field $\alpha(x)$, using the weak-field approximation (\ref{fR-w.f.})
($\alpha_{\rm w.f.}$, dotted lines), or the nonlinear solution of Eq.(\ref{alpha-fR})
($\alpha_{\rm n.l.}$, solid lines). We consider the halo masses
$M=10^{14}h^{-1}M_{\odot}$ (upper panel) and $M=10^{11}h^{-1}M_{\odot}$ (lower panel),
at $z=0$.}
\label{fig-alpha_fR_Hu_z0}
\end{figure}

\begin{figure}
\begin{center}
\epsfxsize=8.5 cm \epsfysize=6. cm {\epsfbox{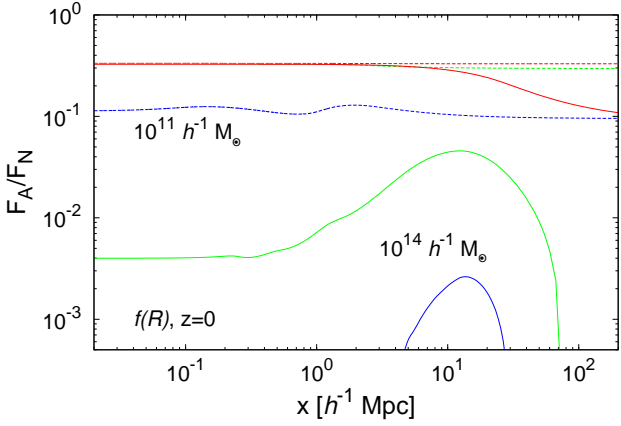}}
\end{center}
\caption{Ratio $F_{\rm A}/F_{\rm N}$ of the fifth force to the Newtonian force, for
$f(R)$ models at $z=0$. The dashed lines are for a halo mass
$M=10^{11}h^{-1}M_{\odot}$ and the solid lines for $M=10^{14}h^{-1}M_{\odot}$.
For both masses, we plot our results for the cases $f_{R_{0}}=-10^{-4},-10^{-5},$
and $-10^{-6}$, from top to bottom.}
\label{fig-FA_FN_fR_Hu_z0}
\end{figure}

We show in Fig.~\ref{fig-deltaLM_fR_Hu_z0} the linear density contrast $\delta_L(M)$
that we obtain at $z=0$, as a function of the halo mass. Because of the fifth force on the
right-hand side in Eq.(\ref{yt-fR}), the collapse is accelerated as compared to the
$\Lambda$-CDM case, and increasingly so for higher $|f_{R_0}|$ and lower masses
(whereas on large scales, we recover General Relativity as $\epsilon(k) \rightarrow 0$ for
$k\rightarrow 0$).
This leads to a linear threshold $\delta_L$, at fixed nonlinear density contrast $\delta=200$,
which is lower than in the $\Lambda$-CDM case and decreases at low mass.

We can check in Fig.~\ref{fig-deltaLM_fR_Hu_z0} that the ``chameleon'' effect, associated
with the nonlinearity of the constraint (\ref{dfR}) or (\ref{alpha-fR}), decreases the
deviation from $\Lambda$-CDM, as compared to the  result which would be obtained using
the weak-field expression (\ref{fR-w.f.}). This is most important on intermediate mass scales,
where nonlinearities overcome spatial gradients and the field $\alpha(x)$ can follow
the matter density field.

This chameleon effect is displayed in Fig.~\ref{fig-alpha_fR_Hu_z0}, where we show the
radial profiles of the density contrast $\delta(x)$ and of the field $\alpha(x)$, using
either the weak-field approximation (\ref{fR-w.f.}) or the nonlinear solution of (\ref{alpha-fR}).
At large radii, we always have $\alpha \simeq \delta \simeq 0$ and the fifth force in
Eq.(\ref{yt-fR})
vanishes. At small radii, because of the factor $k^2$ in Eq.(\ref{fR-w.f.}), associated with
the spatial derivatives in Eq.(\ref{alpha-fR}), the weak-field approximation
$\alpha_{\rm w.f.}$ cannot completely follow the rise of the density profile and
$(\delta-\alpha_{\rm w.f.})$ grows. This gives rise to the fifth force in Eq.(\ref{yt-fR}) and to the
departure from the $\Lambda$-CDM linear density threshold in
Fig.~\ref{fig-deltaLM_fR_Hu_z0}.
The deviation between $\delta$ and $\alpha_{\rm w.f.}$ at the center of the halo increases
for small halo masses. Indeed, because of the Laplacian in Eq.(\ref{fR-delta}), or the
spatial derivatives in Eq.(\ref{alpha-fR}), $\alpha_{\rm w.f.}$ cannot accommodate strong
spatial gradients and cannot follow small-scale density fluctuations.

When we consider the exact solution of the nonlinear Eq.(\ref{alpha-fR}), we can
see in the upper panel of Fig.~\ref{fig-alpha_fR_Hu_z0} that for massive and large
halos the nonlinear field $\alpha(x)$ can follow the rise of the density contrast up to the
halo center, if $|f_{R_0}|$ is not too large. This reduces the fifth force as compared to
the weak-field approximation, in agreement with Fig.~\ref{fig-deltaLM_fR_Hu_z0}.
However, for low-mass halos and small-size objects (at fixed density), the ``cost''
associated with spatial gradients again becomes too important and $\alpha(x)$ cannot
follow the rise of the density field. This implies that within the halo
$\delta-\alpha \simeq \delta$ and the fifth force accelerates the collapse. Moreover,
in this regime the fifth force no longer depends on $\alpha$ and the collapse
follows the weak-field approximation, as seen in Fig.~\ref{fig-deltaLM_fR_Hu_z0}.

These behaviors are illustrated in Fig.~\ref{fig-FA_FN_fR_Hu_z0}, where we show
the ratio $F_{\rm A}/F_{\rm N}$ of the fifth force to the Newtonian force at $z=0$.
Although our density profiles are different from the NFW profiles used in
\cite{Schmidt2010}, we recover the same features.
As explained above,
for the low-mass halo we recover the weak-field limit with a fifth force which is
about one third of the Newtonian force [the high-$k$ limit $\epsilon \rightarrow 1/3$
of Eq.(\ref{eps-fR})], except for the case $f_{R_0} = -10^{-6}$ where the very small
modification of gravity and a stronger chameleon effect yield a smaller fifth force.
For the massive halo, the chameleon effect becomes very important for both
$f_{R_0} = -10^{-5}$ and $-10^{-6}$. This suppresses the fifth force in the
high-density core, while at large radii we recover General Relativity, and the
gravitational force is only modified at about $1 h^{-1}$Mpc for
$M=10^{14}h^{-1}M_{\odot}$.

\subsection{Scalar field models}

In the scalar-tensor theories that we study in this paper, the fifth force is given by
Eq.(\ref{Psi-N-A}),
\beq
\frac{\pl\Psi_{\rm A}}{\pl x} = \frac{c^2}{\MPl} \, \beta(\varphi) \, \frac{\pl\varphi}{\pl x} ,
\eeq
where as in Eqs.(\ref{beta-n-def-scalar}) and (\ref{beta1-kappa2}) we defined
$\beta = \dd A/\dd \varphi$, but the derivative is taken at the local value of $\varphi$
instead of the background $\phib$.
Nevertheless, to express the equations in terms of the function $\beta(a)$ introduced
in Sec.~\ref{Tomography}, we change variables from the field $\varphi(\vx)$ to the field
$\alpha(\vx)$ defined by
\beq
\alpha = a(\varphi) ,
\eeq
where $a(\varphi)$ is the inverse of the function $\phib(a)$, that is, $\alpha(\vx)$ is the scale
factor which was observed when the background value $\phib$ was equal to the present
local value $\varphi(\vx)$.
In particular, from Eq.(\ref{phi-a}) we have
\beq
\frac{\dd \varphi}{\dd\alpha} = \frac{3 \beta_{\alpha} \rhob_{\alpha}}
{c^2\MPl m^2_{\alpha} \alpha} ,
\label{dphi-dalpha}
\eeq
where we note with a subscript $\alpha$ the values of functions taken at point
$\alpha$, such as $\beta_{\alpha}=\beta(\alpha)$, to distinguish from the background
values, such as $\beta=\beta(a)$.
Then, Eq.(\ref{yt}) reads as
\beqa
\hspace{-1cm} \frac{\dd^2 y_M}{\dd\eta^2} + \frac{1 \!-\! 3 w \Ode}{2}
\frac{\dd y_M}{\dd\eta} + \frac{\Om}{2} (y_M^{-3} \!-\! 1) y_M & = & \nonumber \\
&& \hspace{-3cm} \frac{-9\Om a \beta_{\alpha}^2 y_M}{m_{\alpha}^2 \alpha^4 x_M} \;
\frac{\pl\alpha}{\pl x} ,
\label{yt-scalar}
\eeqa
where we again focus on the dynamics of the shell associated with a given mass $M$.
On the other hand, the field $\varphi(x)$ (whence $\alpha(x)$) is given by the
quasistatic Klein-Gordon equation (\ref{Klein-Gordon}).
Using Eq.(\ref{dphi-dalpha}), this reads in spherical symmetry as
\beqa
\hspace{-0.5cm} \frac{\dd^2\alpha}{\dd x^2} + \frac{2}{x} \frac{\dd \alpha}{\dd x}
+ \left[ \frac{\dd\ln\beta_{\alpha}}{\dd\alpha} - 2 \frac{\dd\ln m_{\alpha}}{\dd\alpha}
- \frac{4}{\alpha} \right] \left( \frac{\dd\alpha}{\dd x} \right)^2 & = & \nonumber \\
&& \hspace{-4cm}  \frac{m_{\alpha}^2 \alpha^4}{3 a}
\left[ 1+\delta - \frac{a^3}{\alpha^3} \right] .
\label{alpha-scalar}
\eeqa
Then, to compute the spherical dynamics we numerically solve Eqs.(\ref{yt-scalar}) and
(\ref{alpha-scalar}), using the ansatz (\ref{profile-ansatz}) for the shape of the
density profile.

The ``weak-field'' limit corresponds to linearizing the Klein-Gordon equations
(\ref{Klein-Gordon}) or (\ref{alpha-scalar}) in $\delta\varphi=\varphi-\phib$ or
$\delta\alpha=\alpha-a$.
This gives in Fourier space the weak-field expressions
\beq
\delta \talpha_{\rm w.f.} = \frac{-a^3 m^2}{3(a^2m^2+k^2)} \; \tdelta , \;\;\;\;
\tilde{\Psi}_{\rm A, w.f.} = \epsilon(k) \tilde{\Psi}_{\rm N} ,
\label{scalar-w.f.}
\eeq
where $\epsilon(k)$ was given in Eq.(\ref{eps-scalar}).

On large scales, where the fluctuations are small, we recover the weak-field regime
(\ref{scalar-w.f.}) and we converge to General Relativity in the limit $k\rightarrow 0$
(the spatial gradient and the factor $1/x$ in Eq.(\ref{yt-scalar}) give rise to a factor
$k^2$ as compared to the Newtonian force, which is also seen in the factor
$\epsilon(k)$ in Eq.(\ref{eps-scalar})).

On small scales, a ``screening'' mechanism associated with the nonlinearity
of Eq.(\ref{alpha-scalar}) again ensures that we recover General Relativity in
dense environments, where $\delta\rightarrow +\infty$. However, the details can depend
on the scalar field model.

For dilaton models, where $m(a)$ grows at low $a$, the right-hand side in
Eq.(\ref{alpha-scalar}) makes $\alpha$ converge to $a \delta^{-1/3}$, that is,
$\varphi$ to $\phib(\rhob \rightarrow \rho)$. Indeed, in this limit of large densities the
right-hand side scales as $m_{\alpha}^2 \alpha$ whereas the left-hand side only scales
linearly with $\alpha$. Then, the fifth force on the right-hand side of Eq.(\ref{yt-scalar}) is
suppressed as compared to Newtonian gravity by a factor
$\beta_{\alpha}^2/m_{\alpha}^2$.

For symmetron models with $\hat{m}>1/2$, in dense regions we have
$\alpha \rightarrow a_s$ and more precisely $(\alpha-a_s) \sim \delta^{-1/(2\hat{m}-1)}$.
Then, the fifth force on the right-hand side of Eq.(\ref{yt-scalar}) is
suppressed as compared to Newtonian gravity by a factor
$\delta^{-2\hat{n}/(2\hat{m}-1)}$.
If $\hat{m}<1/2$ we exactly have $\alpha=a_s$ in very dense regions
(with a singular growth at the boundary of the constant-$\alpha$ region of the
form $(\alpha-a_s) \sim (x-x_s)^{2/(1-2\hat{m})}$), and the fifth force is
exactly zero in this domain.

\begin{figure}
\begin{center}
\epsfxsize=8.5 cm \epsfysize=6. cm {\epsfbox{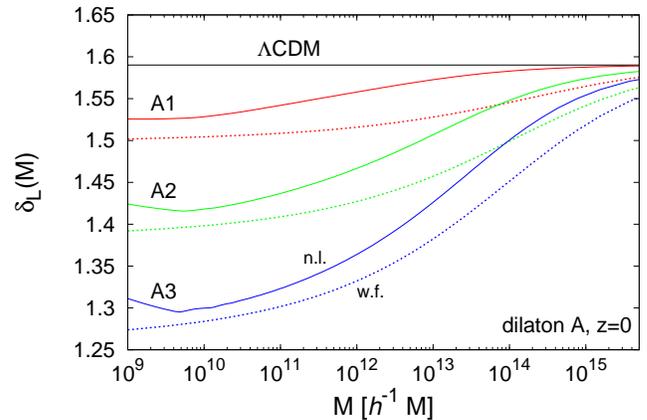}}
\end{center}
\caption{Linear density threshold $\delta_L(M)$, associated with a
nonlinear density contrast $\delta=200$, for some dilaton models at $z=0$
(cases A1, A2, and A3 from top to bottom).
The dotted lines (w.f.) are the weak-field limit (\ref{scalar-w.f.}) and the solid lines (n.l.)
the solution to the fully nonlinear constraint (\ref{alpha-scalar}).}
\label{fig-deltaLM_dilaton_z0}
\end{figure}

\begin{figure}
\begin{center}
\epsfxsize=8.5 cm \epsfysize=6. cm {\epsfbox{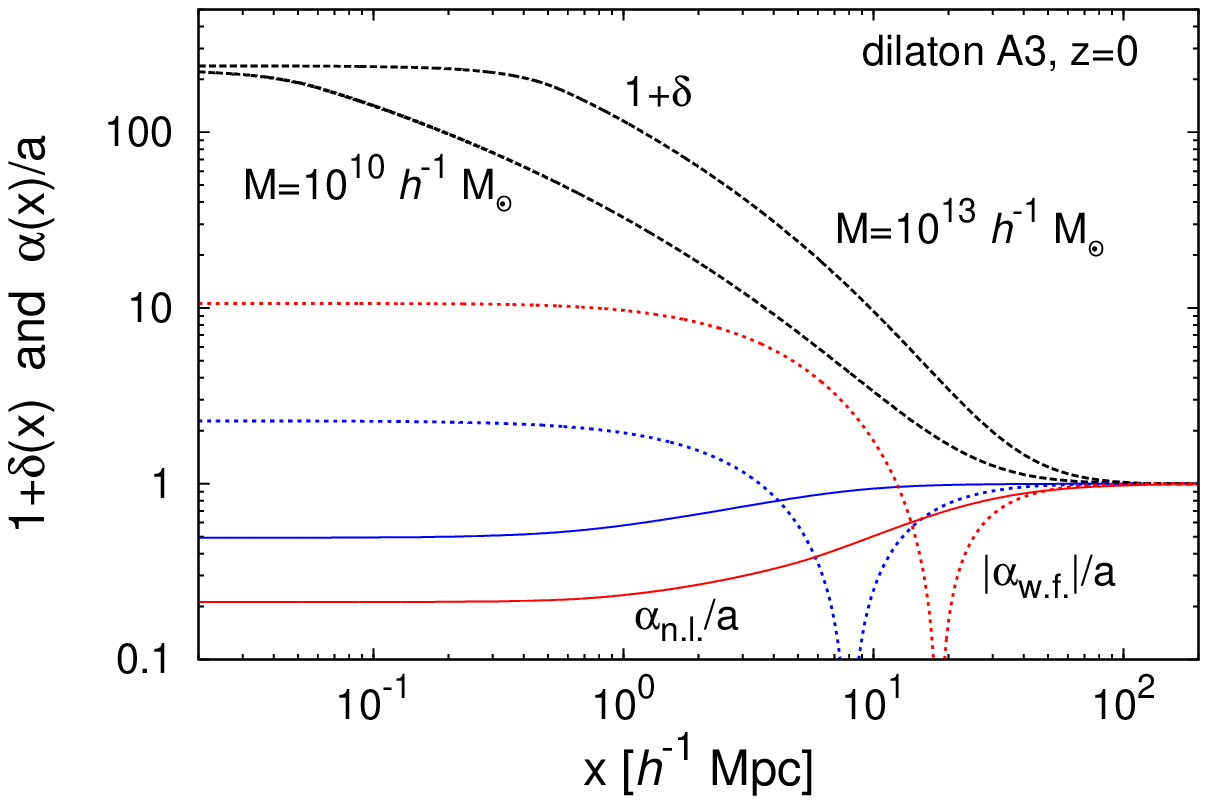}}
\epsfxsize=8.5 cm \epsfysize=6. cm {\epsfbox{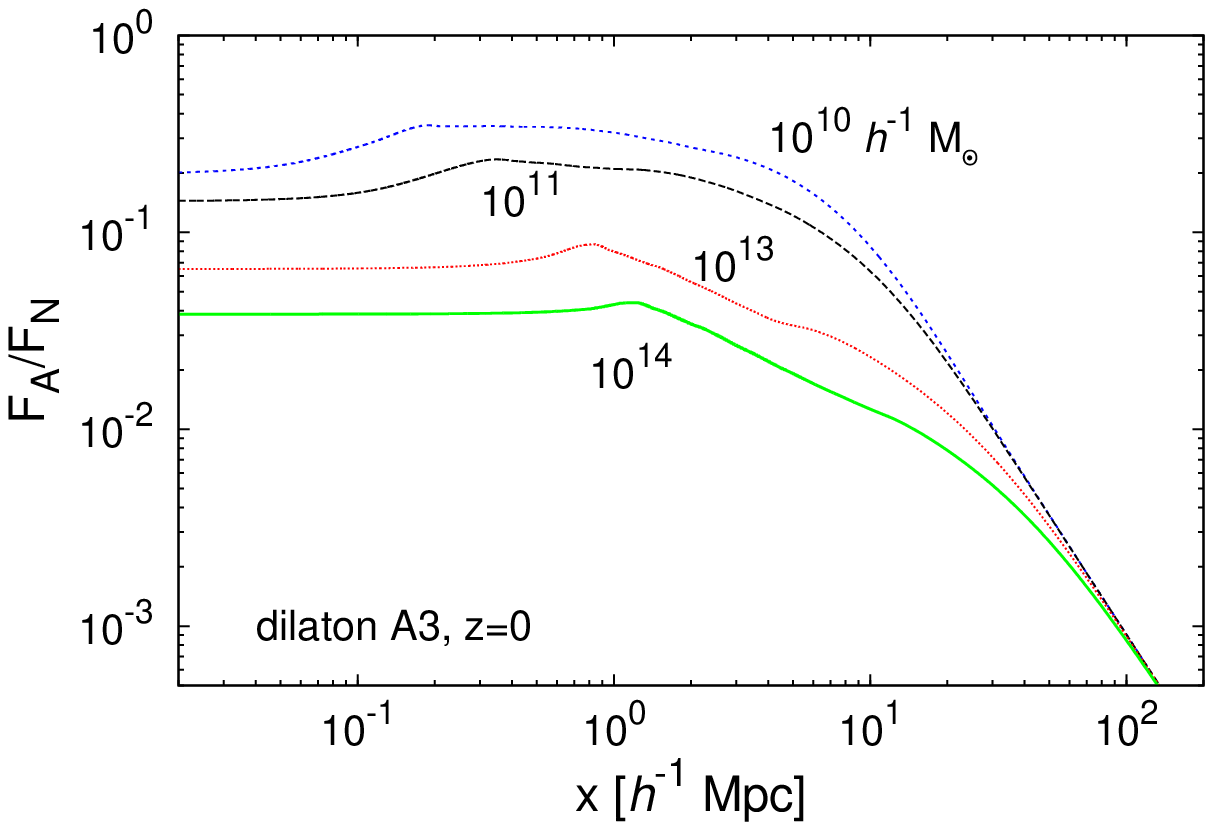}}
\end{center}
\caption{{\it Upper panel:} Radial profile of the nonlinear density contrast $\delta(x)$
(black dashed lines) and of the field $\alpha(x)$, using the weak-field approximation
(\ref{scalar-w.f.}) ($\alpha_{\rm w.f.}$, dotted lines), or the nonlinear solution of
Eq.(\ref{alpha-scalar}) ($\alpha_{\rm n.l.}$, solid lines). We consider the halo masses
$M=10^{13}$ and $10^{10}h^{-1}M_{\odot}$ (where the blue curves for $\alpha/a$
show a smaller deviation from unity, which also appears at a smaller scale), for
the dilaton model A3 at $z=0$.
{\it Lower panel:} Ratio $F_{\rm A}/F_{\rm N}$ of the fifth force to the Newtonian force,
for the dilaton model A3 at $z=0$ (with screening effect).
We show our results for the halo masses $M=10^{10}, 10^{11}, 10^{13}$, and
$10^{14}h^{-1}M_{\odot}$, from top to bottom.}
\label{fig-FA_FN_dilaton_A3_z0}
\end{figure}

\begin{figure}
\begin{center}
\epsfxsize=8.5 cm \epsfysize=6. cm {\epsfbox{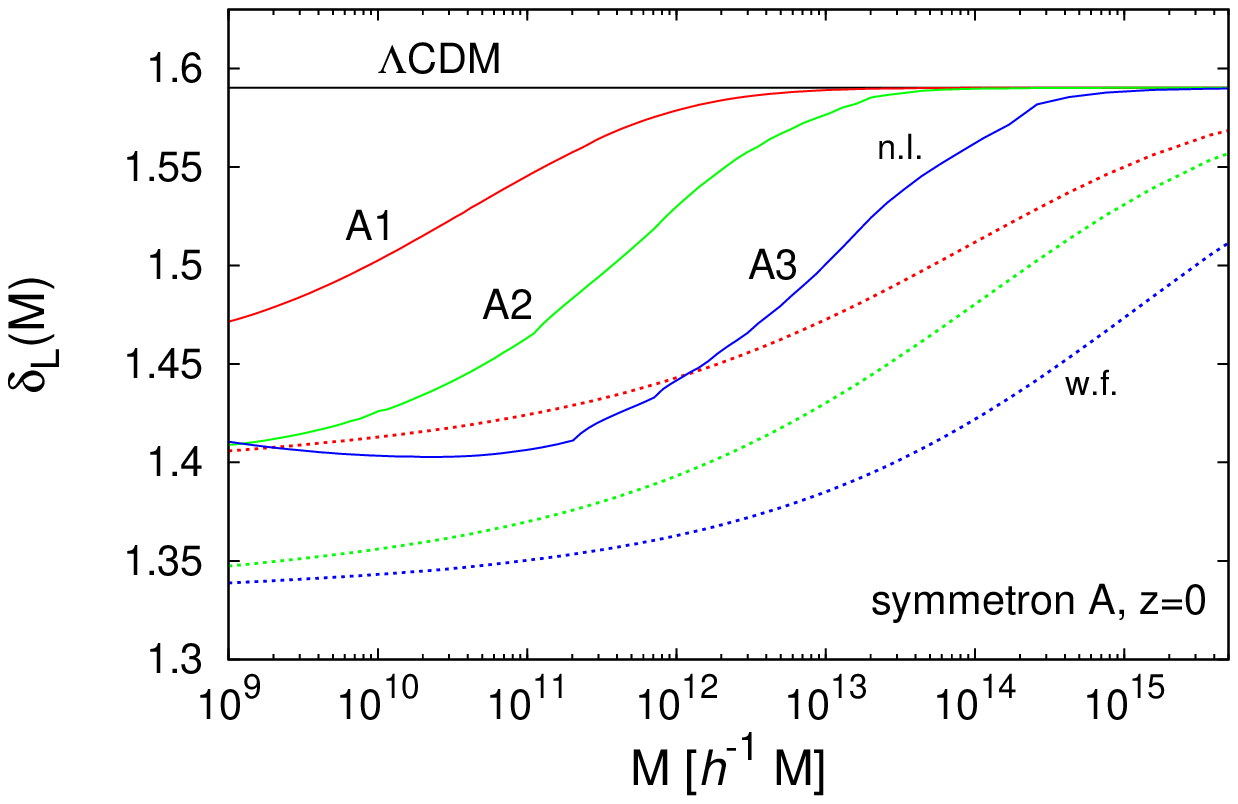}}
\end{center}
\caption{Linear density threshold $\delta_L(M)$, associated with a
nonlinear density contrast $\delta=200$, for some symmetron models at $z=0$
(cases A1, A2, and A3 from top to bottom).
The dotted lines (w.f.) are the weak-field limit (\ref{scalar-w.f.}) and the solid lines (n.l.)
the solution to the fully nonlinear constraint (\ref{alpha-scalar}).}
\label{fig-deltaLM_symmetron_z0}
\end{figure}

\begin{figure}
\begin{center}
\epsfxsize=8.5 cm \epsfysize=6. cm {\epsfbox{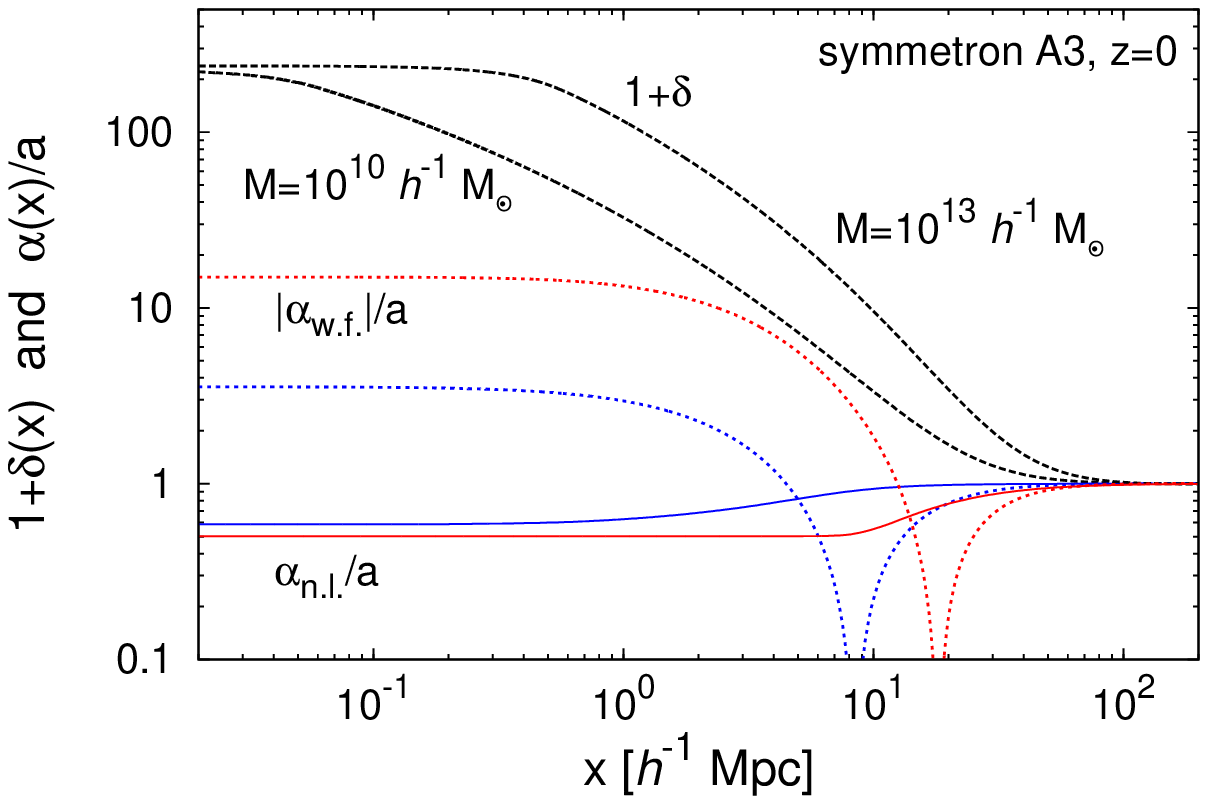}}
\epsfxsize=8.5 cm \epsfysize=6. cm {\epsfbox{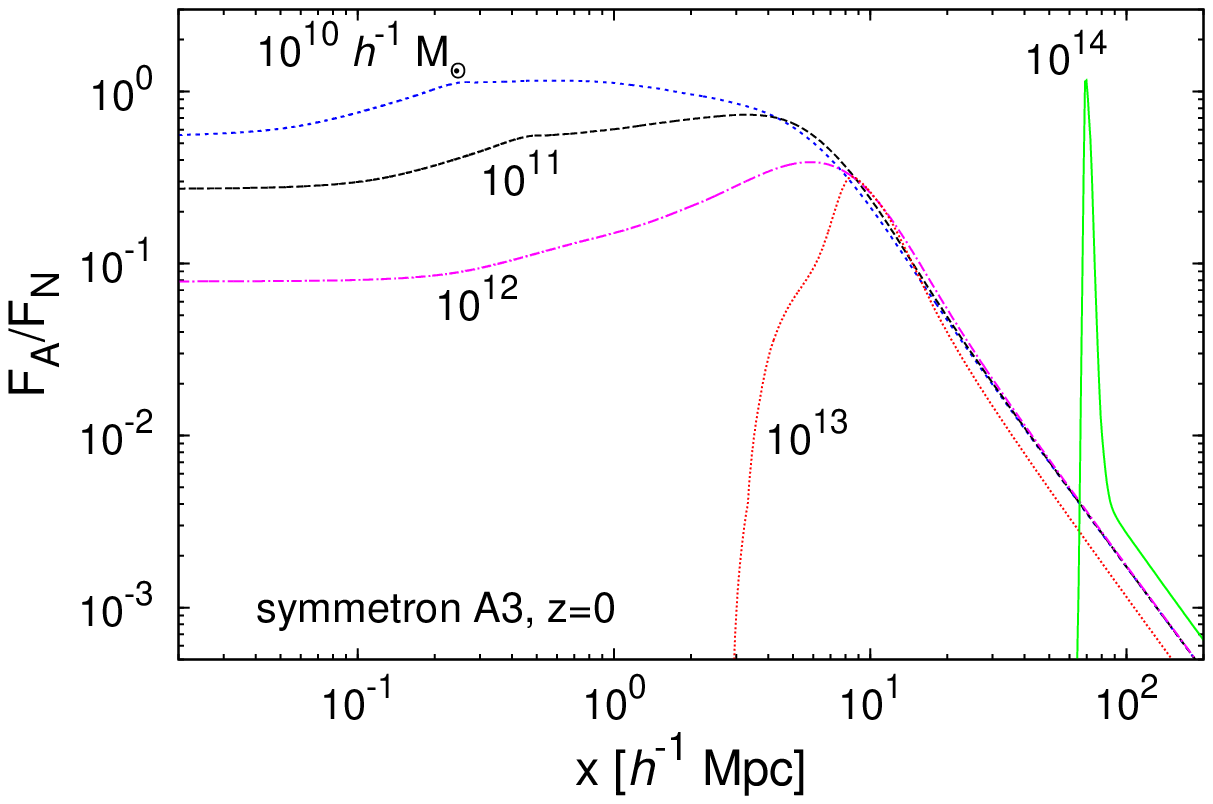}}
\end{center}
\caption{{\it Upper panel:} Radial profile of the nonlinear density contrast $\delta(x)$
(black dashed lines) and of the field $\alpha(x)$, using the weak-field approximation
(\ref{scalar-w.f.}) ($\alpha_{\rm w.f.}$, dotted lines), or the nonlinear solution of
Eq.(\ref{alpha-scalar}) ($\alpha_{\rm n.l.}$, solid lines). We consider the halo masses
$M=10^{13}$ and $10^{10}h^{-1}M_{\odot}$ (where the blue curves for $\alpha/a$
show a smaller deviation from unity, which also appears at a smaller scale), for the
symmetron model A3 at $z=0$.
{\it Lower panel:} Ratio $F_{\rm A}/F_{\rm N}$ of the fifth force to the Newtonian force,
for the symmetron model A3 at $z=0$ (with screening effect).
We show our results for the halo masses $M=10^{10}, 10^{11}, 10^{13}$, and
$10^{14}h^{-1}M_{\odot}$, from top to bottom.}
\label{fig-FA_FN_symmetron_A3_z0}
\end{figure}

We illustrate our results for some dilaton models in Fig.~\ref{fig-deltaLM_dilaton_z0}.
We can check that the fifth force accelerates the collapse and leads
to a smaller linear density threshold $\delta_L(M)$, as compared to the $\Lambda$-CDM
case. Again, the nonlinearities decrease the departure from the $\Lambda$-CDM case,
as compared to the weak-field approximation (\ref{scalar-w.f.}).
In contrast with the results for $f(R)$ theories shown in Fig.~\ref{fig-deltaLM_fR_Hu_z0},
at very low mass we do not converge to the weak-field result but to the $\Lambda$-CDM
threshold. This is due to the fact that the fifth force depends on the fields $\alpha$
in a very different fashion in Eqs.(\ref{yt-fR}) and (\ref{yt-scalar}).
In the $f(R)$ case, a small value of $\alpha$ implies a fifth force which is proportional
to the Newtonian force and no longer depends on the precise value of $\alpha$,
whereas in the scalar field case the fifth force does not relate to the Newtonian force
and remains sensitive to the local value and slope of $\alpha(x)$.

Thus, as seen in the upper panel of Fig.~\ref{fig-FA_FN_dilaton_A3_z0}, the weak-field
approximation yields larger deviations from unity for the ratio $\alpha/a$, which
can become negative on small scales, as compared to the fully nonlinear solution,
which is restricted to $0<\alpha<a$ (for overdense regions).
These constraints imply a smaller range for the nonlinear value and slope of $\alpha$,
which leads to a smaller fifth force. On very small scales, this ensures a convergence
back to General Relativity, which is thus recovered over a broader regime than in $f(R)$
theories.
The lower panel of Fig.~\ref{fig-FA_FN_dilaton_A3_z0} shows how the fifth force
decreases, with respect to the Newtonian force, for larger objects in the range
$M>10^{10}h^{-1}M_{\odot}$.
At large radii it quickly decays as $1/x^2$ as we recover General Relativity
[the factors $1/x_M \, \pl/\pl x$ in Eq.(\ref{yt-scalar}) or $k^2$ in Eq.(\ref{eps-scalar})].
Thus, as compared to Fig.~\ref{fig-FA_FN_fR_Hu_z0}, the modification of gravity is
more localized than in the $f(R)$ models for low-mass halos.
This is because it depends on the local value and slope of the new field
$\varphi(\vx)$, or $\alpha(\vx)$, which makes a fast convergence to General Relativity
possible, following the relaxation of $\varphi$ towards $\phib$.
In contrast, in the $f(R)$ model, if there is a significant modification of gravity in inner
regions, because of a nonzero value of $(\delta-\alpha)$ in Eq.(\ref{yt-fR}) in the core,
its effect at large radii decays in the same manner
as the Newtonian contribution itself (but we still recover the Hubble flow because this
Newtonian force, associated with the overdensity with respect to the mean,
also decays at large distances).

We show our results for some symmetron models in Fig.~\ref{fig-deltaLM_symmetron_z0}.
The general behavior is similar to the one found for dilaton models in
Fig.~\ref{fig-deltaLM_dilaton_z0}, with a linear density threshold $\delta_L(M)$ which is smaller
than the $\Lambda$-CDM one, towards which it converges at large mass.
Again, the result obtained with the exact nonlinear solution of Eq.(\ref{alpha-scalar})
is closer to the $\Lambda$-CDM one, as compared to the weak-field approximation,
and converges back to the $\Lambda$-CDM threshold at very low masses
(this can only be seen for the case A3 in the figure but we checked that at smaller
mass the curves A1 and A2 show the same upturn).
However, as compared to the dilaton models of Fig.~\ref{fig-deltaLM_dilaton_z0},
the difference between the weak-field approximation and the nonlinear result is
much greater. In particular, at high mass the nonlinear result quickly becomes very close
to the $\Lambda$-CDM threshold.

These features are due to the behavior of the field $\alpha(x)$, illustrated in the
upper panel of Fig.~\ref{fig-FA_FN_symmetron_A3_z0}. As noticed above, the screening
mechanism is very efficient because of the lower limit $\alpha \geq a_s$. For the
case $\hat{m}=1/2$ shown in the figure, which is at the boundary between the
regimes $\hat{m}<1/2$ and $\hat{m}>1/2$, the field $\alpha(x)$ in high-density regions
is neither equal to $a_s$ or above $a_s$ by a factor of order $\delta^{-1/(2\hat{m}-1)}$,
but becomes exponentially close as $(\alpha-a_s) \sim e^{-\sqrt{\delta} m_0 (L-x)}$,
where $L$ is the radius of the high-density region. This yields a fifth force which
also decays with $\delta$ as $e^{-\sqrt{\delta} m_0 (L-x)}$.
This behavior is reached for massive halos, where spatial gradients are small and
$\alpha(x)$ can follow the rise of the density contrast until it comes very close to $a_s$.
For low-mass halos, at fixed density, spatial gradients come into play and stop
$\alpha(x)$ before it gets very close to $a_s$.
In both cases, this greatly decreases the fifth force as compared to the weak-field
approximation.

This is also illustrated by the ratio $F_{\rm A}/F_{\rm N}$ shown in the lower panel
of Fig.~\ref{fig-FA_FN_symmetron_A3_z0}. The profile of the ratio $F_{\rm A}/F_{\rm N}$ has already been studied in \cite{Clampitt:2011mx} but in a very different regime
as they considered a symmetry breaking scale factor $a_s=1$.
In such a case, at $z>0$ there is no deviation from $\Lambda$-CDM at all
orders of perturbation theory (because $\varphi=0$ is the single minimum
of the effective potential over a finite range of densities around the background $\rhob$)
nor for the spherical collapse of the overdensity (\ref{profile-ansatz}), which is
typically overdense at all radii.
In \cite{Clampitt:2011mx} they still find a nonzero fifth force because they consider
isolated NFW density profiles, with the boundary condition $\rho \rightarrow 0$
at $x\rightarrow \infty$, whereas our profile (\ref{profile-ansatz}) satisfies
$\rho \rightarrow \rhob$ at large distances, which is more realistic in the early stages
of the collapse.
Nevertheless, these remarks again show that symmetron models with
$a_s \sim 1$ are difficult to describe by analytical means, because they
involve two different phases. An accurate treatment would require a specific method
which explicitly takes into account these two phases but we do not consider it in
this paper as we wish to investigate the general method which applies to generic
modified gravity models.

In our case, where $a_s<1$,
for low-mass halos we recover a behavior which is similar to the one obtained
for dilaton models in Fig.~\ref{fig-FA_FN_dilaton_A3_z0}, because the
field $\varphi(\vx)$, or $\alpha(\vx)$, only probes its regular domain.
For high-mass halos, there is enough room (spatial gradients are less constraining)
for the field $\varphi(\vx)$ to depart from the background value $\phib$ and to
come close to the singular limit $\phib(a_s)$ (i.e., $\alpha=a_s$).
This leads to an almost constant field $\alpha(\vx) \simeq a_s$ in the core
and a vanishing fifth force, as seen by the sharp decay at small radii
in the two cases $M=10^{13}$ and $10^{14}h^{-1}M_{\odot}$.
In the latter case, this gives rise to a localized fifth force at the boundary of the
constant-$\alpha$ region, whereas we always recover as for the dilaton models the
$1/x^2$ decay at large radii. In this case, the symmetron shows features similar to the original chameleon model where a ``thin shell'' entirely responsible for modified gravity develops close
to the surface of the body.
It is likely that this sharp feature is unstable with respect to deviations from spherical
symmetry or gives rise to small-scale perturbations and shell crossings at this radius.
This suggests that in such singular models the collapse may be significantly modified
in localized regions and that the spherical dynamics may not be as efficient as
in the $\Lambda$-CDM cosmology to understand the formation of massive halos.

Thus, we obtain for the spherical collapse of overdensities up to $\delta=200$ the same
trends as those found in Sec.~\ref{numerical-1loop} in the perturbative regime.
The effects of nonlinearities (associated with the chameleon mechanism) are moderate
for the $f(R)$ theories, somewhat greater for the dilaton models, and very large for the
symmetron models.
Then, deviations from the $\Lambda$-CDM dynamics increase at a qualitative level
as we go from $f(R)$ theories to dilaton models, and next to symmetron models.

\section{Matter power spectrum}
\label{power-spectrum}

We have seen in Sec.~\ref{Perturbative} that standard one-loop perturbation theory
does not allow us to go far in the nonlinear regime, where most of the departure from
General Relativity occurs for the models that we consider in this paper.
Therefore, we need a model which applies to a broader range of scales.
In this paper, we use the model developed in \cite{Valageas2013}, which combines
perturbation theory with halo models to provide the matter power spectrum from large
linear scales down to small highly nonlinear scales (see the appendix for details).
As in usual halo models, it splits the matter power spectrum as
\beq
P(k) = P_{\rm 1H}(k) + P_{\rm 2H}(k) ,
\label{Pk-halos}
\eeq
where $P_{\rm 1H}$ is the contribution associated with pairs of particles which belong to the
same halo, whereas $P_{\rm 2H}$ is the contribution associated with pairs of particles which belong to two different halos.

Then, the first contribution reads as
\beq
P_{\rm 1H}(k) = \int_0^{\infty} \frac{\dd\nu}{\nu} f(\nu) \frac{M}{\rhob (2\pi)^3}
\left(  \tu_M(k) - \tW(k q_M) \right)^2 ,
\label{Pk-1H}
\eeq
where $\tu_M(k)$ is the normalized Fourier transform of the halo radial profile,
$\tW(kq_M)$ is the normalized Fourier transform of the top hat of radius $q_M$,
and $f(\nu)$ is the normalized halo mass function, defined as
\beq
n(M) \frac{\dd M}{M} = \frac{\rhob}{M} \, f(\nu) \frac{\dd\nu}{\nu} , \;\;
\mbox{with} \;\; \nu = \frac{\delta_L(M)}{\sigma(M)} .
\label{fnu-def}
\eeq
Here $\sigma(M)$ is the root mean square of the linear density contrast at scale $M$ and
$\delta_L=\cF_M^{-1}(200)$
is the linear density contrast associated with the nonlinear density threshold which defines
collapsed halos, which we choose to be $200$.
As described in Sec.~\ref{spherical-collapse}, $\delta_L(M)$ depends on the mass
because of the scale dependence introduced by the modifications to gravity, and it is
lower than the linear density threshold obtained in the $\Lambda$-CDM case.
This helps the formation of massive halos and increases the one-halo contribution
(\ref{Pk-1H}).
In numerical computations, we use for $f(\nu)$ the fit from
\cite{Valageas2009}, which has been shown to match numerical simulations  while obeying
the asymptotic large-mass tail $f(\nu) \sim e^{-\nu^2/2}$ \cite{Valageas2002b}.
For the halo profiles, we choose the usual NFW profile \cite{Navarro1997}
and the mass-concentration relation from \cite{Valageas2013}.
This means that we neglect the impact of modified gravity on the halo profiles and
we only take into account its effect on the density threshold $\delta_L(M)$.

Next, the two-halo contribution becomes
\beqa
P_{\rm 2H}(k) & = & \int \frac{\dd\Delta\vq}{(2\pi)^3} \; F_{\rm 2H}(\Delta q) \;
\lag e^{\ii\vk\cdot\Delta\vx} \rag^{\rm vir}_{\Delta q} \;
\frac{1}{1+A_1} \nonumber \\
&& \hspace{-1.4cm} \times \; e^{-\frac{1}{2} k^2 (1-\mu^2) \sigma_{\perp}^2} \;
\biggl \lbrace e^{-\varphi_{\parallel}(-\ii k\mu\Delta q \, \sigma^2_{\kappa_{\parallel}})
/\sigma_{\kappa_{\parallel}}^2} + A_1 \nonumber \\
&& \hspace{-1.4cm} \!\! + \!\! \int_{0^+-\ii\infty}^{0^++\ii\infty} \frac{\dd y}{2\pi\ii} \;
e^{-\varphi_{\parallel}(y)/\sigma^2_{\kappa_{\parallel}}}
\left( \! \frac{1}{y} - \frac{1}{y\! + \! \ii k\mu\Delta q \, \sigma^2_{\kappa_{\parallel}}} \! \right)
\!\! \biggl \rbrace  . \nonumber \\
&&
\label{Pk-2H-1}
\eeqa
Let us briefly explain the derivation of Eq.(\ref{Pk-2H-1}) (see the appendix too). It is based on the exact
expression \cite{Schneider1995,Taylor1996},
\beq
P(k) = \int\frac{\dd\Delta\vq}{(2\pi)^3} \, \lag e^{\ii \vk \cdot \Delta\vx} -
e^{\ii\vk\cdot\Delta\vq} \rag ,
\label{Pkxq}
\eeq
which relates the matter power spectrum to the statistics of the Eulerian separation,
$\Delta\vx=\vx_2-\vx_1$, of pairs of particles with initial Lagrangian separation
$\Delta\vq=\vq_2-\vq_1$.
Then, the factor $F_{\rm 2H}$ in Eq.(\ref{Pk-2H-1}) is the probability that a pair of
separation $\Delta\vq$ belongs to two different halos, the factor
$\lag e^{\ii\vk\cdot\Delta\vx} \rag^{\rm vir}_{\Delta q}$ is the contribution to
$e^{\ii\vk\cdot\Delta\vx}$ due to internal motions within each halo,
the factor $e^{-\frac{1}{2} k^2 (1-\mu^2) \sigma_{\perp}^2}$ is the contribution associated
with large-scale motions transverse to the initial separation $\Delta\vq$
(which are taken from Lagrangian linear theory, whence the Gaussian result),
and the factor $e^{-\varphi_{\parallel}(-\ii k\mu\Delta q \, \sigma^2_{\kappa_{\parallel}})
/\sigma_{\kappa_{\parallel}}^2}$ is the contribution associated with large-scale longitudinal
motions.
Here $\sigma_{\perp}^2$ and $\sigma_{\kappa_{\parallel}}^2$ are the variances of the
transverse and longitudinal relative displacements as given by linear theory (up to
normalization factors).
The factor $A_1$ (which depends on $\varphi_{\parallel}$) and the complex integral
in the last term arise from an adhesion-like regularization to mimic the formation of
pancakes, see \cite{Valageas2013} for details.

For our purposes, the main point is that the expression (\ref{Pk-2H-1})
satisfies the following constraints:

(a) It has a perturbative expansion in integer powers of $P_L$, as in standard
perturbation theory (but it also includes some nonperturbative contributions of the
form $e^{-1/\sigma^2}$).

(b) It is consistent with linear theory.

(c) It is consistent with one-loop perturbation theory, when the skewness $S_3$
of the scale-dependent characteristic function $\varphi_{\parallel}$ is given by
\beqa
S_3(\Delta q) & = & - \frac{24\pi}{\sigma_{\kappa_{\parallel}}^4}
\int_0^{\infty} \dd k \; \frac{P_{\rm 1loop}(k) - P^{\rm Z}_{\rm 1loop}(k)}{(\Delta q)^4 k^2}
\nonumber \\
&& \times \left[ 2 + \cos(k \Delta q) - 3 \frac{\sin(k\Delta q)}{k\Delta q} \right] ,
\label{S3-1loop}
\eeqa
where $P^{\rm Z}_{\rm 1loop}$ is the one-loop power spectrum associated with the
Zel'dovich dynamics \cite{ZelDovich1970} while $P_{\rm 1loop}$ is the true
one-loop power spectrum (thus, this is also a measure of the deviation from the
simple Zel'dovich dynamics, which is recovered at all perturbative orders when
$S_3=0$).
In practice, as in \cite{Valageas2013}, this is implemented by choosing for the
characteristic function $\varphi_{\parallel}$ the ansatz
\beq
\varphi_{\parallel}(y) = \frac{1-\alpha}{\alpha} \left(1+\frac{y}{1-\alpha}\right)^{\alpha}
- \frac{1-\alpha}{\alpha} ,
\label{phi-alpha-def}
\eeq
where the scale-dependent parameter $\alpha(\Delta q)$ is given by
\beq
\alpha(\Delta q) = \frac{2-S_3}{1-S_3} .
\label{S3-alpha}
\eeq

(d) The underlying non-Gaussian probability distribution $\cP(\Delta x_{\parallel})$
is everywhere positive, normalized to unity, and satisfies the constraint
$\lag \Delta x_{\parallel} \rag = \Delta q$.

(e) It is well behaved at high $k$, where it remains positive while becoming subdominant
with respect to the one-halo contribution.

Standard perturbation theory clearly satisfies points (a) to (c) but not points (d) and (e).
In particular, it is well known that when truncated at a finite order it can lead to
power spectra which become unphysically large or negative at high $k$. In contrast,
Eq.(\ref{Pk-2H-1}) is built as a regularization of perturbation theory which always remains
consistent with some physical constraints such as point (d), which ensures that this
contribution to the matter power spectrum remains well behaved at high $k$ (positive
with typically a $k^{-2}$ decay).
Together with the one-halo contribution (\ref{Pk-1H}), this provides a realistic
description of the matter power spectrum from large to small scales,
which has been compared to numerical simulations for the $\Lambda$-CDM cosmology
in \cite{Valageas2013}.

This model (\ref{Pk-halos}) can be at once applied to the modified gravity scenarios
that we consider in this paper. For the two-halo contribution (\ref{Pk-2H-1}), we
need to provide the linear power spectrum (\ref{P-tree}) and the one-loop contribution
(\ref{P-1loop}), which determines the characteristic function $\varphi_{\parallel}$
through Eqs.(\ref{S3-1loop})-(\ref{S3-alpha}).
For the one-halo contribution (\ref{Pk-1H}), we need to provide the threshold
 $\delta_L(M)$, obtained in Sec.~\ref{spherical-collapse}.
 As compared to the $\Lambda$-CDM case, the main new sources of inaccuracy
 are that we neglect the impact of modified gravity on the low-mass slope of the
 halo mass function and on the shapes of halo profiles (whereas in the
 $\Lambda$-CDM case these parameters have already been fitted to numerical
 simulations, for instance by choosing the NFW profile).

\subsection{$f(R)$ theories}
\label{Pk-fR}

\begin{figure}
\begin{center}
\epsfxsize=8.5 cm \epsfysize=6. cm {\epsfbox{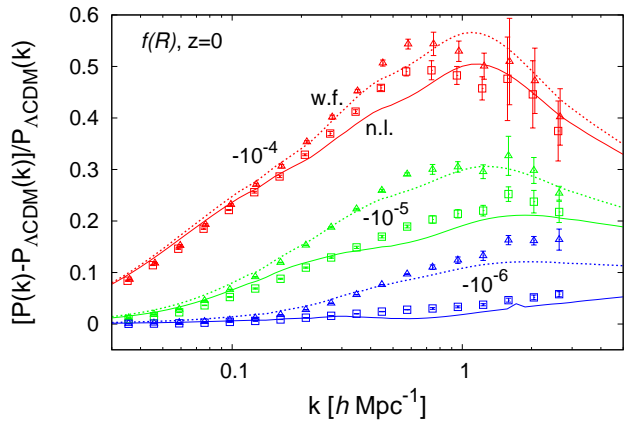}}
\end{center}
\caption{Relative deviation from $\Lambda$-CDM of the power spectrum in $f(R)$
theories, at redshift $z=0$, for $n=1$ and $f_{R_0}=-10^{-4}, -10^{-5}$, and $-10^{-6}$.
In each case, the triangles and the squares are the results of the ``no-chameleon'' and
``with-chameleon'' simulations from \cite{Oyaizu2008}, respectively.
We plot the relative deviation of the nonlinear power power spectrum
without chameleon effect (w.f., dotted lines) and with chameleon effect (n.l., solid lines).}
\label{fig-dPlk_fR_Hu_z0}
\end{figure}

We show our results for the deviation from $\Lambda$-CDM of the nonlinear
matter density power spectrum in Fig.~\ref{fig-dPlk_fR_Hu_z0}, for $f(R)$ theories
at $z=0$.
For each $f(R)$ model, we plot both the ``no-chameleon'' and ``with-chameleon''
cases studied in \cite{Oyaizu2008} through numerical simulations.

The ``no-chameleon'' case corresponds to the weak-field approximation discussed in
Secs.~\ref{Perturbative} and \ref{spherical-collapse}: the constraint equation
(\ref{fR-delta}) is linearized in the fluctuation $\delta R$ of the Ricci scalar.
This means that in the perturbative approach which provides the power
spectrum (\ref{Ptree+1loop}), up to one-loop order, we only include the factor
$\epsilon(k,\eta)$ which modifies the linear matrix $\cO$ in Eq.(\ref{O-mod}) and we neglect
the new quadratic and cubic vertices $\gamma^s_{2;11}$ and $\gamma^s_{2;111}$.
Next, in the computation of the spherical collapse which provides the linear density
threshold $\delta_L(M)$, we use the same linearization in $\delta R$, which corresponds
to the weak-field expression (\ref{fR-w.f.}) for the fifth force.
In other words, the ``no-chameleon'' case corresponds to using the linear approximation
in $\delta\rho$ for the fifth force, i.e. truncating the expansion (\ref{Psi-n}) at $n=1$,
[but $\delta\rho$ itself is nonlinear, in the sense of the expansion (\ref{psi-n-def})].

The ``with-chameleon'' case corresponds to keeping the fully nonlinear constraint
equation (\ref{fR-delta}). In the perturbative approach at one-loop order, this means that
we include the new quadratic and cubic vertices $\gamma^s_{2;11}$ and
$\gamma^s_{2;111}$, in addition to the linear kernel $\epsilon$, in the equation of
motion (\ref{O-Ks-def}).
(As noticed in Sec.~\ref{1loop-fR}, the cubic vertex $\gamma^s_{2;111}$ can actually be
neglected at this order, but not the quadratic vertex $\gamma^s_{2;11}$.)
In the spherical-collapse dynamics we solve the exact nonlinear constraint equation
(\ref{alpha-fR}).

\begin{figure*}
\begin{center}
\epsfxsize=8.5 cm \epsfysize=6. cm {\epsfbox{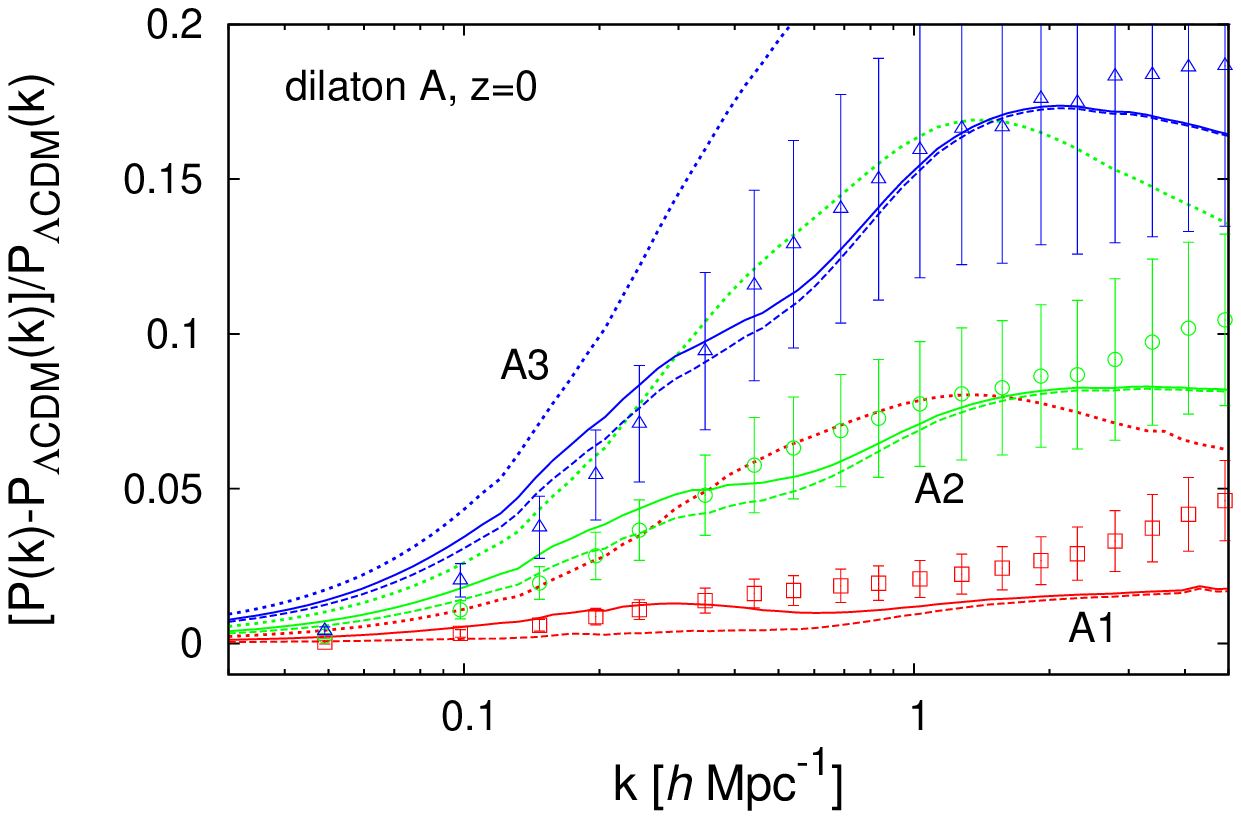}}
\epsfxsize=8.5 cm \epsfysize=6. cm {\epsfbox{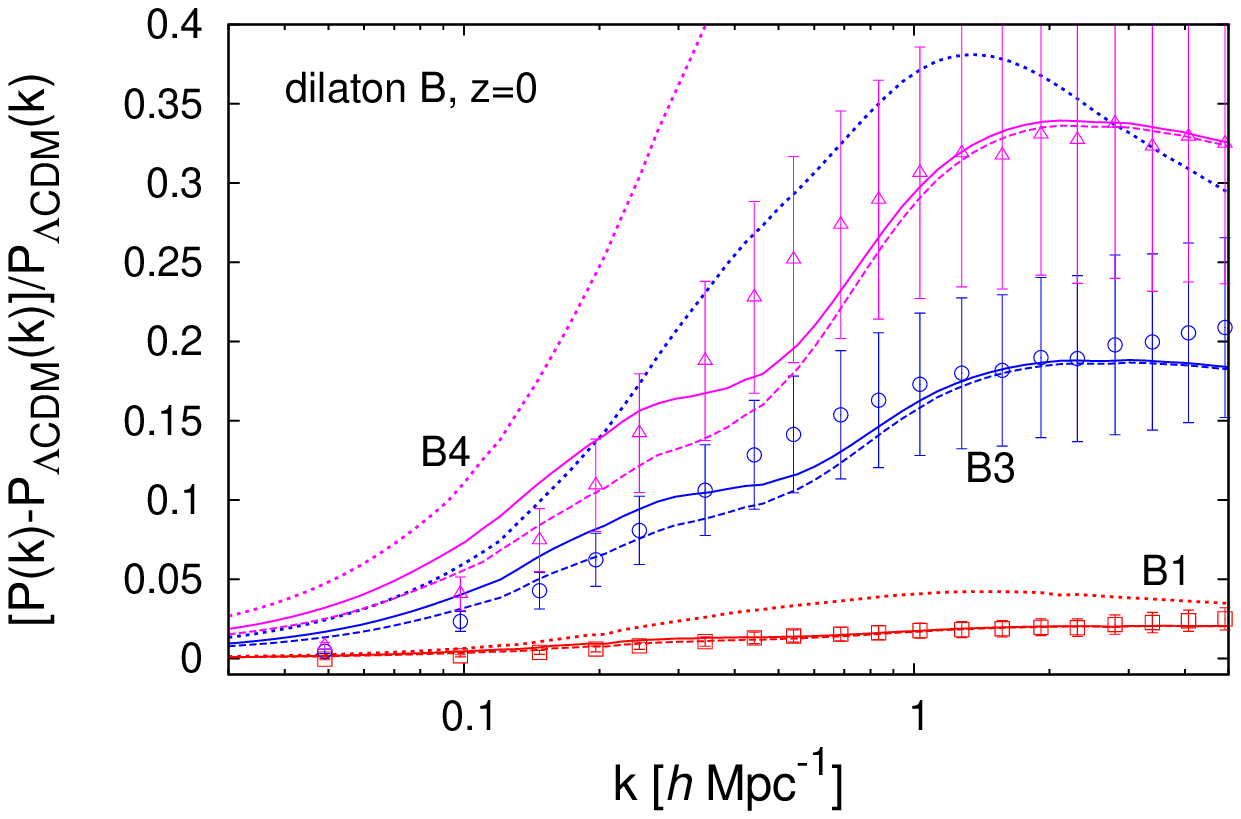}}
\epsfxsize=8.5 cm \epsfysize=6. cm {\epsfbox{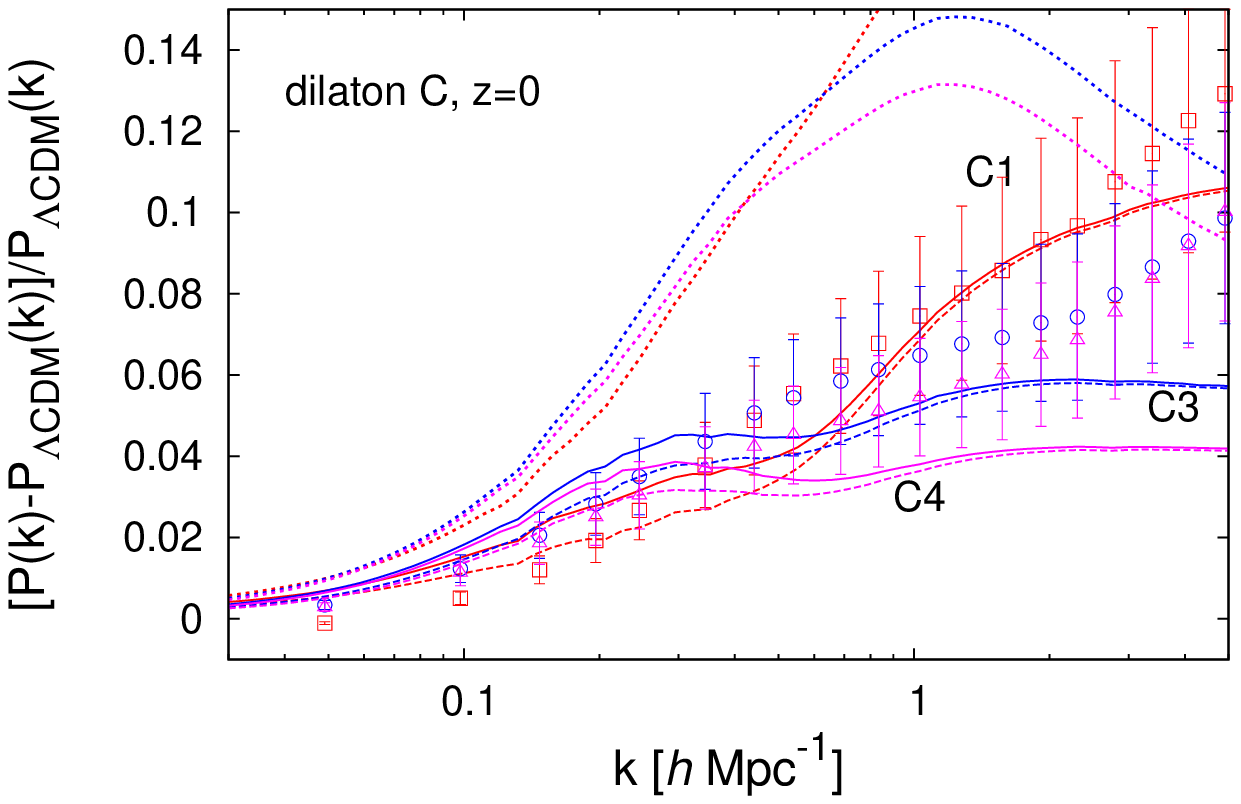}}
\epsfxsize=8.5 cm \epsfysize=6. cm {\epsfbox{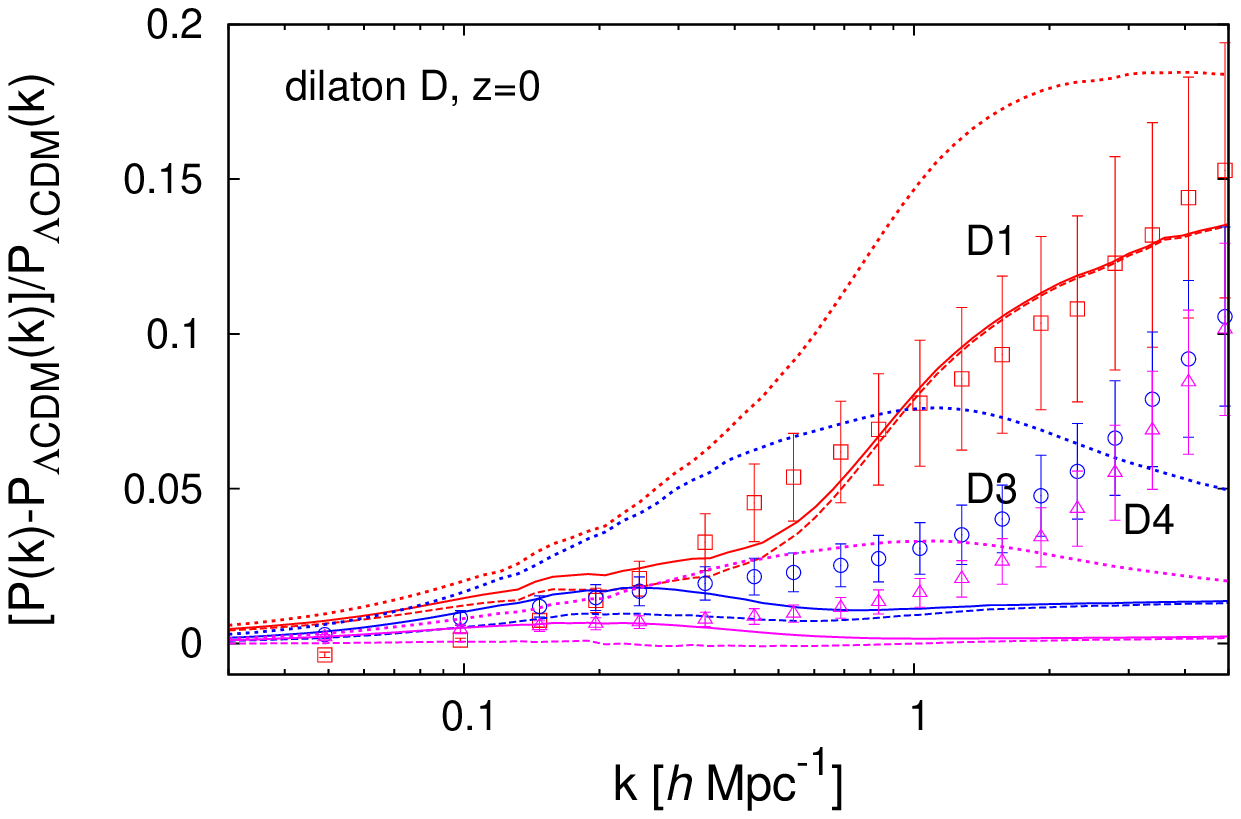}}
\end{center}
\caption{Relative deviation from $\Lambda$-CDM of the power spectrum in dilaton models,
at redshift $z=0$.
The symbols are the results from the simulations in \cite{Brax2012}, with the full nonlinear
screening effect.
We plot the relative deviation of the nonlinear power spectrum without the screening
effect (dotted lines), and with the screening effect, where we only include the
quadratic vertex $\gamma^s_{2;11}$ (dashed lines) or also the cubic
vertex $\gamma^s_{2;111}$ (solid lines) in the one-loop power spectrum.}
\label{fig-dPlk_dilaton_z0}
\end{figure*}

We can see in Fig.~\ref{fig-dPlk_fR_Hu_z0} that our approach is able to reproduce
reasonably well the deviations from the $\Lambda$-CDM power spectrum up to
$k \sim 3 h$Mpc$^{-1}$. In particular, it captures both the dependence on $f_{R_0}$
and the impact of the chameleon mechanism.
We do not have simulation results on smaller scale, to which
we may compare our predictions, and the agreement may deteriorate at higher $k$.
Indeed, on small scales the power spectrum is sensitive to the shape of halo profiles
and their mass-concentration relation, which are expected to be modified at some level
as compared to  $\Lambda$-CDM. Then, if these changes are large enough
they cannot be neglected as in this paper, if one is interested in small scales.
On the other hand, it may be possible to improve our modelization if one could build
a reliable model to predict such modifications to halo profiles.

As compared with the PPF approximation introduced in \cite{Hu2007}, which
interpolates between the linear regime, where the modification of gravity is
taken into account at the linear level without chameleon effect, and the nonlinear
regime where one uses the $\Lambda$-CDM prediction, our framework
does not introduce additional interpolation parameters.
Moreover, the convergence to General Relativity on smaller scales is obtained by
explicitly taking into account the chameleon mechanism (at one-loop order in
the perturbative regime and exactly in the spherical dynamics used in the one-halo
term). Therefore, the rate of convergence is truly governed by this nonlinear
effect, which depends on the modified gravity model, rather than by an independent
parametrization which requires some tuning (on the coefficient $c_{\rm nl}$ or
the function $\Sigma^2(k)$ that enter the interpolation \cite{Hu2007,Koyama2009}).

In any case, the comparison with Fig.~\ref{fig-dPk_fR_Hu_PT_z0} shows that our
simple approach, which combines one-loop perturbation theory with the halo model,
is already able to go significantly beyond the perturbative regime.
Indeed, the range of the agreement with the simulations increases from
$k \sim 0.2$ to $k \sim 3 h$Mpc$^{-1}$ at least, as we go from
Fig.~\ref{fig-dPk_fR_Hu_PT_z0} to Fig.~\ref{fig-dPlk_fR_Hu_z0}.
This is especially important as most of the signal occurs on the mildly nonlinear
scales $k \sim 1 h$Mpc$^{-1}$. Moreover, smaller, highly nonlinear, scales suffer from
other sources of uncertainties, which already appear in the $\Lambda$-CDM case, due to
the inaccuracy of the halo profiles and concentrations, and to the impact of the
baryon physics.

\subsection{Scalar-tensor models}
\label{Pk-scalar}

\begin{figure}
\begin{center}
\epsfxsize=8.5 cm \epsfysize=6. cm {\epsfbox{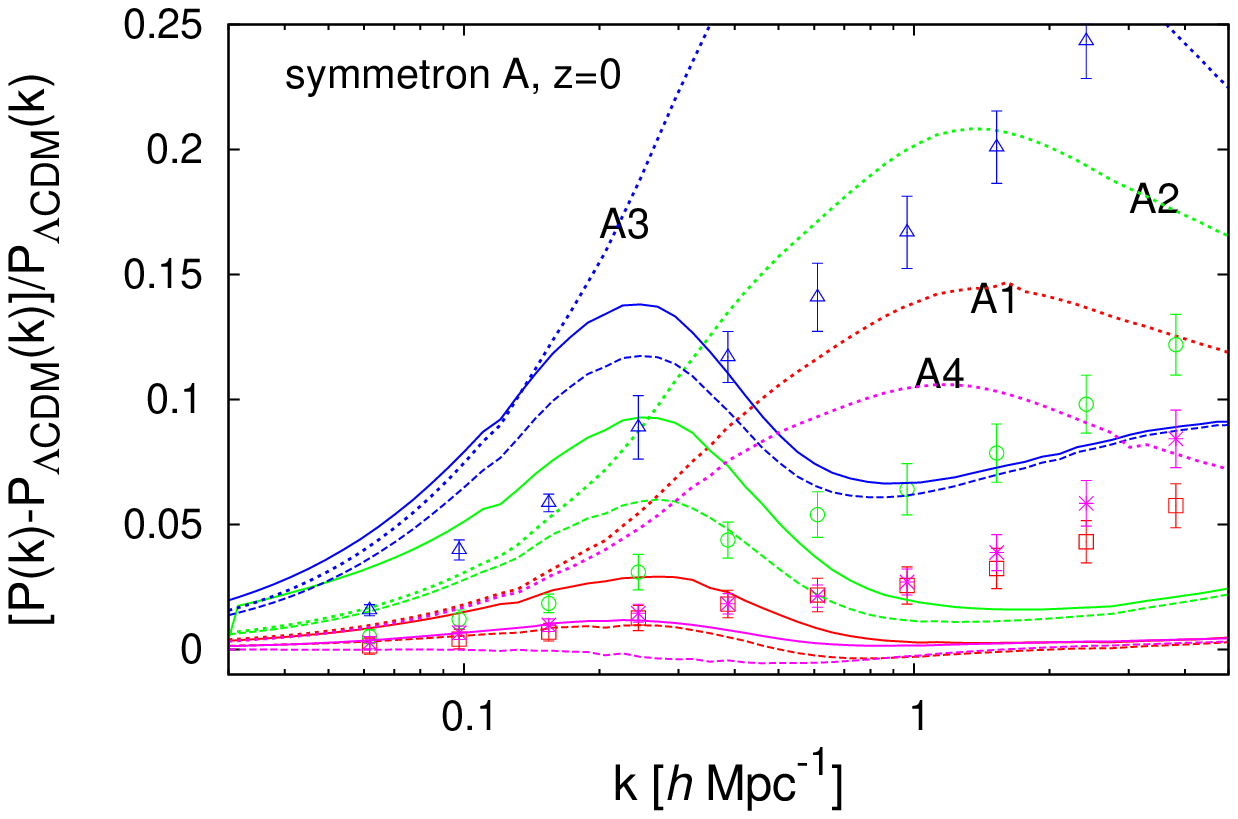}}
\epsfxsize=8.5 cm \epsfysize=6. cm {\epsfbox{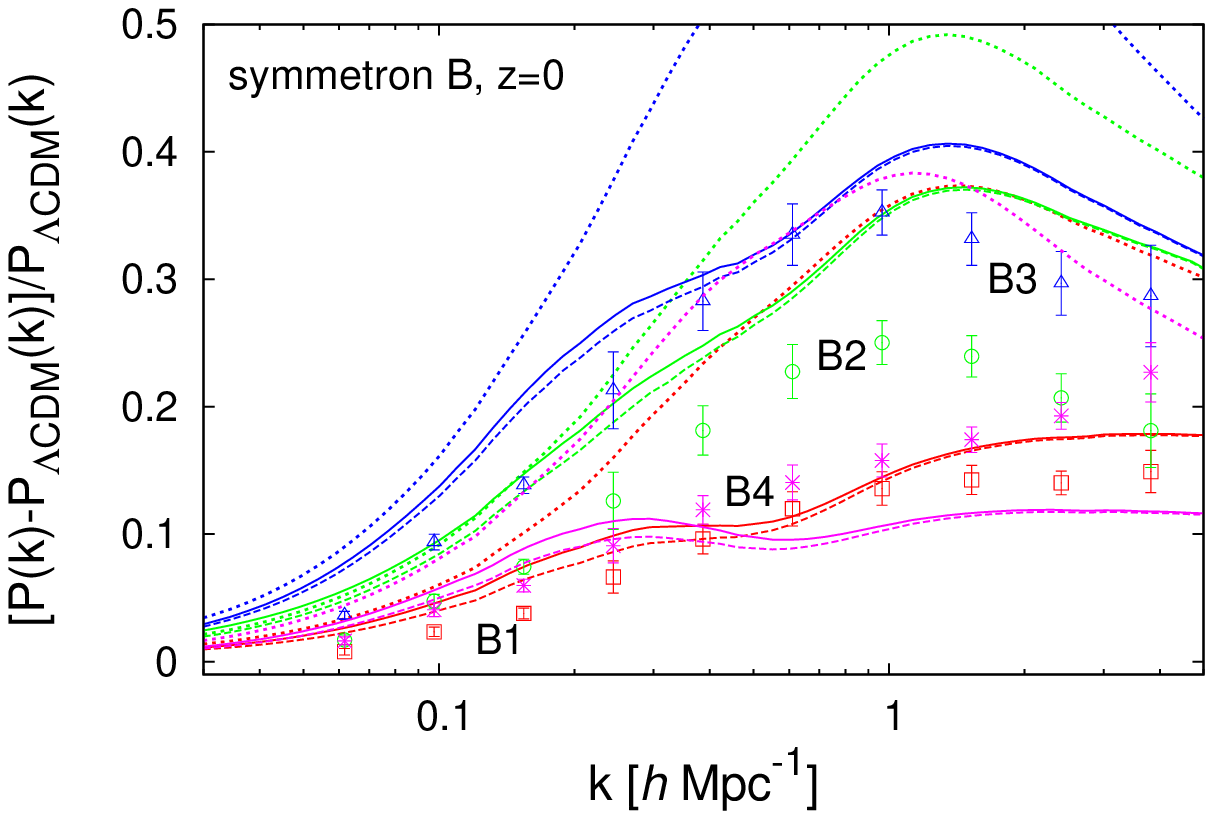}}
\end{center}
\caption{Relative deviation from $\Lambda$-CDM of the power spectrum in symmetron
models, at redshift $z=0$.
The symbols are the results from the simulations in \cite{Brax2012}, with the full nonlinear
screening effect.
We plot the relative deviation of the nonlinear power spectrum without the screening
effect (dotted lines), and with the screening effect, where we only include the
quadratic vertex $\gamma^s_{2;11}$ (dashed lines) or also the cubic
vertex $\gamma^s_{2;111}$ (solid lines) in the one-loop power spectrum.}
\label{fig-dPlk_symmetron_z0}
\end{figure}

We show our results for the deviation from $\Lambda$-CDM of the nonlinear power
spectrum for dilaton models at $z=0$ in Fig.~\ref{fig-dPlk_dilaton_z0}.
Although we only have results from simulations which use the fully nonlinear Klein-Gordon
equation (\ref{Klein-Gordon}), as in Fig.~\ref{fig-dPlk_fR_Hu_z0} for the
$f(R)$ theories, we plot both our ``no-screening'' and ``with-screening'' predictions.

Again, the ``no-screening'' result corresponds to truncating the expansion (\ref{Psi-n}) at $n=1$,
that is, using the linear approximation in $\delta\rho$ of the fifth force or the
linearized Klein-Gordon equation. This approximation is used for both the perturbative
one-loop power spectrum and the spherical-collapse threshold $\delta_L(M)$.

The ``with-screening'' result solves the exact nonlinear Klein-Gordon equation
(\ref{alpha-scalar}) in the spherical collapse. In the perturbative part, we consider the
results obtained when we only include the new quadratic vertex $\gamma^s_{2;11}$
(in addition to the linear factor $\epsilon$), or both the quadratic and cubic
vertices $\gamma^s_{2;11}$ and $\gamma^s_{2;111}$ (higher-order vertices do not
contribute at one-loop order).
Indeed, as seen in Sec.~\ref{1loop-scalar}, in contrast with the case of $f(R)$ theories,
the cubic vertex $\gamma^s_{2;111}$ is not negligible on perturbative scales.

In agreement with the behaviors found in Sec.~\ref{numerical-1loop} at the perturbative
level and in Sec.~\ref{spherical-collapse} for the spherical collapse, the comparison of
Fig.~\ref{fig-dPlk_dilaton_z0} with Fig.~\ref{fig-dPlk_fR_Hu_z0} shows that the impact
of the screening effect is greater for these dilaton models than for the $f(R)$ theories.
This greatly reduces the deviation of the power spectrum from the $\Lambda$-CDM case.
We can check that our approach is able to recover this effect and to provide a reasonable
match with the numerical simulations. At high $k$ we tend to underestimate the deviation
from $\Lambda$-CDM. This may be due to our neglect of modifications to the halo profiles.
This discrepancy appears at a larger scale, $k \sim 1 h$Mpc$^{-1}$, for the models
C4, D3, and D4, which are those where our model fares worse. However, they correspond
to very small deviations from the $\Lambda$-CDM power spectrum, a few percents
at $k \sim 1 h$Mpc$^{-1}$, which is at the limit of the accuracy of our modelization and
amplifies the errors associated with our approximations (such as keeping NFW profiles).
Nevertheless, even in these difficult cases we recover the order of magnitude of the
deviation from $\Lambda$-CDM and of the screening effect.
In particular, we again significantly extend the range of validity of the analytical predictions,
as compared to the one-loop perturbative results shown in
Fig.~\ref{fig-dPk_dilaton_PT_z0}, from $k \sim 0.2$ to $k \sim 1 h$Mpc$^{-1}$
(the precise values depend somewhat on the dilaton model).
Although we can distinguish the effect of the cubic vertex $\gamma^s_{2;111}$ on weakly
nonlinear scales, its impact remains rather small and could be neglected in view of the
overall accuracy of our modelization.

We show our results for symmetron models in Fig.~\ref{fig-dPlk_symmetron_z0}, in
the same fashion as in Fig.~\ref{fig-dPlk_dilaton_z0}.
As in Sec.~\ref{1loop-scalar}, we can see on perturbative scales that the screening
effect has not converged yet at one-loop order for the cases A2, A3, and to a small
extent B2. Indeed, for these cases, on large scales, whereas including the first
nonlinear (quadratic) vertex $\gamma^s_{2;11}$ decreases the deviation from
$\Lambda$-CDM as compared to the ``no-screening'' prediction, including the
next (cubic) vertex $\gamma^s_{2;111}$ over-corrects and yields a larger deviation
than the ``no-screening'' prediction.
This leads to an overestimation of the deviation from $\Lambda$-CDM on perturbative
scales.
To improve the modelization for these difficult cases, it may be necessary
to go beyond one-loop order in the perturbative part, and more precisely up to the order
where the screening effect is seen to converge.
In practice, this requires heavier computations, especially since the time and space integrations
do not factorize (in contrast with the $\Lambda$-CDM case where this is true up to a very good
approximation).
Moreover, the perturbative expansion of the screening effect may not converge
very well (for instance, because of the singularity of the coupling functions
$\beta_n(a)$ and $\kappa_n(a)$ at $a_s$).

Then, especially for the models A2 and A3 where these effects are the  largest,
our model gives a spurious oscillation for $\Delta P(k)/P(k)$ at $k \sim 0.4 h$Mpc$^{-1}$
with a significant underestimation of the signal at $k>1 h$Mpc$^{-1}$.
As for the other models, which are reasonably well reproduced by our approach, some
of this discrepancy may be due to the changes of halo profiles.
As we discuss in Sec.~\ref{2-halo-1-halo} below, this underestimation at high $k$
for the A models is related to the strong effect of the singular boundary $\phib(a_s)$
on the behavior of the scalar field $\varphi(\vx)$ and of the fifth force $F_{\rm A}$
noticed in Fig.~\ref{fig-FA_FN_symmetron_A3_z0}.
Then, numerical simulations may be the only tool to obtain accurate predictions for these
models.
Nevertheless, even for these difficult cases our approach provides the correct order of
magnitude of the deviation from $\Lambda$CDM and of the screening effect for
$k \lessapprox 1 h$Mpc$^{-1}$.
For the other models, A1, A4, B1, B3, and B4, our predictions show a reasonable
agreement with the simulations, up to about $1 h$Mpc$^{-1}$.

\begin{figure*}
\begin{center}
\epsfxsize=8.5 cm \epsfysize=6. cm {\epsfbox{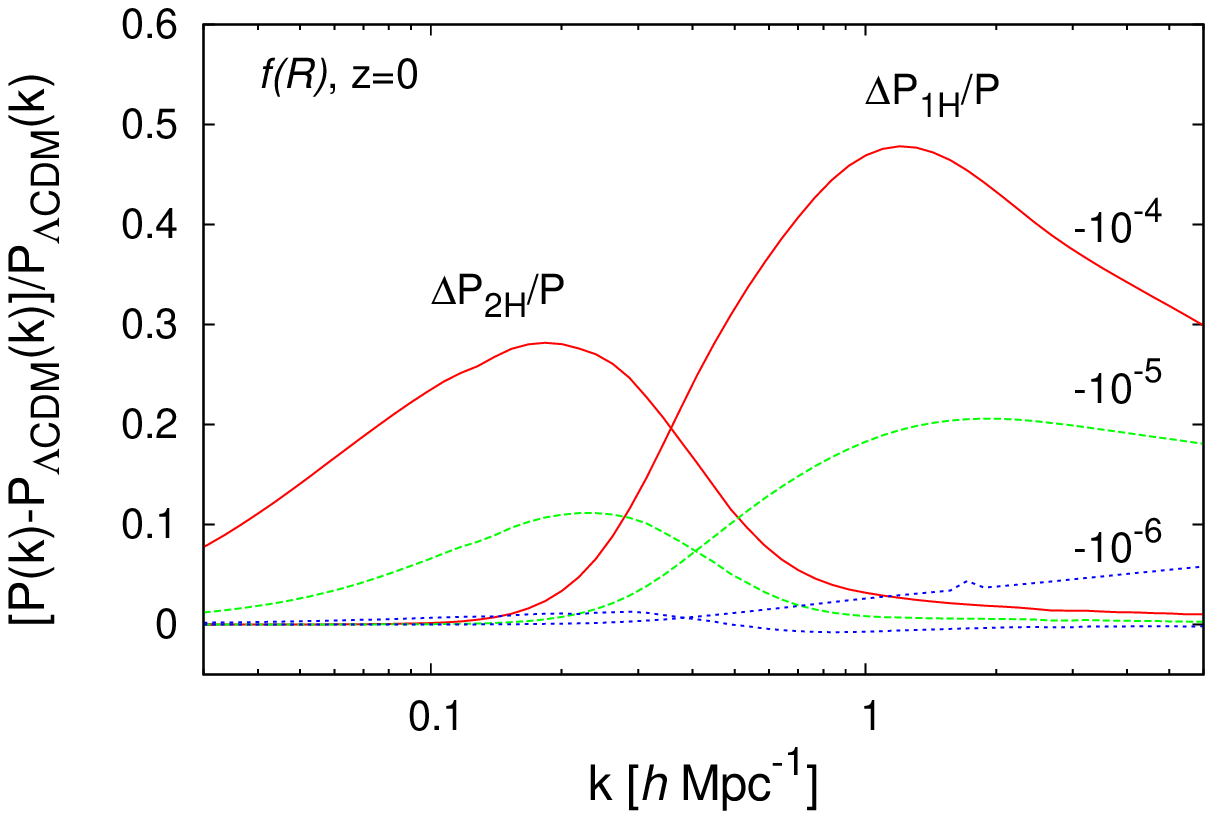}}
\epsfxsize=8.5 cm \epsfysize=6. cm {\epsfbox{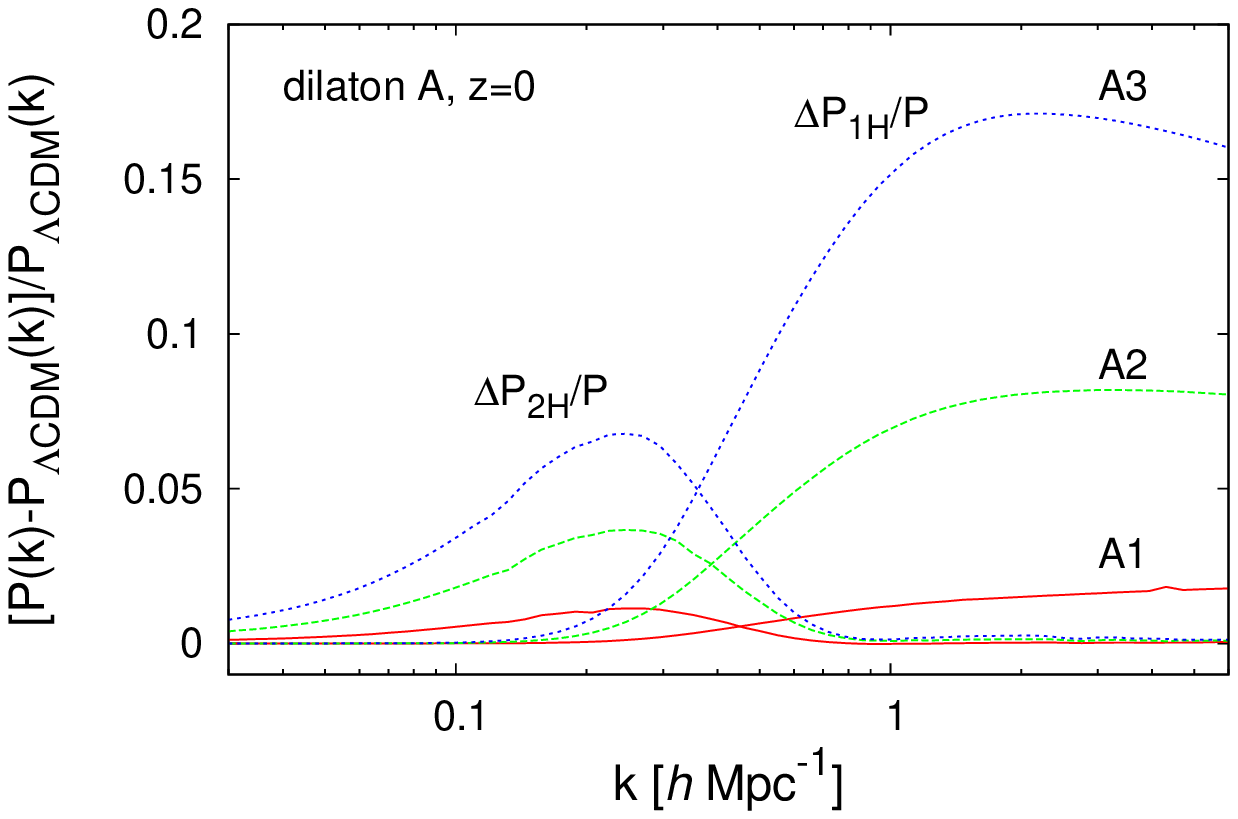}}
\epsfxsize=8.5 cm \epsfysize=6. cm {\epsfbox{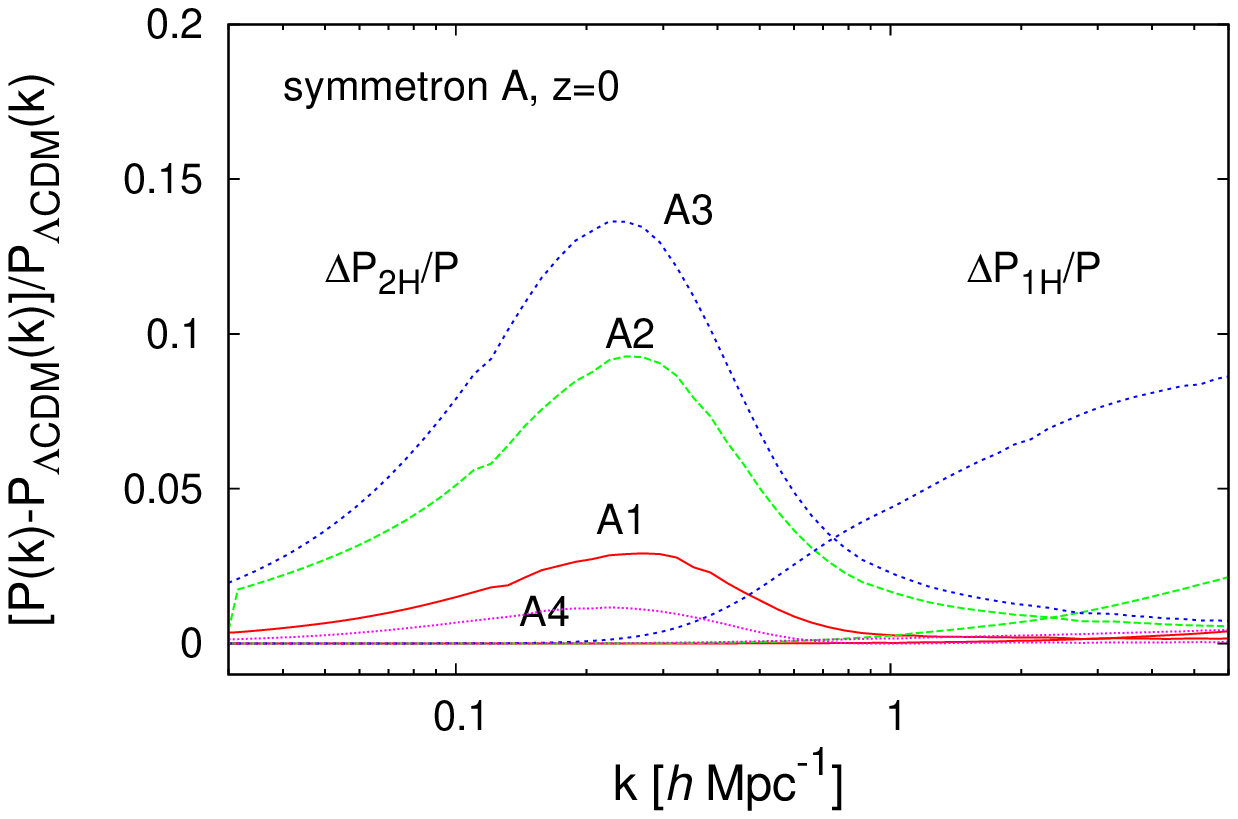}}
\epsfxsize=8.5 cm \epsfysize=6. cm {\epsfbox{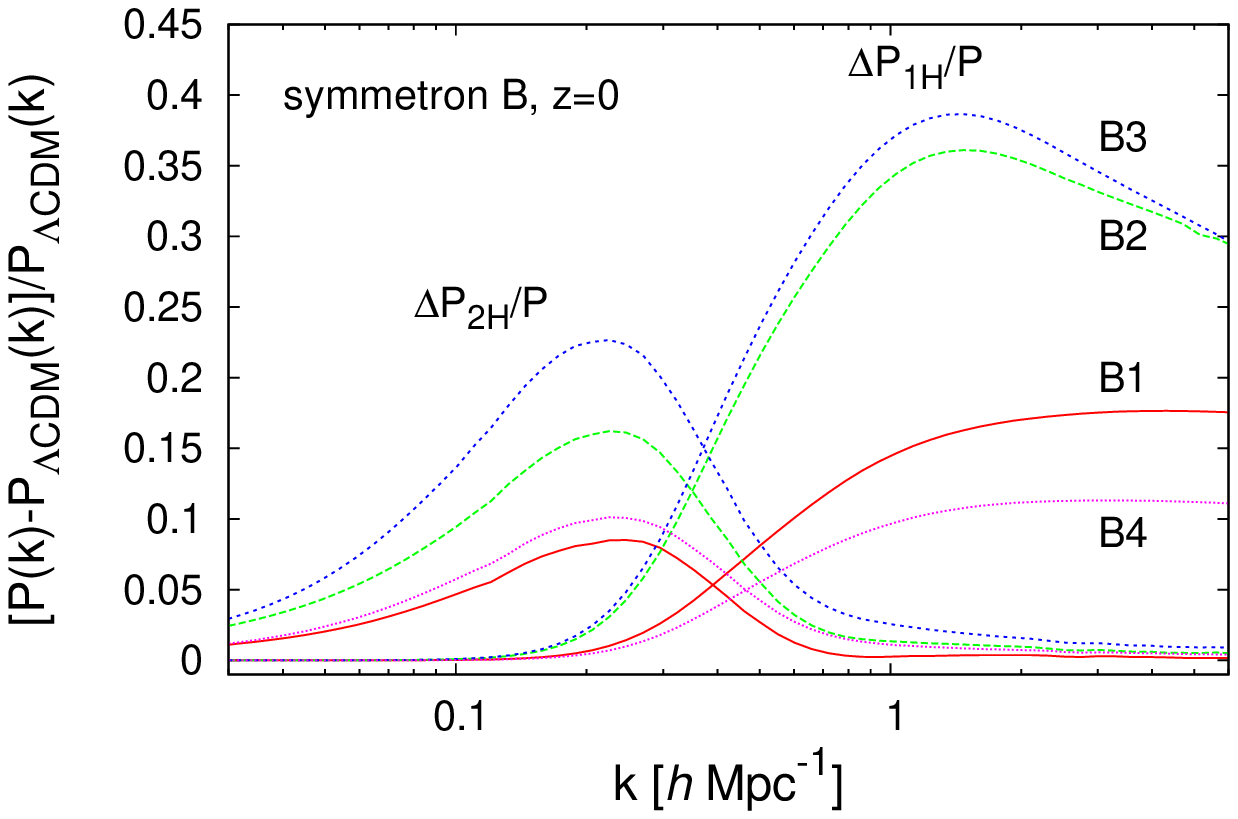}}
\end{center}
\caption{Relative deviation from $\Lambda$-CDM of the power spectrum in
$f(R)$, dilaton, and symmetron models, at redshift $z=0$.
For each model, we show the contribution from the modification to the two-halo term,
$\Delta P_{\rm 2H}/(P_{\rm 2H}+P_{\rm 1H})$ (curves with a peak around
$k \sim 0.2 h$Mpc$^{-1}$), and the contribution from the modification to the one-halo
term, $\Delta P_{\rm 1H}/(P_{\rm 2H}+P_{\rm 1H})$ (curves with a peak around
$k \sim 2 h$Mpc$^{-1}$ or which keep growing at high $k$).
We only consider the results with the full chameleon or screening effects.}
\label{fig-dPlk_Halo_z0}
\end{figure*}

\subsection{Two-halo and one-halo contributions}
\label{2-halo-1-halo}

One use of semi-analytic approaches like ours is to distinguish between the different
ingredients which build up the matter power spectrum. This allows us for instance
to compare the contributions from the large-scale perturbative regime and those from
the small-scale nonperturbative regime.
Thus, we show in Fig.~\ref{fig-dPlk_Halo_z0} the contributions to the difference
from the $\Lambda$-CDM power spectrum which arise from either the two-halo
or one-halo terms. For each model, the sum of the two curves,
$\Delta P_{\rm 2H}/P+\Delta P_{\rm 1H}/P=\Delta P/P$, gives
back the results shown in Figs.~\ref{fig-dPlk_fR_Hu_z0}-\ref{fig-dPlk_symmetron_z0}.

The contribution from the modification to the two-halo term peaks on weakly nonlinear
scales, $k \sim 0.2 h$Mpc$^{-1}$, because on very large scales we recover
General Relativity whereas on small scales the two-halo term gives a
negligible contribution to the full power spectrum.
For the same reason, the modification to the one-halo term only plays a role on
nonlinear scales, $k \gtrsim 0.5 h$Mpc$^{-1}$, where the one-halo term
becomes the dominant contribution to the power spectrum.
Therefore, systematic perturbative expansions can only describe the modifications
to the power spectrum below $k \lesssim 0.5 h$Mpc$^{-1}$ (and actually, slightly below,
because, as explained in the appendix, our two-halo term already contains some
small nonperturbative contributions associated with pancake formation).
At higher $k$, one must rely on more phenomenological approaches as we probe
the inner shells of virialized halos.
This also means that the theoretical accuracy of the full power spectrum is higher
and better controlled for $k \lesssim 0.5 h$Mpc$^{-1}$, down to a few percent
\cite{Valageas2013}, than for $k \gtrsim 0.5 h$Mpc$^{-1}$, where it should be about
$10\%$.
However, as seen in Sec.~\ref{1loop-scalar} and \ref{Pk-scalar}, this accuracy, which
holds for $\Lambda$-CDM-like cosmologies, is not reached in peculiar cases such
as some symmetron models because of the screening mechanism.
Indeed, we have seen that this involves an additional expansion scheme (as compared
with $\Lambda$-CDM) which can converge more slowly than the usual expansion
in the linear density and velocity fields. This can be the limiting factor of the
perturbative approach but as shown in Sec.~\ref{1loop-scalar} this can be
detected from the comparison between different orders (i.e., as we include
successive vertices $\gamma^s_{2;1..1}$).

The decomposition displayed in Fig.~\ref{fig-dPlk_Halo_z0} shows that
the behavior above $k \gtrsim 0.5 h$Mpc$^{-1}$ is due to the one-halo term,
hence to the spherical-collapse threshold $\delta_L(M)$ studied in
Sec.~\ref{spherical-collapse}, because this is the only effect that we include
in this regime. At smaller scales, $k \gtrsim 3 h$Mpc$^{-1}$, we can expect
modifications to the halo profiles (e.g., to the mass-concentration relation)
to come into play \cite{Valageas2013}.
Nevertheless, our results already explain the behaviors found in
Figs.~\ref{fig-dPlk_fR_Hu_z0} and \ref{fig-dPlk_dilaton_z0}, where it is seen
that in $f(R)$ theories with $|f_{R_0}| \gtrsim 10^{-5}$ the deviation from
$\Lambda$-CDM of the power spectrum decreases with $k$ in the range
$1 < k < 5 h$Mpc$^{-1}$, whereas it is roughly constant for the dilaton models.
Indeed, as noticed in Sec.~\ref{spherical-collapse} from the comparison
of Figs.~\ref{fig-deltaLM_fR_Hu_z0} and \ref{fig-deltaLM_dilaton_z0},
the dilaton screening mechanism is more efficient than the $f(R)$ chameleon
effect (in this regime), and the linear density threshold $\delta_L(M)$ is significantly
lower for these $f(R)$ models than for these dilaton models, on mass scales
$10^{14} < M < 10^{16} h^{-1} M_{\odot}$.
In particular, $\delta_L(M)$ converges more slowly to the $\Lambda$-CDM threshold
at large mass for the $f(R)$ models with $|f_{R_0}| \gtrsim 10^{-5}$ than for these
dilaton models.
Then, because of the exponential factor $e^{-\delta_L(M)^2/(2\sigma_M^2)}$ of the
large-mass tail of the halo mass function, the deviation from $\Lambda$-CDM
of $P_{\rm 1H}(k)$ grows faster at lower $k$ (which corresponds to more massive
and larger halos) in these $f(R)$ models.
This leads to the faster increase of the deviation from $\Lambda$-CDM of the full
power spectrum at lower $k$ in the range $1 < k < 5 h$Mpc$^{-1}$, until
the one-halo contribution becomes subdominant at $k \lesssim 0.5 h$Mpc$^{-1}$.
In contrast, for the case $f_{R_0}=-10^{-6}$ the deviation due to the one-halo term
keeps increasing with $k$ in the range $1 < k < 5 h$Mpc$^{-1}$, because at large
mass the linear threshold $\delta_L(M)$ is very close to the $\Lambda$-CDM
result, as seen in Fig.~\ref{fig-deltaLM_fR_Hu_z0}.
Thus, the behavior of the deviation from $\Lambda$-CDM in this range of
wave numbers depends on the balance between the increased sensitivity
at large mass of the exponential factor $e^{-\delta_L(M)^2/(2\sigma_M^2)}$
and the convergence to General Relativity.
This cannot be predicted a priori for a given class of models and we must
evaluate this effect by explicit computations, as in Fig.~\ref{fig-dPlk_Halo_z0}.

The two lower panels of Fig.~\ref{fig-dPlk_Halo_z0} also explain part of the
discrepancy found for symmetron models in Fig.~\ref{fig-dPlk_symmetron_z0}.
Indeed, we can see that the one-halo contribution seems too small as compared to
the two-halo contribution for the ``symmetron-A'' models, if we compare with the
``symmetron-B'' models and the $f(R)$ and dilaton models.
This is due to the very efficient screening effect noticed in
Figs.~\ref{fig-deltaLM_symmetron_z0} and \ref{fig-FA_FN_symmetron_A3_z0},
where we found that at large mass the
linear threshold $\delta_L(M)$ becomes very close to the $\Lambda$-CDM prediction.
This was due to the lower bound $\alpha_s$ (or $\phib(a_s)$) which
diminishes the range of values of the new degree of freedom $\alpha(\vx)$
and greatly reduces the fifth force, as compared to the weak-field limit where this
constraint is discarded.
This is not the case for the ``symmetron-B'' models (in this regime) because
their phase transition takes place earlier, at $a_s=0.33$ instead of $0.5$.
This gives more room for the new degree of freedom $\alpha(\vx)$ within the
regular domain $a_s < \alpha \leq a$, and we checked that their linear threshold
$\delta_L(M)$ is halfway between the $\Lambda$-CDM and weak-field results.
In other words, the scalar field $\varphi$ is not so strongly pinned down to the singular
value $\phib(a_s)$ and a significant fifth force can appear.
This explains why our one-halo contribution is greater for the ``B'' models than for
the ``A'' models in the lower panels of Fig.~\ref{fig-dPlk_Halo_z0}.
This also explains why we obtained smooth curves in the lower panel of
Fig.~\ref{fig-dPlk_symmetron_z0}, with a reasonable agreement with the simulations,
whereas we obtained a spurious oscillation in the upper panel with a significant
underestimation of the power spectrum at high $k$.
This discrepancy for the ``A'' models suggests that in these cases our approach
of the spherical-collapse dynamics is not sufficient to give an accurate account
of the impact of this modified gravity on virialized objects. It is likely that deviations
from spherical symmetry and perturbations to the radial density profile break down
the fast relaxation towards $\phib(a_s)$ observed in our very symmetric case,
because of spatial gradients. This would increase the fifth force and explain the rise
with $k$ measured in the simulations for the deviation from $\Lambda$-CDM
of the power spectrum.
Fortunately, these difficult cases could be detected a priori from
Figs.~\ref{fig-deltaLM_symmetron_z0} and \ref{fig-FA_FN_symmetron_A3_z0},
by looking for the cases where such singular
behaviors (sticking to a singular value) occur.

In any case, Fig.~\ref{fig-dPlk_Halo_z0} shows that it remains useful to consider
modified gravity models that, even though not fully general, are well defined at
the nonlinear level while covering a broad range of models. Indeed, to obtain
reliable estimates of the deviations from $\Lambda$-CDM and to assess the range
of validity of these results, it is useful to go beyond linear theory and even beyond the
perturbative regime. This allows us to take into account key screening mechanisms and
to estimate the relative importance of different contributions, which show different
degrees of accuracy (depending on whether they derive from systematic perturbative
expansions or more phenomenological halo models).

\section{Conclusion}
\label{Conclusion}

In this paper, we have studied semi-analytically  the screening mechanisms which are necessary to make  modified gravity
models consistent with observations on small scales.
As these mechanisms rely on the nonlinearity of the equations of motion, it is necessary
to go beyond linear theory. We have presented a general approach, using perturbation
theory (which applies to large scales) and the spherical-collapse dynamics (which allows us
to handle mildly nonlinear scales with the help of a halo model), to tackle these effects.
Our approach applies to a large class of modified gravity scenarios, including $f(R)$
theories and scalar-tensor theories, such as dilaton and symmetron models.

The new degree of freedom, as compared to
the $\Lambda$-CDM case, which may be associated with the fluctuations $\delta R$ of the Ricci
scalar (in $f(R)$ theories) or of the new scalar field, $\delta\varphi$, in scalar-tensor theories,
gives rise to a fifth force which can be written as a new contribution $\Psi_{\rm A}$ to the
gravitational potential.
This new field is nonlinearly coupled to the matter density, for instance through a Klein-Gordon
equation with an effective potential which depends on $\rho$.
Fortunately, in many cases this equation can be simplified by using the quasistatic
approximation, so that this new degree of freedom is fully determined by the current
density field (and one does not need to keep track of the past history of the field).
We have checked that this approximation is valid for the cases that we consider in this
paper. It would break down for very singular coupling functions [such as $\beta(a) \sim
(a-a_s)^{\hat{n}}$ with $\hat{n}<0.25$].
Then, the equation for the new degree of freedom takes the form of a constraint
equation (i.e., without time derivatives) and implicitly determines the new field, whence
the fifth force, as a nonlinear functional of the density field.

First, we have described how to compute the matter density power
spectrum within a perturbative approach. One first solves the constraint equation for the
new field as a perturbative expansion in the nonlinear density fluctuation $\delta\rho$.
This also yields the fifth-force potential $\Psi_{\rm A}$ as a perturbative expansion in
powers of $\delta\rho$. Then, using the usual single-stream approximation, which applies
on large perturbative scales, one solves the new equations of motion for the density and
velocity fields as a second perturbative expansion in powers of the linear density
fluctuation $\delta_L$. Because the fifth force is a nonlinear, nonpolynomial functional of
the density field, the Euler equation is no longer quadratic but contains vertices of all
orders.
In this paper, we have computed the matter density power spectrum up to one-loop order,
which corresponds to third order in the fields. At this order, only three new vertices are
relevant, $\epsilon$, $\gamma^s_{2;11}$, and $\gamma^s_{2;111}$, which arise from
the linear, quadratic, and cubic terms in $\delta\rho$ of $\Psi_{\rm A}$.
The linear vertex $\epsilon$ corresponds to the weak-field limit, where we linearize
in the fluctuations of the new degree of freedom. The quadratic and cubic vertices
$\gamma^s_{2;11}$ and $\gamma^s_{2;111}$ contain the first signs of the screening
mechanism.

Thus, for the modified gravity models that we investigate here, the quadratic vertex
$\gamma^s_{2;11}$ decreases the deviation from the $\Lambda$-CDM dynamics, as compared
with the weak-field approximation. In fact, if we use this truncated equation of motion
up to high densities, for instance within a spherical-collapse study, we find that this
quadratic vertex even stops the collapse at finite density. The next nonlinear vertex,
the cubic one $\gamma^s_{2;111}$, is a higher-order correction that somewhat diminishes
the amplitude of this screening mechanism.
We have found that for the $f(R)$ theories, the cubic vertex can actually be safely neglected
on the perturbative scales described by one-loop-order perturbation theory.
For the dilaton models, the cubic vertex makes a small but noticeable correction.
For the symmetron models, the situation is less favorable and in some cases, associated
with the most singular coupling functions, the screening mechanism has not converged
yet at this order, on these large scales. Indeed, it may happen that the cubic vertex
$\gamma^s_{2;111}$ overcorrects the screening mechanism and yields a deviation from
$\Lambda$-CDM which is larger than the one obtained at linear order.
It is not obvious whether the expansion would show a good convergence at higher orders.
In all cases, the one-loop perturbative predictions only apply to rather large scales,
$k \lesssim 0.15 h$Mpc$^{-1}$ at $z=0$, where the signal is not very large.

Next, to go beyond the perturbative regime, we have studied the spherical-collapse
dynamics. More precisely, we have focused on the linear density threshold $\delta_L(M)$
that is required to reach a fixed nonlinear density contrast of $200$, which we use
to define virialized halos. Because the fifth force accelerates the collapse (in the models
studied here), this threshold is lower than the $\Lambda$-CDM prediction, especially for
small or moderate masses. We have also studied in detail the behavior of the new degree
freedom and the differences between the $f(R)$, dilaton, and symmetron models.

As in the perturbative regime, we find that the impact of the screening mechanism is
smallest for the $f(R)$ theories. In particular, for small halos, which correspond to small
scales at fixed density, the chameleon mechanism is no longer relevant. This is because
in this regime the fifth force is not sensitive to the exact value of the rescaled field $\alpha$,
which is much smaller than $\delta$. More precisely, for small halos the density at the virial
radius is not large enough to overcome the effect of spatial gradients. This prevents
the chameleon mechanism from taking place (the field cannot follow the rise of the density
field) and the total gravitational force is equal to the Newtonian force, multiplied by a factor
$4/3$ as in the weak-field limit.

In the dilaton models, the screening mechanism is more efficient and can become important
again for very low-mass halos. This is because the fifth force builds up in a very different
manner in scalar-tensor theories as compared to $f(R)$ models. It is no longer
produced by the integral over smaller radii of the difference between the new field and
the matter density contrast, but by the local spatial derivative of the new field.
Then, for small objects at finite density, the new rescaled field $\alpha$ again becomes
small and flat (because it cannot accommodate gradients that are too strong), but instead of a
large fifth force this now yields a small fifth force.
The situation is similar in symmetron models, with an even stronger screening mechanism
because of the singularity of the coupling functions, which pinpoints  the new field $\alpha$
(close) to the singular value $a_s$.

Finally, we have combined the perturbative expansion with the spherical collapse dynamics
to obtain a modelization of the matter power spectrum from large linear scales to mildly
nonlinear scales, using a recently developed approach which uses the halo model.
At this stage, we cannot expect to describe very small, highly nonlinear scales, because
we neglect the impact of modified gravity on halo shapes.
We again find that our approach fares best for $f(R)$ theories,
where it reproduces both ``no-chameleon'' and ``with-chameleon'' simulations.
This allows us to extend the validity of semi-analytical predictions up to
$k \sim 3 h$Mpc$^{-1}$ at least, at $z=0$. This is a significant improvement over the linear
or one-loop perturbative results, which are restricted to $k \lesssim 0.15 h$Mpc$^{-1}$.

For the dilaton models, the accuracy of our modelization depends somewhat on the
model, but we usually obtain a reasonable agreement with simulations up to
$k \sim 1 h$Mpc$^{-1}$. In particular, we capture the impact of the screening mechanism.
On small scales, $k > 1 h$Mpc$^{-1}$, we tend to underestimate the power spectrum.
This may be due to our neglect of any change to the halo profiles.
The amplitude of this discrepancy worsens for models where the deviations from
$\Lambda$-CDM are small, which are more sensitive to such approximations.

Again, the situation appears most difficult for the symmetron models, especially in those
cases where the screening mechanism had not converged at one loop or the spherical
collapse of massive halos is governed by the singular value $\phib(a_s)$ of the scalar
field.
There, our predictions can differ from the simulations by a factor of 2, in the range
$k \leq 1 h$Mpc$^{-1}$.
Nevertheless, we still predict the correct order of magnitude of the deviation of the power
spectrum from the $\Lambda$-CDM one. In more favorable cases, we obtain a reasonable
agreement. Fortunately, the difficult cases can be detected a priori from the bad behavior
of the perturbative expansion and the impact of the singular boundary $\phib(a_s)$,
or from a spurious oscillation in the prediction for the power spectrum (which is related
to these two problems).

Therefore, we have found that it is possible to build an efficient semi-analytical modelization
of the matter density power spectrum, which takes into account the nonlinear screening
mechanism, for a large class of modified gravity theories.
This is most accurate for the $f(R)$ theories, where the chameleon effect is moderate
(but this still requires a fully nonlinear analysis).
This is also due to the fact that in these theories the gravitational potential remains of the same
form as the Newtonian one in the two asymptotic regimes, with a multiplicative factor of
unity on large scales and of $4/3$ on small scales. This is not far from a moderately varying
effective Newton's constant, and we can expect the dynamics (e.g., halo shapes) to remain
similar to the $\Lambda$-CDM ones.
Our model remains valid for scalar-tensor theories, where the fifth force shows a very different
behavior and the screening mechanism is stronger, except for some symmetron models
associated with singular coupling functions.

Such semi-analytic modelizations, which go beyond linear theory, are necessary because
the linear regime is restricted to very large scales where the deviations from
$\Lambda$-CDM are small. They allow us to probe a broader range of scales, up to the
mildly nonlinear regime where the departure from $\Lambda$-CDM is largest (e.g., for
some $f(R)$ models) or more significant (especially as smaller scales are less reliable
because of the lower accuracy of theoretical predictions, for instance because of the
impact of baryon physics). Moreover, we can compare the relative contributions
from perturbative and nonperturbative terms and detect features associated with
phase transitions as in some symmetron models. This allows one to estimate the validity
of the predictions. This is an advantage of modified gravity models that are fully defined
at the nonlinear level, while covering a broad range of models.

In order to improve our modelization, it may be useful to consider the impact of modified gravity
on halo profiles. However, it is not obvious a priori how to devise a robust analytical
approach. This would probably require detailed numerical simulations, to see for instance
whether these effects may be described through a small set of parameters or to serve
as a guideline for analytical modeling.

In order to handle the problematic cases of some symmetron models, where one-loop perturbation
theory has not converged yet, one should devise more efficient methods to take into
account the screening mechanism. One may either go to two-loop (or higher) order,
to see whether the expansion starts converging, or introduce some resummation schemes
which manage to accurately describe the screening mechanism.
The spherical collapse itself should also be improved.
An accurate treatment of such models, which involve two different phases
around a critical density $\rho_s$, would probably require a specific method
that explicitly takes into account these two phases.

Another topic would be the study of models where the effective potential of the
new scalar field shows several degenerate minima. This could lead to interesting features
associated with topological defects. Then, one would need to embed our approach within
a more general framework which is able to describe several domains and the nonlinear
physics at their interface.
We leave these issues for future works.

\begin{acknowledgments}

We thank W. Hu, B. Li, and H. A. Winther for sending us the numerical data for
$f(R)$, dilaton, and symmetron models.
This work is supported in part by the French Agence Nationale de la Recherche under Grant ANR-12-BS05-0002.

\end{acknowledgments}

\appendix

\section*{Appendix: Combining the halo model with one-loop perturbation theory}

In this appendix we provide some more details about the halo model and the calculation of the power spectrum in the nonlinear regime.
In the Lagrangian-space framework, particles and their motion are followed unlike in the Eulerian-space approach where fields on fixed grids, such as the velocity field, are used. Particles follow trajectories
$\vx(\vq,t)$, labelled by their initial position $\vq$, and we can write $\vx(\vq,t)=\vq+ \Psi (\vq,t)$ where $\Psi$ is the displacement field. At linear order, the variances of the relative displacement $\Delta\Psi=\Psi_2-\Psi_1$ of two particles $1$ and $2$, in the transverse and longitudinal directions with respect to the initial
separation vector $\Delta\vq=\vq_2-\vq_1$, are
\beq
\sigma^2_{\parallel}(\Delta q) = 2 \int \dd\vk \; [1-\cos (k_\parallel \Delta q)] \frac{k^2_\parallel}{k^4} P_L (k) ,
\eeq
\beq
\sigma^2_{\perp}(\Delta q) =  2 \int \dd\vk \; [1-\cos (k_\parallel \Delta q)] \frac{k^2_\perp}{k^4} P_L (k) ,
\eeq
where $k_{\perp}$ is the component along one of the two transverse directions.
The power spectrum in the Zel'dovich approximation is obtained by considering the displacement field at linear order.
For Gaussian initial conditions, this yields a Gaussian distribution for the relative
displacements and, using the exact expression (\ref{Pkxq}), this leads to the power spectrum
\be
P^Z(k)= \int\frac{\dd\Delta\vq}{(2\pi)^3} \; e^{\ii k\mu \Delta q -\frac{1}{2} k^2 \mu^2 \sigma^2_\parallel -\frac{1}{2} k^2 (1-\mu^2) \sigma_\perp^2} ,
\label{PZ-def}
\ee
where $\mu= (\vk\cdot\Delta \vq)/(k\Delta q)$.
By expanding up to second order in $P_L$ and defining $P^Z (k)= P_L(k) + P^Z_{\rm 1 loop}(k) +\dots$, we have
\begin{eqnarray}
P^Z_{\rm 1 loop}(k)&=& -k^2 \sigma^2_v P_L(k) +\int \dd\vk_1 \dd\vk_2 \;
\delta_D( \vk_1 +\vk_2 -\vk)\nonumber \\
&&\times  \frac{(\vk\cdot\vk_1)^2 (\vk\cdot\vk_2)^2}{2k_1^4k_2^4} P_L(k_1) P_L(k_2) ,
\end{eqnarray}
where $\sigma_v^2= \frac{4\pi}{3} \int_0^\infty \dd k \, P_L(k) j_0(qk)$.
We go beyond the Zel'dovich approximation by including nonlinearities in the distribution of the parallel displacement field, while we keep linear theory for the transverse one. Thus, introducing the rescaled longitudinal displacement
$\kappa_{\parallel}$ and its linear variance,
\beq
\kappa_{\parallel} = \frac{\Delta x_{\parallel}}{\Delta q} , \;\;\;
\sigma^2_{\kappa_{\parallel}} = \frac{\sigma_{\parallel}^2}{(\Delta q)^2} ,
\eeq
we define its cumulant generating function $\varphi_\parallel (y)$ by
\be
\lag e^{-y \kappa_{\parallel}/\sigma^2_{\kappa_{\parallel}}} \rag_{\parallel}
= e^{-\varphi_\parallel (y)/\sigma^2_{\kappa_{\parallel}}} ,
\label{phi-par-def}
\ee
where the average is over the parallel displacements.
The ansatz (\ref{phi-alpha-def}) used for $\varphi_{\parallel}$, which depends on the
scale $\Delta q$, agrees with the expansion
\be
\varphi_\parallel (y) = y-\frac{y^2}{2} +S_3 \frac{y^3}{6}+\dots ,
\ee
where $S_3(\Delta q)$ is constructed from
\begin{eqnarray}
S_3(\Delta q) &=& -\frac{24\pi}{\sigma_{\kappa_\parallel}^4} \int_0^{\infty} \dd k \;
\frac{P_{\rm 1 loop}(k)- P^Z_{\rm 1 loop}(k)}{(\Delta q)^4 k^2}\nonumber\\
&& \times \left[ 2+ \cos(k\Delta q) - 3 \frac{\sin (k\Delta q)}{k\Delta q} \right] ,
\label{S3-def}
\end{eqnarray}
and $P_{\rm 1 loop}(k)$ is the exact one-loop power spectrum constructed with
perturbation theory.
This ensures that the associated power spectrum is exact up to one-loop order and
it reads as
\beq
P^{\parallel}(k)= \int\frac{\dd\Delta\vq}{(2\pi)^3} \;
e^{-\varphi_{\parallel}(-\ii k\mu \Delta q \sigma_{\kappa_{\parallel}}^2)/
\sigma_{\kappa_{\parallel}}^2} \; e^{-\frac{1}{2} k^2 (1-\mu^2) \sigma_\perp^2} .
\label{Pk-par}
\eeq
If we truncate $\varphi_{\parallel}$ at quadratic order, $\varphi_{\parallel} = y-y^2/2$,
we recover the Zel'dovich power spectrum (\ref{PZ-def}). Thus, the power spectrum
(\ref{Pk-par}) is a generalization of the Zel'dovich power spectrum. It is consistent
with the exact perturbative expansion up to one-loop order (i.e., $P_L^2$), whereas
the Zel'dovich power spectrum only agrees at linear order, and it also contains
some perturbative terms at all higher orders in both Eulerian and Lagrangian spaces
[generated through the nonpolynomial function $\varphi_{\parallel}$ and the exponential
in Eq.(\ref{Pk-par})].

The generating function (\ref{phi-par-def}) also defines the probability distribution
function of $\kappa_\parallel$,
\be
{\cal P}^{\parallel}(\kappa_\parallel)= \int_{-\ii\infty}^{\ii\infty} \frac{\dd y}{2\pi \ii
\sigma^2_{\kappa_\parallel}} \; e^{[\kappa_\parallel y -\varphi_\parallel(y)]/
\sigma_{\kappa_\parallel}^2} .
\label{Pkap-par}
\ee
The perturbative expressions (\ref{Pk-par}) and (\ref{Pkap-par}) do not take into
account nonperturbative phenomena such as shell crossings, which can be
approximated using a simplified adhesion model whereby particles coalesce when
$\kappa_\parallel<0$. This is described by modifying the probability distribution
\be
{\cal P}^{\rm ad.}(\kappa_\parallel)= a_1 \, \Theta (\kappa_\parallel >0)
{\cal P}^{\parallel}(\kappa_\parallel)+ a_0 \, \delta_D(\kappa_\parallel) ,
\label{Pkap-ad}
\ee
where $a_{0,1}$ are determined by the constraints
$\lag 1 \rag_\parallel= \lag \kappa_\parallel \rag_\parallel=1$.
This provides a simplified account of the formation of pancakes (the first
nonperturbative structures on large scales, such as the ``walls'' around cosmic
voids or underdense regions), and it leads to the ``cosmic web'' power spectrum
\beqa
P_{\rm c.w.}(k) & = & \int \frac{\dd\Delta\vq}{(2\pi)^3} \;
\frac{1}{1+A_1} \nonumber \\
&& \hspace{-1.4cm} \times \; e^{-\frac{1}{2} k^2 (1-\mu^2) \sigma_{\perp}^2} \;
\biggl \lbrace e^{-\varphi_{\parallel}(-\ii k\mu\Delta q \, \sigma^2_{\kappa_{\parallel}})
/\sigma_{\kappa_{\parallel}}^2} + A_1 \nonumber \\
&& \hspace{-1.4cm} \!\! + \!\! \int_{0^+-\ii\infty}^{0^++\ii\infty} \frac{\dd y}{2\pi\ii} \;
e^{-\varphi_{\parallel}(y)/\sigma^2_{\kappa_{\parallel}}}
\left( \! \frac{1}{y} - \frac{1}{y\! + \! \ii k\mu\Delta q \, \sigma^2_{\kappa_{\parallel}}} \! \right)
\!\! \biggl \rbrace \nonumber \\
&& \label{Pk-cw}
\eeqa
where $A_1=(1-a_1)/a_1$. The power spectra (\ref{Pk-par}) and (\ref{Pk-cw})
are identical to all orders of perturbation theory, and only differ by nonperturbative
corrections of the form $e^{-1/\sigma^2}$ associated with the adhesion-like
modification (\ref{Pkap-ad}).

To go to highly nonlinear scales, we use the halo model and the power spectrum
is split over one-halo and two-halo components as in Eq.(\ref{Pk-halos}).
Then, the probability that two particles belong to the same halo is \cite{Valageas2011d}\be
F_{\rm 1H}(\Delta q)= \int_{\nu_{\Delta q/2}}^\infty \frac{d\nu}{\nu} f(\nu) \frac{(2q_M - \Delta q)^2(4q_M +\Delta q)}{16 q_M^3} ,
\ee
where $\nu= \delta_L(M)/\sigma_M$ as in Eq.(\ref{fnu-def}) and
$M= 4\pi\bar\rho q_M^3 /3$. The linear density contrast $\delta_L(M)$, which is the
one which leads to a halo of nonlinear density contrast 200, depends on $M$ as the modified gravity dynamics is scale dependent. The lower bound of the integral corresponds to the mass enclosed within a radius $\Delta q/2$. The probability of belonging to two halos is $F_{\rm 2H}=1-F_{\rm 1H}$.
Finally, the average of the component of the particle displacements which is associated
with small-scale virialized motions within halos reads as
\be
\lag e^{\ii \vk\cdot\Delta \vx} \rag^{\rm vir}_{\Delta q}= \left[ \frac{\int_0^{\nu_{\Delta q/2}} \frac{d\nu}{\nu} f(\nu) \tilde u_M(k)}{\int_0^{\nu_{\Delta q/2}} \frac{d\nu}{\nu} f(\nu)}
\right]^2 ,
\label{virial}
\ee
because we assume that virialized motions within two different halos are uncorrelated.
We have defined the Fourier transform of the halo profile as
\be
\tilde u_M(k)= \frac{\int \dd\vx \; e^{-i\vk\cdot\vx} \rho_M(x)}{\int \dd\vx \; \rho_M(x)} ,
\ee
where $M=\int \dd\vx \, \rho_M (x)$.

Then, the two-halo part $P_{\rm 2H}(k)$ of the power spectrum is given by
Eq.(\ref{Pk-2H-1}), where we recognize the ``cosmic web'' power spectrum
(\ref{Pk-cw}), to which we have added the factor $F_{\rm 2H}$, to avoid
double-counting with the one-halo term, and the small-scale motions factor
(\ref{virial}), to take into account the finite width of halos.
The one-halo part $P_{\rm 1H}(k)$ is given as usual by Eq.(\ref{Pk-1H}),
with the counterterm $\tW^2$ associated with mass and momentum conservation,
which ensures that $P_{\rm 1H}(k) \propto k^4$ at low $k$.
Again, this gives a ``halo-model'' power spectrum (\ref{Pk-halos}) which is identical
to Eq.(\ref{Pk-par}) at all orders of perturbation theory. In particular, thanks to the
choice (\ref{S3-def}), it agrees with standard perturbation theory up to one-loop
order (and contains partial terms at all higher orders, generated through the
function $\varphi_{\parallel}(y)$, as well as nonperturbative terms of the
form $e^{-1/\sigma^2}$).
These are all the ingredients which are necessary to evaluate the power spectrum in
our improved formulation of the halo model.

\bibliography{ref1}   

\end{document}